\documentclass[aip,pof,reprint,amsmath,amssymb]{revtex4-1}
\usepackage[pdftex,bookmarks=true,colorlinks=true,linkcolor=blue,citecolor=blue]{hyperref}
\usepackage{dcolumn}
\usepackage{bm}
\usepackage{ifpdf}
\ifpdf
\usepackage{etex}
\usepackage[pdftex]{graphicx}
\usepackage[update]{epstopdf}
\else
\usepackage[dvips]{graphicx}
\fi
 \usepackage{upgreek}

\usepackage{xspace}

\usepackage{textcomp}

\newcommand{\mrd}{\mathrm{d}}      
\DeclareMathOperator{\sgn}{sgn}

\newcommand{\tlag}{\Delta_t}               

\newcommand{\deltanu}{\delta_\nu}
\newcommand{\rhof}{\rho_f}

\newcommand{\Deltax}{\Delta_x}
\newcommand{\Deltaxp}{\Delta_{x,p}}
\newcommand{\Deltaz}{\Delta_z}
\newcommand{\Deltazp}{\Delta_{z,p}}

\newcommand{\xvec}{\mathbf{x}}     
\newcommand{\xvecp}{\mathbf{x}_p}  

\newcommand{\Reb}{\mathrm{Re}_b}

\newcommand{\Retau}{\mathrm{Re}_\tau}

\newcommand{\ynull}{y_{0}}

\newcommand{\omegac}{\omega_{c}}
\newcommand\coline{\mbox{\protect\rule[0.5ex]{1cm}{.5pt}}}
\newcommand\daline{\mbox{\bf -- -- --}}
\newcommand\dadoline{\mbox{\bf-- $\cdot$ --}}

\newcommand\dodoline{\mbox{ $\mathbf{ \cdots \cdots}$}}

\newcommand{\Uc}{U_{c}}
\newcommand{\uc}{u_c}

\newcommand{\Ubh}{U_{bh}}
\newcommand{\utau}{u_\tau}
\newcommand{\uprime}{u^\prime}
\newcommand{\vprime}{v^\prime}
\newcommand{\wprime}{w^\prime}
\newcommand{\pprime}{p^\prime}

\newcommand{\phihat}{\widehat{\phi}}
\newcommand{\phihatast}{\widehat{\phi}^\ast}

\newcommand{\Rmax}{\left|R_{\phi\psi}\right|_{\rm max}}
\newcommand{\Rhat}{\widehat{R}}
\newcommand{\period}{\mathcal{T}}

 \usepackage{verbatim}

\usepackage{tikz}
\usetikzlibrary{shapes,arrows,positioning,calc,through}
\tikzstyle{every picture}+=[remember picture]
\usetikzlibrary{patterns}

\definecolor{colorccb}{rgb}{0,0,1}
\definecolor{colormgv}{rgb}{0,0.33,0}
\definecolor{colormu}{rgb} {1,0,0}

\newcommand{\CCB}[1]{{\color{colorccb}#1}}

\newcommand{\CCBc}[1]{{\tt\color{colorccb}[#1]}}

\newcommand{\MGVc}[1]{{\tt\color{colormgv}[#1]}}
\newcommand{\MUc} [1]{{\tt\color{colormu} [#1]}}
\newcommand{\revision}[2]{#2}
\newcommand{\revisions}[1]{}

\date{\today}
\begin{document}
%
%
\title[%
Scales of force and torque on wall-mounted spheres in channel flow]{%
  Spatial and temporal scales of force and torque acting on
  wall-mounted spherical particles in open channel flow}  
\author{C.\ Chan-Braun}
\affiliation{Institute for Hydromechanics, Karlsruhe Institute of 
  Technology, 76131 Karlsruhe, Germany}
\author{M.\ Garc\'{\i}a-Villalba}
\affiliation{Bioingenier\'{\i}a e Ingenier\'{\i}a Aeroespacial, Universidad
  Carlos III de Madrid, Legan\'es 28911, Spain}
\author{M.\ Uhlmann$^1$}
%
\thanks{Email address for correspondence: markus.uhlmann@kit.edu}
%
%
%
\begin{abstract}
Data from direct numerical simulation of open channel flow over a
geometrically rough wall at a bulk Reynolds number of $\Reb=2900$, 
generated by Chan-Braun et al.\ 
\citep{chanbraun_garciavillalba_uhlmann_JFM_2011} 
[``Force and torque acting on particles in a transitionally rough
open-channel flow'', J.\ Fluid Mech.\ 684, 441--474 (2011),
10.1017/jfm.2011.311]
are further 
analysed with respect to the time and length scales of force and torque  
acting on the wall-mounted spheres. 
For the two sizes of spheres in a square arrangement (11 and 49 wall
units in diameter, yielding hydraulically smooth and transitionally
rough flow, respectively), the spatial structure of drag, lift and
spanwise torque is investigated. 
The auto-correlation and spectra in time as well as 
the space-time correlation and convection velocities are 
presented and discussed.
It is found that the statistics of 
spanwise particle torque 
are similar to  
those of shear stress at a smooth wall.
Particle drag and lift are shown to differ from spanwise particle
torque, exhibiting considerably smaller time and length scales; the
convection velocities of drag and lift are somewhat larger than those
of spanwise torque.  
Furthermore, correlations between the flow field and
particle-related quantities are presented. 
%
The spatial structure of the correlation between streamwise velocity
and drag/spanwise torque features elongated shapes reminiscent of
buffer-layer streaks. 
The correlation between the pressure field and the particle drag
exhibits two opposite-signed bulges on the upstream and downstream
sides of a particle. 
\end{abstract}


\maketitle
\section{Introduction}
The present research is motivated by the problem of sediment
erosion in open channel flows such as rivers. 
\revision{%
  Sediment erosion is of interest 
  from the point of view of fundamental research 
  as well as for
  applied engineers wishing to predict 
  its onset, 
  e.g.\ in order to design pier protection.}
{%
  Other applications of processes involving sediment erosion can be
  found e.g.\ in aeolian transport (dune formation), 
  in the chemical industry (pneumatic conveying) and 
  in biological systems (blood flow, flow through the respiratory
  tract).  
  In order to be able to understand such complex systems a number of
  fundamental questions remain open.%
} 
At the core of the problem is the interaction between
a turbulent flow field and sediment particles.  
On the one hand turbulent flow induces a hydrodynamic force and torque
on a particle which can lead to the onset of particle motion, while on
the other hand the sediment alters the turbulence structure by acting
as a rough wall boundary.   
\revision{}{%
  Despite a long tradition of research on this subject 
  a detailed description 
  of the mechanisms responsible for the onset of erosion in turbulent
  flow is currently not available.  
}

The fundamental physics of the fluid-particle interaction have
commonly been studied in simplified systems, for example by
considering mono-sized spheres instead of naturally shaped gravel.
\revision{}{%
  Some researchers assume that the sediment particles are fixed in
  space and, using measurements of the hydrodynamic forces, attempt to 
  analyze the processes eventually leading to erosion. 
}
Data describing the force acting on submerged particles as part of a
rough wall are in fact relatively scarce
\citep{Einstein_Elsamni_RMP_1949, Hall_JFM_1988,
  Mollinger_Nieuwstadt_JFM_1996, Singh_sandman_williams_JHR_07}. 
Recent studies  present approximations of lift and drag on cubes,
spheres and naturally 
shaped stones by local pressure measurements
\citep{Hofland_etal_JHE_2005, Hofland_Battjes_JHE_2006, dwivedi_melville_shamseldin_2010, Detert_Weitbrecht_Jirka_JHE_2010}.
%
In 
Ref.~\onlinecite{chanbraun_garciavillalba_uhlmann_JFM_2011} 
high-fidelity data of open channel flow over an array of wall-mounted
spheres was generated by means of direct numerical simulation. 
\revision{%
  The focus in that study, where the flow was in the hydraulically
  smooth and transitionally rough regimes, was on the characterization
  of force and torque acting on the particles by means of analysis of
  the single-point statistics. 
  It was shown that some aspects of particle torque can
  be explained by an analogy to the force acting on a smooth-wall
  surface element of comparable size. 

  In the present work we determine 
  the spatial structure and the temporal characteristics of the
  hydrodynamic force and torque acting on wall-mounted spherical
  particles in open channel flow. 
  For this purpose we further analyze the data 
  of Ref.~\onlinecite{chanbraun_garciavillalba_uhlmann_JFM_2011}. 
  Since the particle force and torque are 
  generated by the time-varying velocity and pressure fields around the
  immersed objects, the force and torque signals should reflect some
  of the spatio-temporal characteristics of the turbulent flow in their
  vicinity. 
  As a consequence, it can be expected that the 
  particle force/torque and the flow field are substantially
  correlated. 
  Here we compute and analyze this correlation in an attempt to reveal
  the flow structures which significantly contribute to the generation
  of particle force and torque.  
}{%
  The focus in that study, 
  where two particle sizes were simulated (one leading to
  hydraulically smooth flow, one to transitionally rough flow),  
  was on the characterization of force and torque acting on the
  particles by means of an analysis of single-point and single-time
  statistics.  
  It was shown that some aspects of the statistics of particle torque
  (the shape of the probability density function and the magnitude of
  its moments) 
  can be explained by an analogy to the force/torque acting on a
  surface element of comparable size in an equivalent flow over a
  smooth wall.   
  The success of this simplified model indicates that 
  the spheres are acting as a filter with respect to the size of the
  flow scales which can effectively generate torque fluctuations.  
  %
  In Ref.~\onlinecite{chanbraun_garciavillalba_uhlmann_JFM_2011},
  however, several aspects of the interaction between the
  turbulent flow and the wall-mounted particles have not been
  addressed: 
  %
  the particle force and torque data was not analyzed with
  respect to its spatial and temporal correlation; 
  the relation between force/torque acting on the particles
  and the surrounding flow field was not investigated.
}

The structure of wall-bounded turbulent flow 
in general 
has been extensively
studied in the past.  
The role of coherent flow structures, in particular, has received
considerable attention 
\citep[cf.][]{Robinson_ARFM_1991, Panton_PAS_2001, 
  mckeon_sreenivasan_PTRS_2007, jimenez_ARFM_2012} 
as it has greatly improved our understanding of the underlying
flow dynamics. 
The particular effect of roughness on wall-bounded flows has been
reviewed 
in Ref.~\onlinecite{Jimenez_ARFM_2004}. 
The similarity between smooth-wall flows and those over rough walls 
-- when considering regions away from the wall -- 
has been investigated in detail in recent years  
\citep{Wu_Christensen_POF_2007,
  Volino_Schultz_Flack_JFM_2007,
  schultz_flack_POF_2009,
  Hong_Katz_Schultz_JFM_2011,
  Monty_allen_lien_chong_EIF_2011, 
  krogstad_efros_POF_2012}.  
On the other hand it is now generally accepted that the flow is
strongly affected by wall-roughness within a wall-distance of several
multiples of the roughness height.  

For smooth walls, several studies have tackled the problem of the
structure and propagation speed of the wall shear-stress and the wall
pressure  
\citep{Kim_JFM_1989, 
  choi_moin_POF_90,
  kravchenko_choi_moin_POF_1993,  kim_hussain_POF_1993, 
  jeon_etal_pof_99, 
  quadrio_luchini_pof_03, 
  delAlamo_Jimenez_JFM_2009, Hutchins_etal_JFM_2011}.
For flows over rough walls, similarly precise data 
on the spatial structure and propagation speed of 
the signature of flow quantities at the wall (e.g.\ the forces acting
on the roughness elements) 
are missing in the literature. 
Therefore, a considerable uncertainty exists regarding the flow
structures which contribute to the generation of these unsteady
forces. 

\revision{}{%
  With the objective to provide additional insight into the
  interaction between turbulent flow and near-wall particles 
  we determine in the present work 
  the spatial structure and the temporal characteristics of the
  hydrodynamic force and torque acting on wall-mounted spherical
  particles in open channel flow. 
  For this purpose we further analyze the data 
  of Ref.~\onlinecite{chanbraun_garciavillalba_uhlmann_JFM_2011}.
  Since the particle force and torque are 
  generated by the time-varying velocity and pressure fields around the
  immersed objects, the force and torque signals should reflect some
  of the spatio-temporal characteristics of the turbulent flow in their
  vicinity. 
  As a consequence, it can be expected that the 
  particle force/torque and the flow field are substantially
  correlated. 
  Here we compute and analyze this correlation in an attempt to reveal
  the flow structures which significantly contribute to the generation
  of particle force and torque.  
  %
  Both aspects of the process (the spatio-temporal correlation of
  particle force/torque and the correlation between force/torque
  and the surrounding flow field) are difficult to investigate in
  laboratory experiments. The present contribution fills this gap by
  providing previously unavailable data 
  which contributes to a better understanding of the action of
  turbulent flow upon complex shaped elements of a solid boundary. 
}

The present article is structured as follows. 
In \S~\ref{sec:num_setup} the setup of the simulations is described together
with a brief discussion of the numerical method. 
In \S~\ref{sec_struc} the structure of the force and torque acting on
the particles is presented. 
Two-point correlations are discussed in \S~\ref{sec_spatial}, 
temporal auto-correlations are discussed in \S~\ref{sec_temporal}, and
spatio-temporal correlations and convection velocities are discussed
in \S~\ref{sec_spatiotemporal}. In \S~\ref{sec_corr_flow_for}
correlations between the flow field and the 
force/torque acting on the particles are presented. 
Conclusions are given in \S~\ref{sec_concl}. 

\section{Flow configuration and notation} 
\label{sec:num_setup}
\begin{table}
\begin{center}
\begin{tabular}{lccccccccccc}
Case &   $U_{bh} / u_\tau $  & $Re_{b}$ &
$Re_\tau$ & $D^+$ & $D/h$ & $D/\Deltax $ & 
$\Deltax^+$ & $N_p$ &  $\uptau_c U_{bH} / H$
&
\revision{}{$N_{t1}$}&
\revision{}{$N_{t2}$}
\\ 
F10 &  15.2 & 2870 & 188 & 10.7 & 0.057& 14 & 0.77 & 9216 &
$120$ 
&\revision{}{$12750$}&\revision{}{$67$}
\\ 
F50 &  12.2 & 2880 & 235 & 49.3 & 0.210 & 46 & 1.07 & 1024 &
$120$ 
&\revision{}{$16500$}&\revision{}{$114$}
\end{tabular}
\caption{\label{tab:setup_param}%
  Setup parameters of simulations;
  $U_{bH}$ is the bulk velocity based on
  the domain height $H$, $U_{bh}$ is the bulk velocity based on
  the effective channel height $h$ defined as $h=H-0.8D$,
  $u_\tau$ is the friction velocity, 
  $Re_{b}=U_{bH}H/\nu$ is the bulk Reynolds number, $Re_\tau=u_\tau h
  / \nu$ is the friction Reynolds number, $D^+= D u_\tau/\nu$
  is the particle diameter in viscous units, $D / \Deltax$ 
  is the resolution of a particle, $\Deltax^+$ is the grid
  spacing in viscous units,
  $N_p$ is the total number of
  particles, 
  $\uptau_c$ is the time over which statistics were collected, 
  \revision{}{%
    $N_{t1}$ and $N_{t2}$ are the number of snapshots used in the
    computation of the force/torque correlations
    (\ref{eqn:corr_funct}) and in the computatgion of correlations
    between force/torque and the flow field
    (\ref{eqn:fix_scales_paflo_corr}), respectively.   
  }
} 
\end{center}
\end{table}
The flow configuration is identical to the one described in
Ref.~\onlinecite{chanbraun_garciavillalba_uhlmann_JFM_2011} 
and consists of 
pressure-driven open channel flow over a
geometrically rough wall.
%
The  wall is formed by one layer of fixed spheres 
packed in a square arrangement above a rigid wall located at 
$y=0$. 
\revision{}{%
  The number of particles in this layer is henceforth denoted as
  $N_p$.} 
This rigid wall is additionally roughened by spherical caps
(cf.\ figure~1 in
Ref.~\onlinecite{chanbraun_garciavillalba_uhlmann_JFM_2011}). 
%
The flow is studied through direct numerical
simulation of the incompressible Navier-Stokes equations. 
The equations are discretised in space by a second-order
finite-difference scheme on a staggered, equidistant grid.  
In time a three-step Runge-Kutta scheme is employed for the convective
terms, while the viscous terms are treated by a Crank-Nicolson
scheme. 
A fractional-step method is used to enforce the divergence-free
constraint. 
The particles forming the rough wall are represented through an
immersed boundary method \citep{Uhlmann_JCP_2005}.

As in 
Ref.~\onlinecite{chanbraun_garciavillalba_uhlmann_JFM_2011},  
a coordinate system is used such that $x$, $y$ and $z$ denote the
streamwise, wall-normal and spanwise direction respectively; 
\revision{}{%
  the velocity components in these three coordinate directions are
  $u$, $v$ and $w$. 
}
In the following, normalisation with wall units will be denoted by a
superscript $+$. A prime denotes the fluctuation 
with respect to an average over wall-parallel planes and time
(indicated by angular brackets), e.g.\ considering a quantity
$\phi(\mathbf{x},t)$, its fluctuation is defined as  
$\phi^\prime(\mathbf{x},t)=\phi(\mathbf{x},t)-\langle \phi \rangle(y)$. 
\revision{}{%
  The hydrodynamic force vector has the Cartesian components $F_x$,
  $F_y$ and $F_z$, while the components of the particle torque are
  denoted as $T_x$, $T_y$ and $T_z$.}

The physical and numerical
parameters of the simulations are provided 
in Table~\ref{tab:setup_param}. 
As in 
Ref.~\onlinecite{chanbraun_garciavillalba_uhlmann_JFM_2011},  
the value of the friction velocity $u_\tau$ is defined by
extrapolation of the total shear stress in the fluid 
away from the wall down to the location of the virtual wall, which is
defined at the plane $\ynull=0.8D$. 
The effective channel height is defined as $h=H-\ynull$ where $H$
denotes the domain height. 
Two cases are considered, differing only in the size of the
wall-mounted spheres: the spheres' diameter measures $D/H=0.055$ in case
F10, while it is set to $D/H=0.18$ in case F50. The corresponding value
normalized with wall units is $D^+=10.7$ ($D^+=49.3$) in case F10
(F50).  
%
The bulk velocity based on the domain height is defined
as $U_{bH}=1/ H \int_0^H \langle u \rangle \mathrm{d}y$ and the bulk
velocity based on the effective flow depth is defined as $U_{bh}= 1 /
h \int_{y_0}^H \langle u \rangle \mathrm{d}y$. 
\revision{%
  The Reynolds number computed from the bulk velocity and the 
  domain height, $Re_{b}=U_{bH}H/\nu$, is practically identical in both
  cases, measuring $2870$ and $2880$, respectively.}
{%
  In the present simulations the bulk velocity $U_{bH}$ was kept
  constant in time.  
  The chosen values of the Reynolds number defined from the bulk
  velocity and the domain height, $Re_{b}=U_{bH}H/\nu$, were
  practically identical in both cases, measuring $2870$ and $2880$,
  respectively. 
  On the other hand the Reynolds number based upon the friction
  velocity, $Re_\tau=hu_\tau/\nu$, takes a value close to that of an
  equivalent smooth-wall flow in case F10 ($Re_\tau=188$), while it is
  significantly higher in case F50 ($Re_\tau=235$).}

\begin{figure}
  \begin{center}
    \hspace{-.75\linewidth}
  \begin{minipage}{.90\linewidth}
    \includegraphics[width=1.\linewidth]{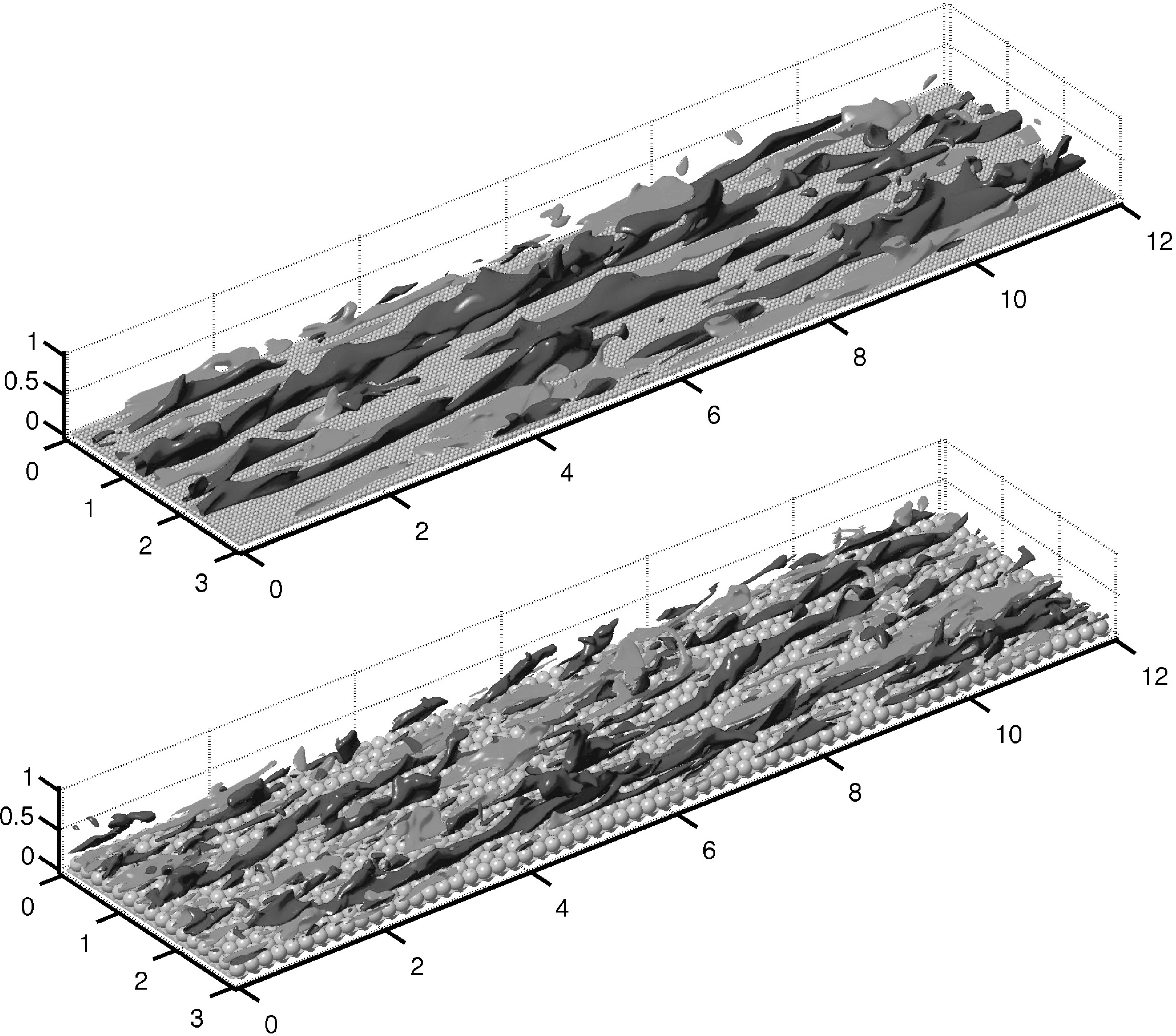}%
    \hspace*{-1.0\linewidth}
    \raisebox{0.3\linewidth}{($b$)}\hspace{-3ex}
    \raisebox{0.65\linewidth}{($a$)}\hspace{-3ex}
    \hspace{-.1\linewidth}
    \raisebox{0.175\linewidth}{$y/H$}\hspace{-5ex}
    \raisebox{0.55\linewidth}{$y/H$}\hspace{-6ex}
    \hspace{.66\linewidth}
    \raisebox{0.11\linewidth}{$x/H$}\hspace{-6ex}
    \raisebox{0.485\linewidth}{$x/H$}
    \hspace{-.65\linewidth}
    \raisebox{0.06\linewidth}
    {\rotatebox{-37}{$z/H$}\hspace{-6ex}
      \rotatebox{37}{}}
    \raisebox{0.42\linewidth}
    {\rotatebox{-37}{$z/H$}
      \rotatebox{37}{}}
  \end{minipage}

  \end{center}
  \caption{Instantaneous flow field in case F10 ($a$)
    and case F50 ($b$).
    Light (dark) surfaces show iso-surfaces of
    the streamwise velocity fluctuation at values $+ 3 u_\tau$
    ($-3u_\tau$).
  }  
  \label{fig:snapshot} 
\end{figure}

\begin{figure}
    \begin{minipage}{4ex}
      \rotatebox{90}{\hspace{4ex}$z/H$}
    \end{minipage}
    \begin{minipage}{.9\linewidth}
      ($a$)\\
      \includegraphics[width=1.\linewidth]
      {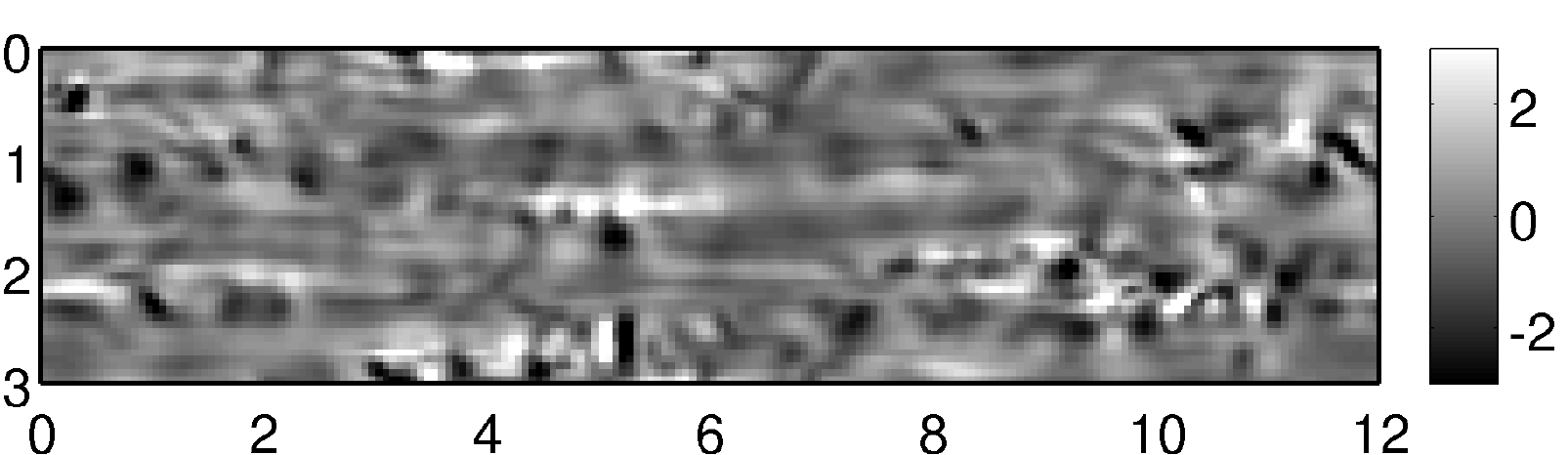}  
      \centerline{$x/H$}
    \end{minipage}
    \begin{minipage}{2ex}
      $\displaystyle\frac{F_x^\prime}{\sigma_{F_x}}$
    \end{minipage}
    \\
    \begin{minipage}{4ex}
      \rotatebox{90}{\hspace{4ex}$z/H$}
    \end{minipage}
    \begin{minipage}{.9\linewidth}
      ($b$)\\
      \includegraphics[width=1.\linewidth]
      {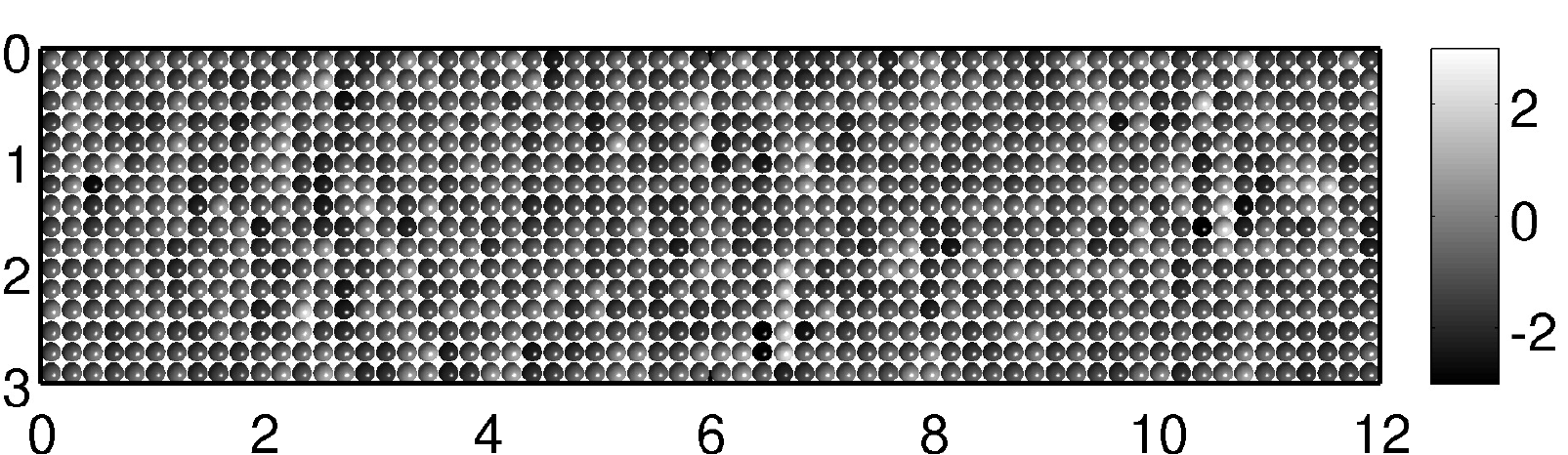}
      \centerline{$x/H$}
    \end{minipage}
    \begin{minipage}{2ex}
      $\displaystyle\frac{F_x^\prime}{\sigma_{F_x}}$
    \end{minipage}
  \caption{\label{fig:snapshot_part}%
    Instantaneous values of particle drag fluctuations 
    in case F10 ($a$) and case F50 ($b$).
    Note that in $(a)$ space-filling shaded squares are shown (for
    clarity), while in $(b)$ a projection of shaded spheres is
    represented. 
    %
  }  
\end{figure}

Fig.~\ref{fig:snapshot} contains snapshots of the 
flow fields 
for both particle sizes, showing iso-surfaces of streamwise velocity
fluctuations.  
In case F10 the spheres are small with respect to the domain 
dimensions and the scales of turbulent motion.  
In case F50 the spheres are approximately three times larger when
scaled in outer units, but they remain moderately small with respect
to the scales of motion.  
Fig.~\ref{fig:snapshot_part} illustrates the instantaneous particle
drag fluctuations in case F10 and case F50. In case F10 (small
spheres) 
the drag is found to correlate over several sphere
diameters in the streamwise and the spanwise directions 
(Fig.~\ref{fig:snapshot_part}$a$). 
In case F50 the picture is different, and a 
similarly 
strong correlation
cannot be inferred from Fig.~\ref{fig:snapshot_part}($b$). 
In both cases, extreme events appear very localised, i.e.\ events with
high particle drag fluctuations coexist next to events with high
negative particle drag fluctuations. 
As will be discussed below this leaves a footprint on the spatial
and temporal correlation functions.
%

\section{Spatial and temporal correlations of particle-related
  quantities}  
\label{sec_struc}
In the following the scales of spanwise torque, drag and lift are 
studied by correlation functions in time, space and space-time. 
The correlation function for particle
quantities related to an array of spheres in square
arrangement can be defined as
\revision{%
  \begin{equation}
    \label{eqn:corr_funct}
    R_{\phi\psi} (\Deltaxp,\Deltazp,\tlag) = \frac{1}{N_t N_p}
    \sum_{i=1}^{N_t} \sum_{l=1}^{N_p}  
    \phi^\prime(x^{(l)}_p, z^{(l)}_p,t_i) \
    \psi^\prime(x^{(l)}_p+\Deltaxp,z^{(l)}_p+\Deltazp, t_i+\tlag)\,, 
  \end{equation}
}{%
  \begin{equation}
    \label{eqn:corr_funct}
    R_{\phi\psi} (\Deltaxp,\Deltazp,\tlag) = \frac{1}{N_{t1} N_p}
    \sum_{i=1}^{N_{t1}} \sum_{l=1}^{N_p}  
    \phi^\prime(\mathbf{x}^{(l)}_p,t_i) \
    \psi^\prime(x^{(l)}_p+\Deltaxp,y^{(l)}_p,z^{(l)}_p+\Deltazp, t_i+\tlag)\,, 
  \end{equation}%
}
where 
$\phi^\prime(\mathbf{x}^{(l)}_p,t)$ and
$\psi^\prime(\mathbf{x}^{(l)}_p,t)$ are
fluctuations of force or torque components on 
the $l$th sphere,
centred at $\mathbf{x}^{(l)}_p$ at time, $t$.
Separation in time is denoted by $\tlag$, separations in the spanwise 
and the streamwise direction by  $\Deltaxp$ and $\Deltazp$,
respectively.  
%
The subscript $p$
is used to stress the discrete nature of separations in space which
can only be multiples of the particle distance. 
\revision{%
  The number of time steps and particles under consideration are $N_t$
  and  $N_p$ respectively. 
  The correlations are based on 
  particle data collected
  during run-time 
  $\uptau_c$ (cf.\ Table~\ref{tab:setup_param}) and consists of a number
  of samples of the order of $10^4$ for each particle, 
  i.e.\ a total of 12750 (16500) samples in case F10 (F50). 
}{%
  $N_{t1}$ denotes the number of snapshots used in the evaluation of
  (\ref{eqn:corr_funct}), with the values for each case listed
  in Table~\ref{tab:setup_param}. 
}
\subsection{Spatial correlation of spanwise torque, drag and lift
 on an array of particles} \label{sec_spatial}
\label{sec:fix_scorr}
\begin{figure}
    \begin{minipage}{3ex}
      \rotatebox{90}{$\Deltazp / h$}
    \end{minipage}
    \begin{minipage}{0.40\linewidth}
      ($a$)
      \hspace*{.2\linewidth}
      $\Deltaxp / D$
      \\
      \includegraphics[width=1.0\linewidth]
      {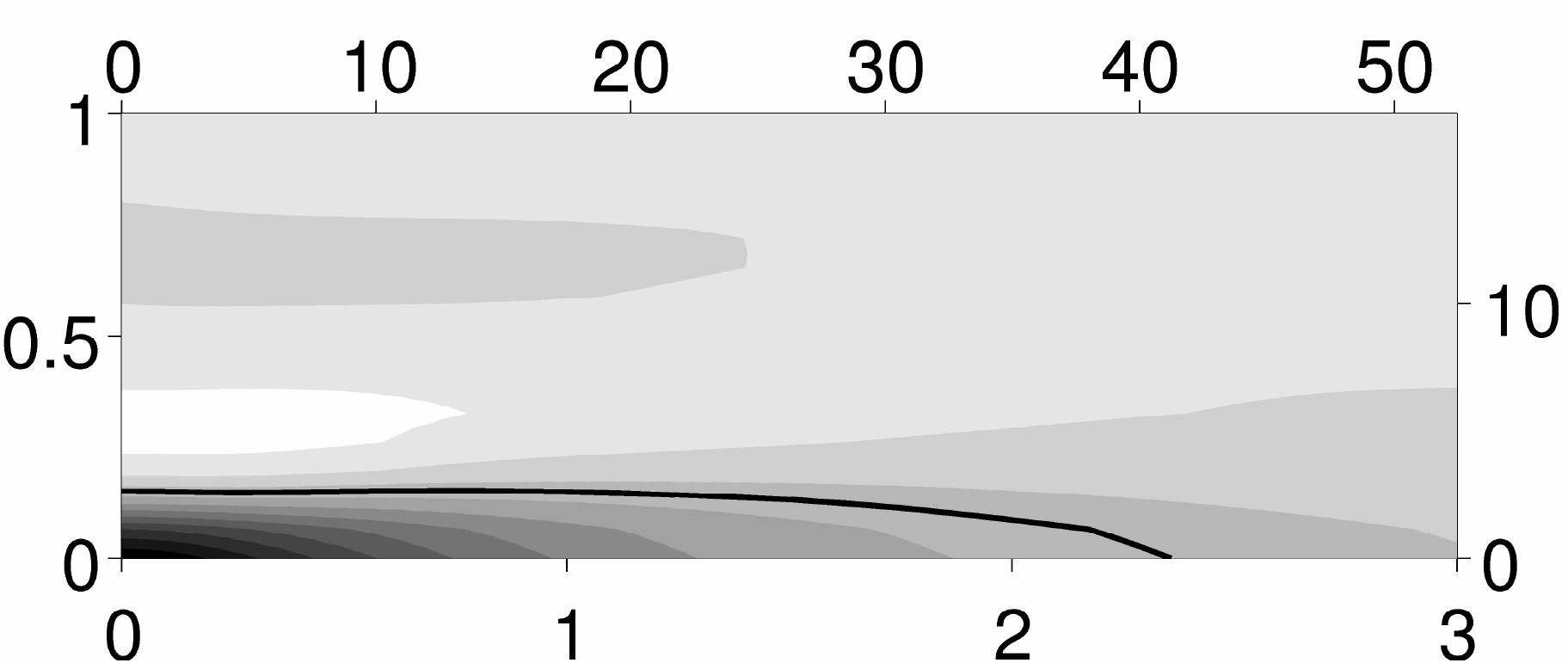}
    \end{minipage}
    \begin{minipage}{3ex}
      \rotatebox{90}{$\Deltazp / D$}
    \end{minipage}
    \hfill
    \begin{minipage}{3ex}
      \rotatebox{90}{$\Deltazp / h$}
    \end{minipage}
    \begin{minipage}{0.40\linewidth}
      ($b$)
      \hspace*{.2\linewidth}
      $\Deltaxp / D$
      \\
      \includegraphics[width=1.0\linewidth]
      {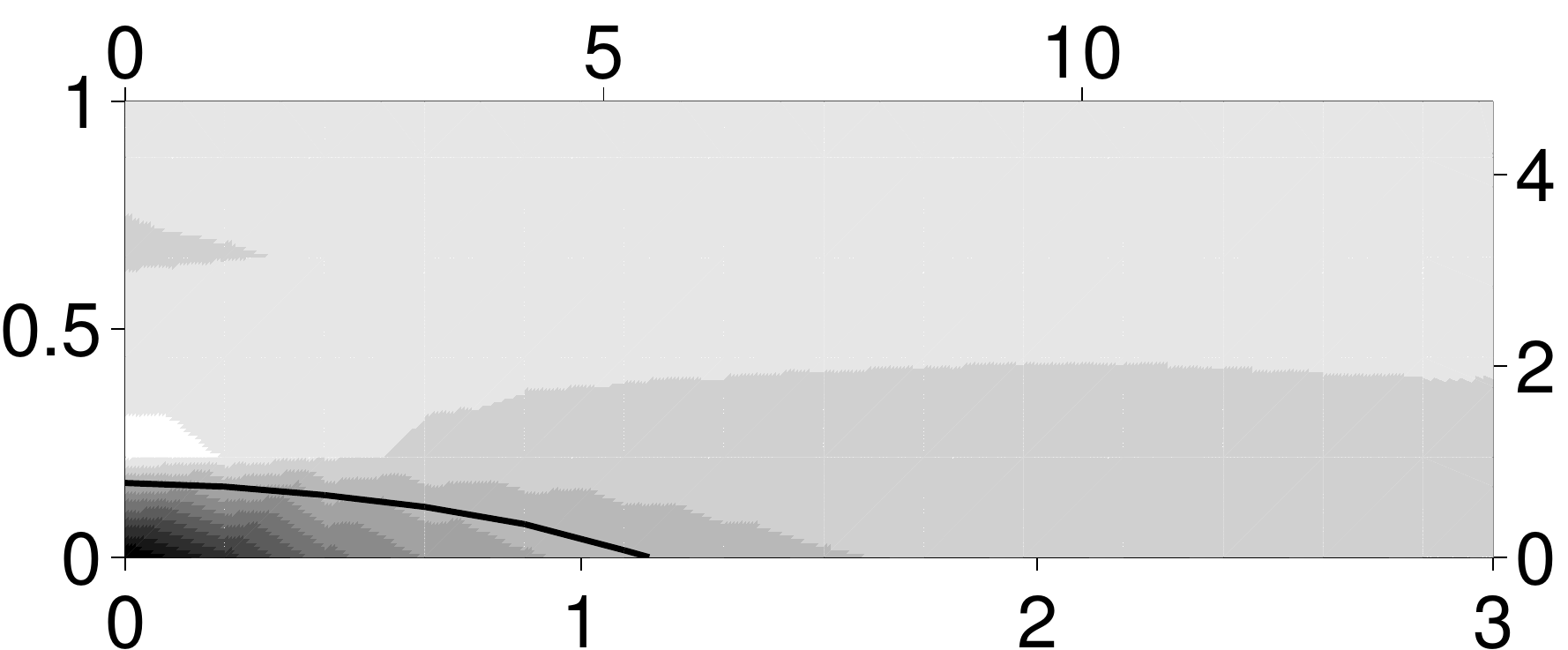}
    \end{minipage}
    \begin{minipage}{3ex}
      \rotatebox{90}{$\Deltazp / D$}
    \end{minipage}
    \\[3ex]
    \begin{minipage}{3ex}
      \rotatebox{90}{$\Deltazp / h$}
    \end{minipage}
    \begin{minipage}{0.40\linewidth}
      ($c$)
      \\
      \includegraphics[width=1.0\linewidth]
      {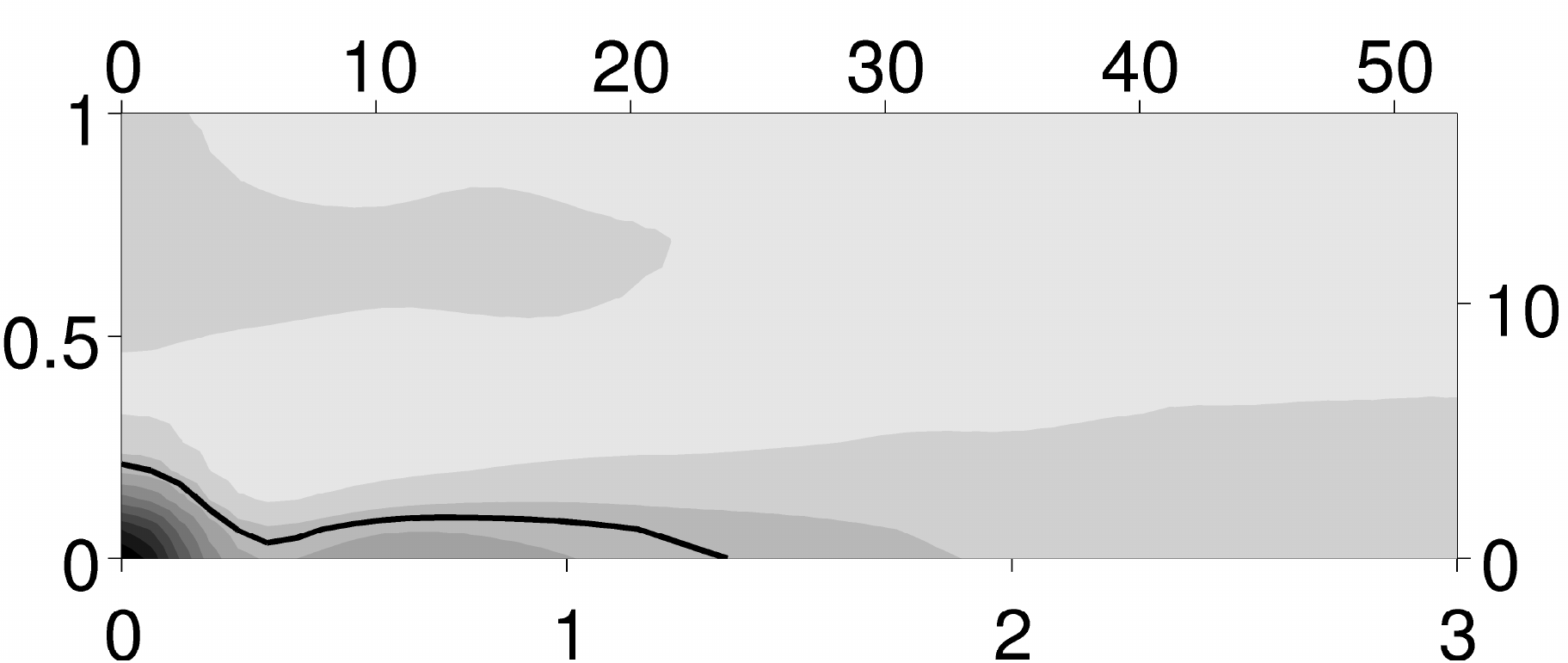}
    \end{minipage}
    \begin{minipage}{3ex}
      \rotatebox{90}{$\Deltazp / D$}
    \end{minipage}
    \hfill
    \begin{minipage}{3ex}
      \rotatebox{90}{$\Deltazp / h$}
    \end{minipage}
    \begin{minipage}{0.40\linewidth}
      ($d$)
      \\
      \includegraphics[width=1.0\linewidth]
      {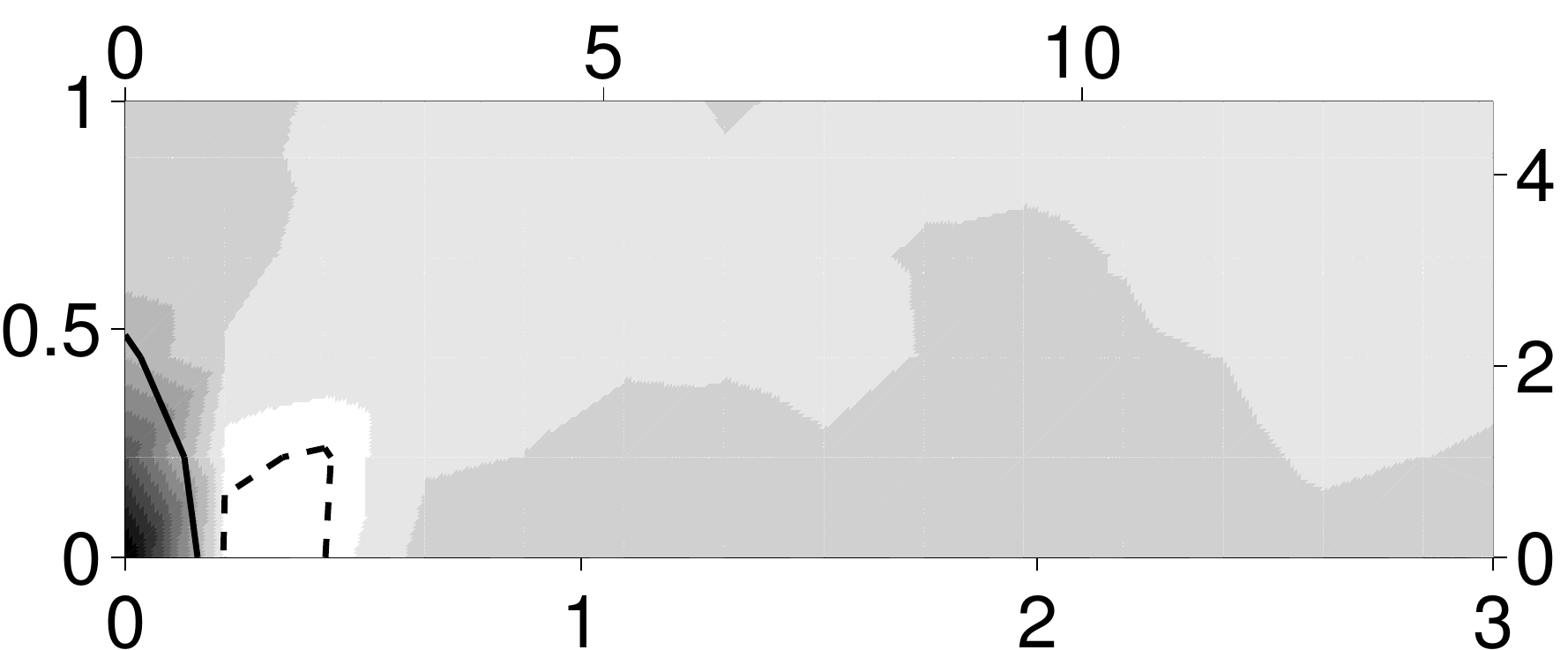}
    \end{minipage}
    \begin{minipage}{3ex}
      \rotatebox{90}{$\Deltazp / D$}
    \end{minipage}
    \\[3ex]
    \begin{minipage}{3ex}
      \rotatebox{90}{$\Deltazp / h$}
    \end{minipage}
    \begin{minipage}{0.40\linewidth}
      ($e$)
      \\
      \includegraphics[width=1.0\linewidth]
      {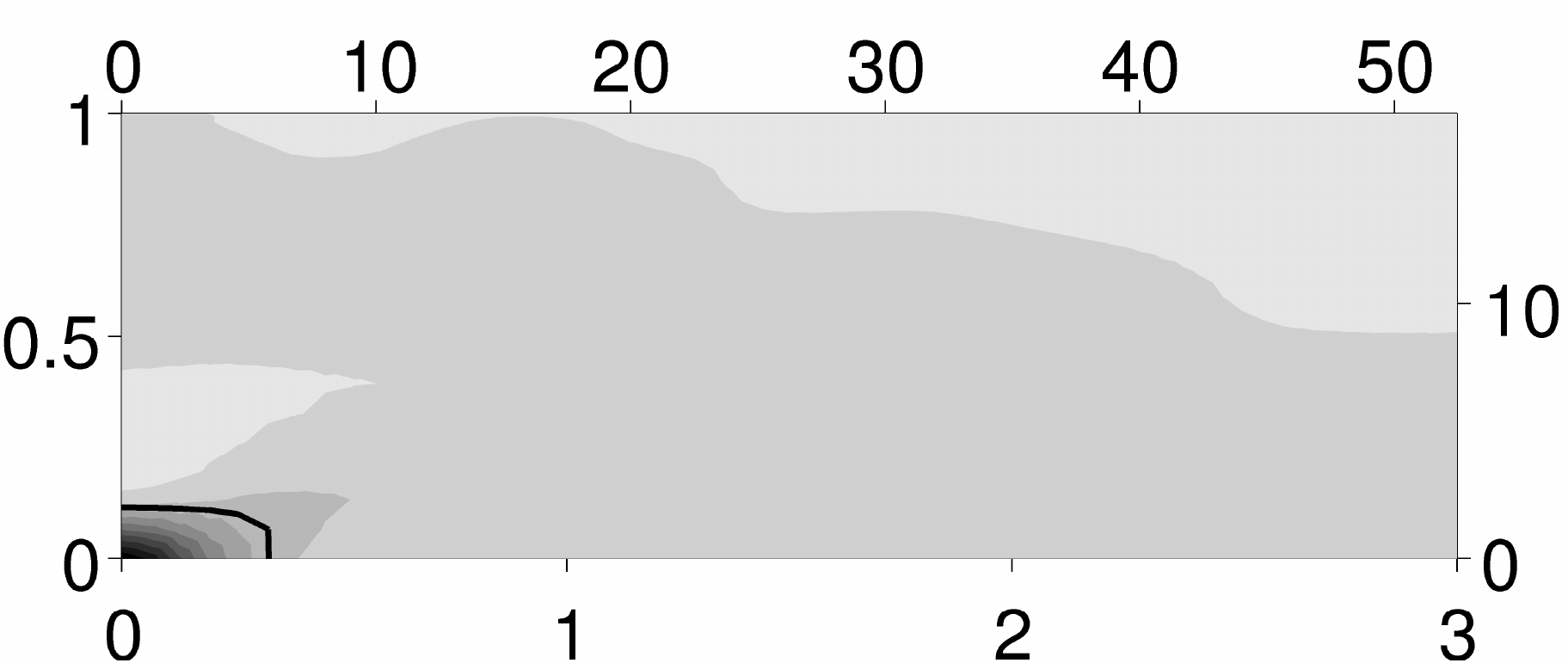}
      \centerline{$\Deltaxp / h$}
    \end{minipage}
    \begin{minipage}{3ex}
      \rotatebox{90}{$\Deltazp / D$}
    \end{minipage}
    \hfill
    \begin{minipage}{3ex}
      \rotatebox{90}{$\Deltazp / h$}
    \end{minipage}
    \begin{minipage}{0.40\linewidth}
      ($f$)
      \\
      \includegraphics[width=1.0\linewidth]
      {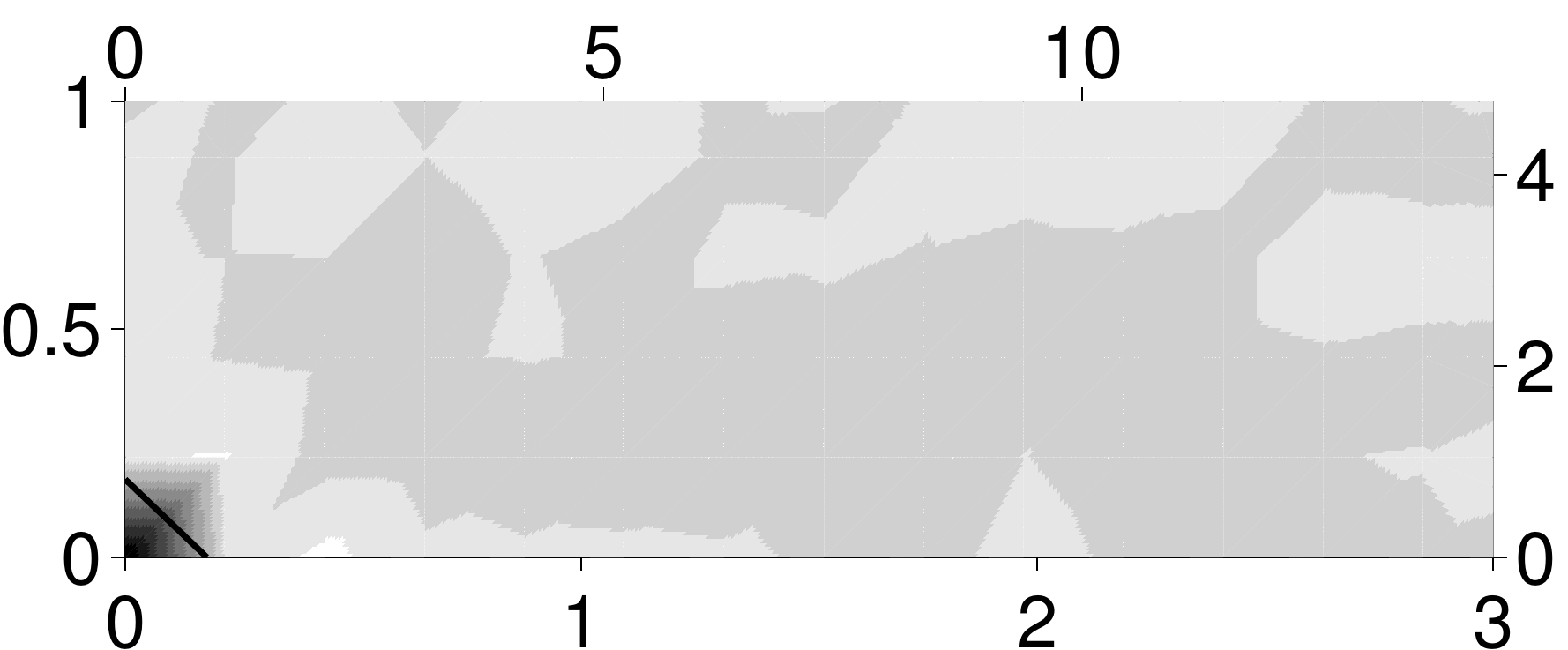}
      \centerline{$\Deltaxp / h$}
    \end{minipage}
    \begin{minipage}{3ex}
      \rotatebox{90}{$\Deltazp / D$}
    \end{minipage}

    \begin{minipage}{0.40\linewidth}
      \includegraphics[width=1.0\linewidth, clip, viewport=0 0  500 60]
      {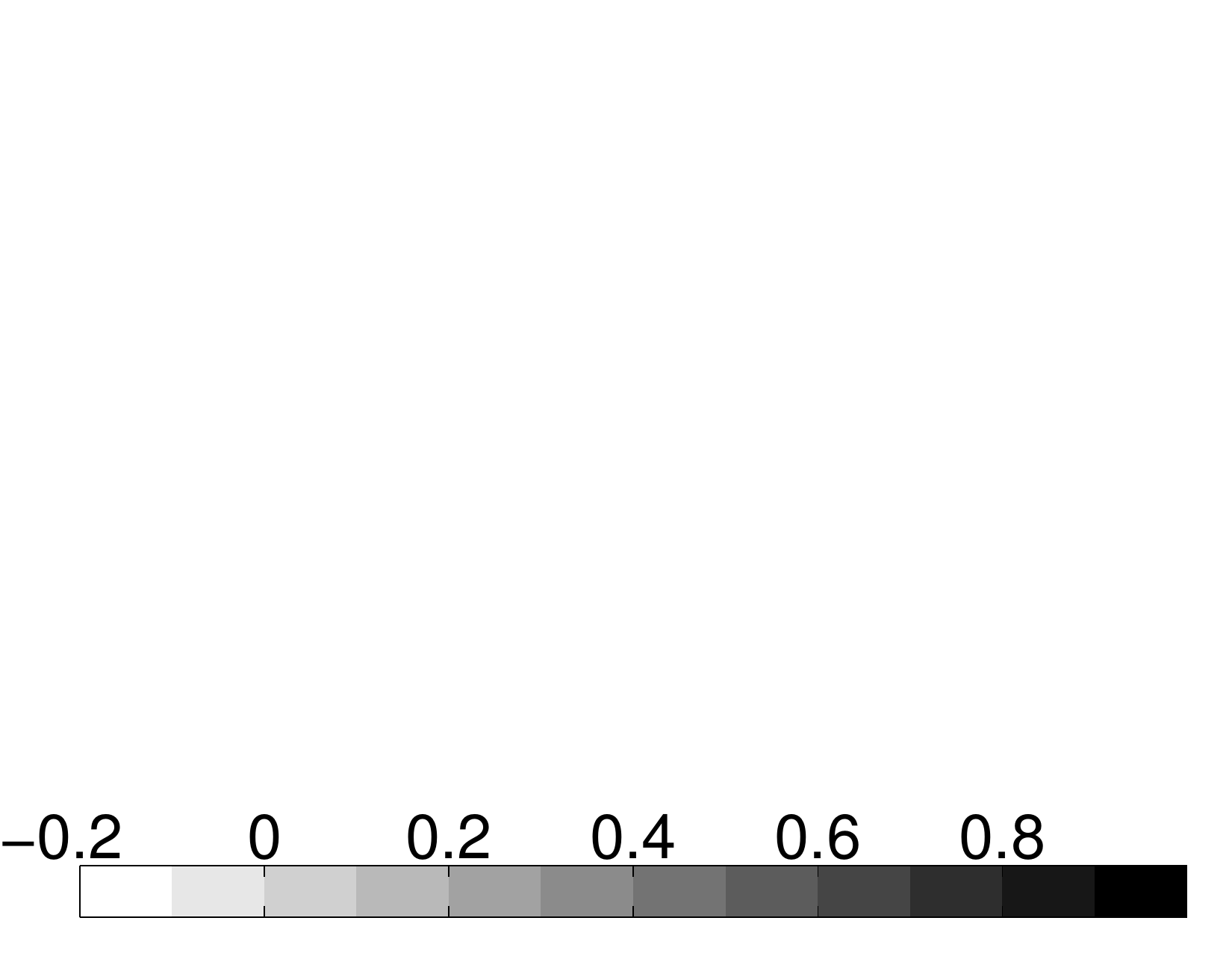}
      \centerline{$R_{\phi\phi}(\Deltaxp)/\sigma_\phi^2$}
    \end{minipage}
    \caption{\label{fig:auto_corr_2D}%
      Two-point correlation, $R_{\phi\phi}(\Deltaxp)/\sigma_\phi^2$, of
      particle torque and force fluctuation as a function of 
      space lag, $\Deltaxp$ and $\Deltazp$ scaled by $h$ and by $D$. 
      The panels show results of 
      case F10 ($a$,$c$,$e$) and case F50 
      ($b$,$d$,$f$). 
      Spanwise torque fluctuation, $T^\prime_z$ ($a$,$b$), 
      drag fluctuation, $F^\prime_x$ ($c$,$d$), 
      and lift fluctuations, $F^\prime_y$ ($e$,$f$).
      {The shading shows values from $-0.2$ to $1$ from white to
        black.}
      Lines show iso-contour at $0.15$ (\coline) and $-0.15$
      (\daline).
      %
    }
\end{figure}
\begin{figure}
    \begin{minipage}{3ex}
      \rotatebox{90}{$R_{\phi\phi}(\Deltaxp)/\sigma_\phi^2$}
    \end{minipage}
    \begin{minipage}{.45\linewidth}
      \centerline{$\Deltaxp/D$}
      \includegraphics[width=0.95\linewidth]
      {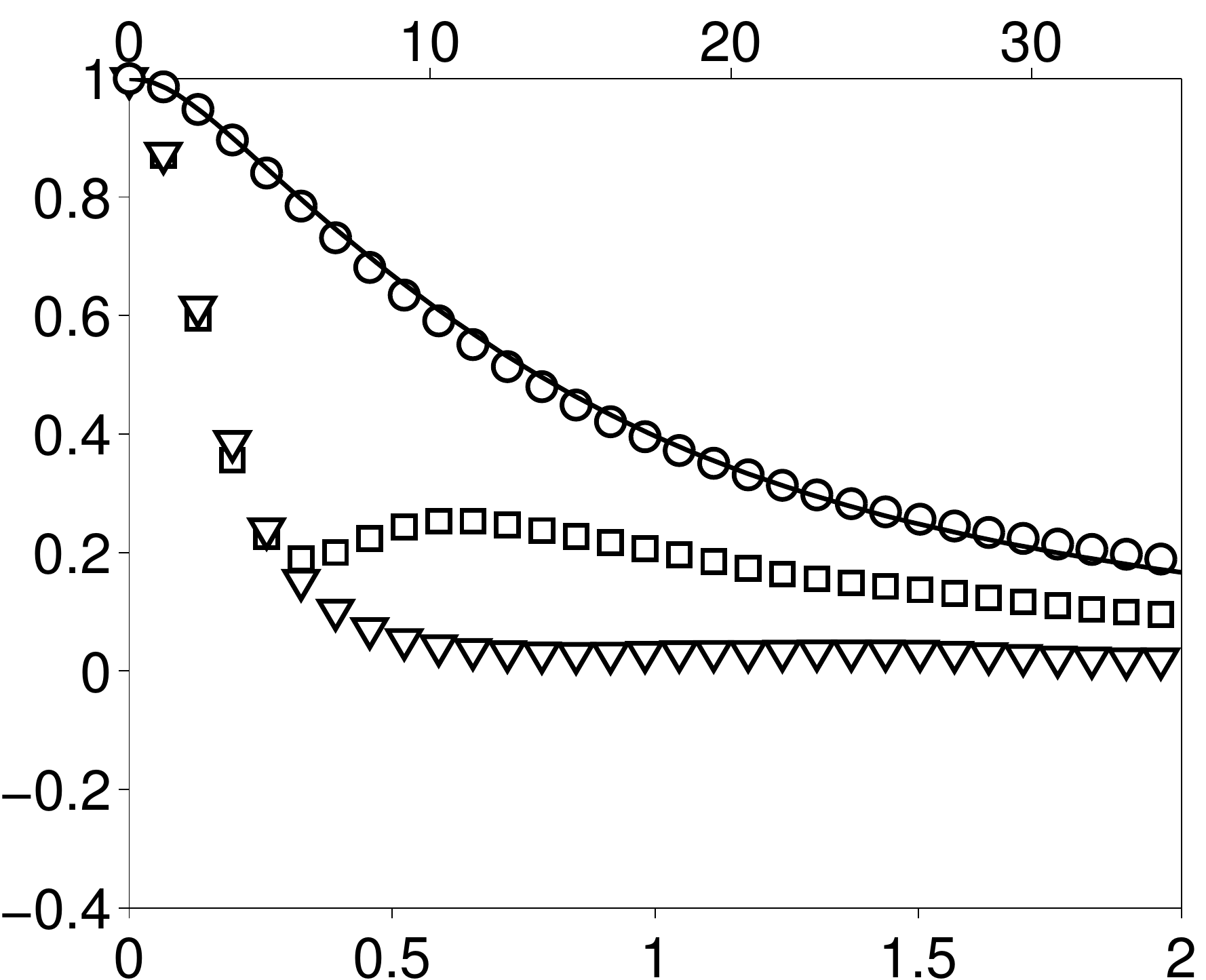}
      \hspace{-0.2\linewidth}\raisebox{0.55\linewidth}{($a$)}
      \\
      \centerline{$\Deltaxp / h$} 
    \end{minipage}
    \begin{minipage}{.45\linewidth}
      \centerline{$\Deltaxp/D$}
      \includegraphics[width=0.95\linewidth]
      {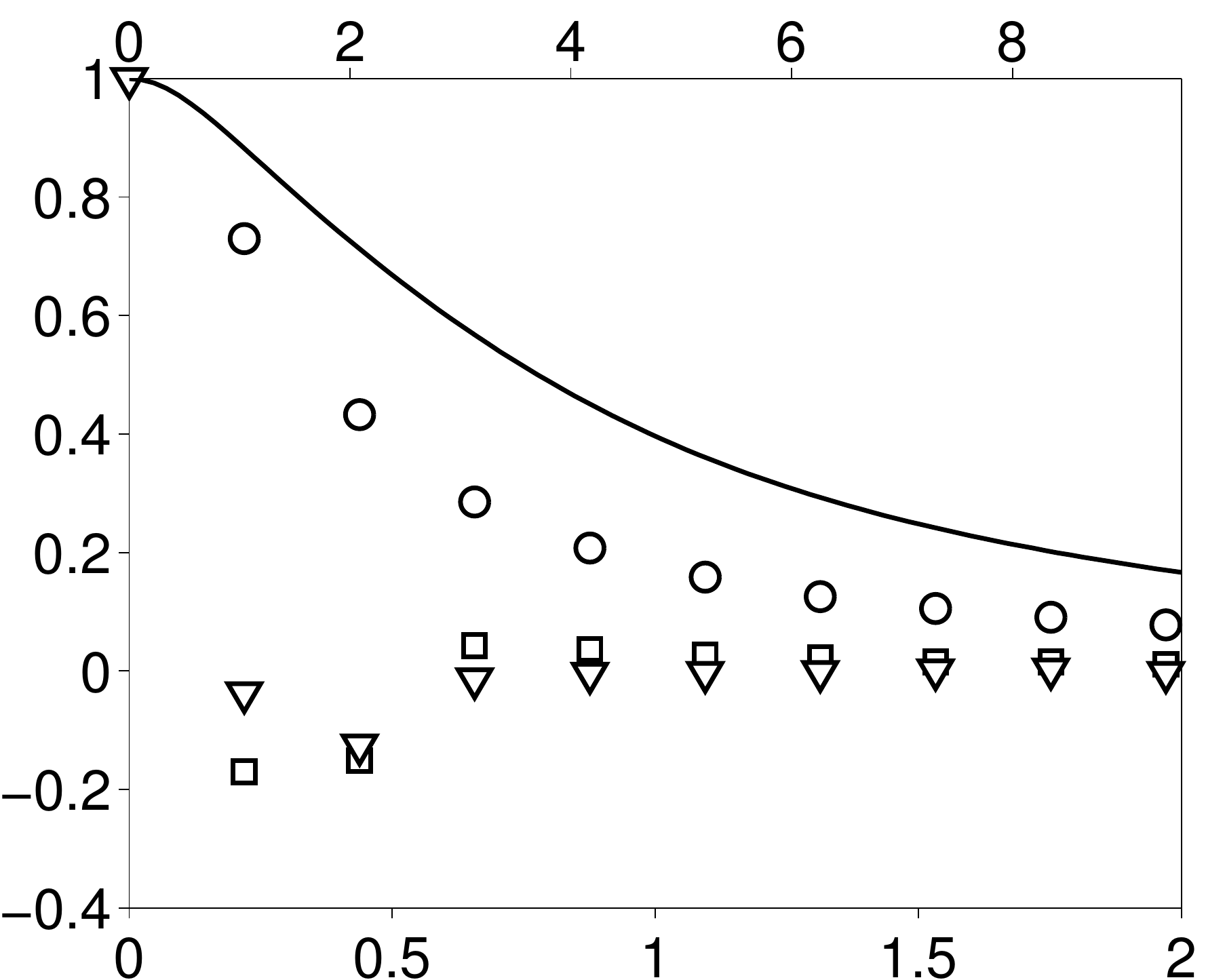}
      \hspace{-0.2\linewidth}\raisebox{0.55\linewidth}{($b$)}
      \\
      \centerline{$\Deltaxp / h$} 
    \end{minipage}
    \caption
    {Two-point correlation, $R_{\phi\phi} (\Deltaxp)/\sigma_\phi^2$,
      of drag ($F^\prime_x$, $\square$), 
      lift ($F^\prime_y$, $\triangledown$) 
      and spanwise torque ($T^\prime_z$, $\circ$) 
      fluctuations on a particle for
      streamwise separations $\Deltaxp$ in case of  
      case F10 ($a$) and case F50 ($b$).
      As a reference, the continuous line shows the 
      two-point correlation of shear stress fluctuation
      $\tau^\prime_{12}=\mu \left.\partial
        u^\prime /\partial y\right|_{y=0}$ 
      at a smooth wall as a function of
      $\Deltaxp / h$.} 
    \label{fig:auto_corr_x}
\end{figure}
\begin{figure}
    \begin{minipage}{3ex}
      \rotatebox{90}{$R_{\phi\phi}/\sigma_\phi^2$}
    \end{minipage}
    \begin{minipage}{.45\linewidth}
      \centerline{$\Deltazp / D$} 
      \includegraphics[width=0.95\linewidth]
      {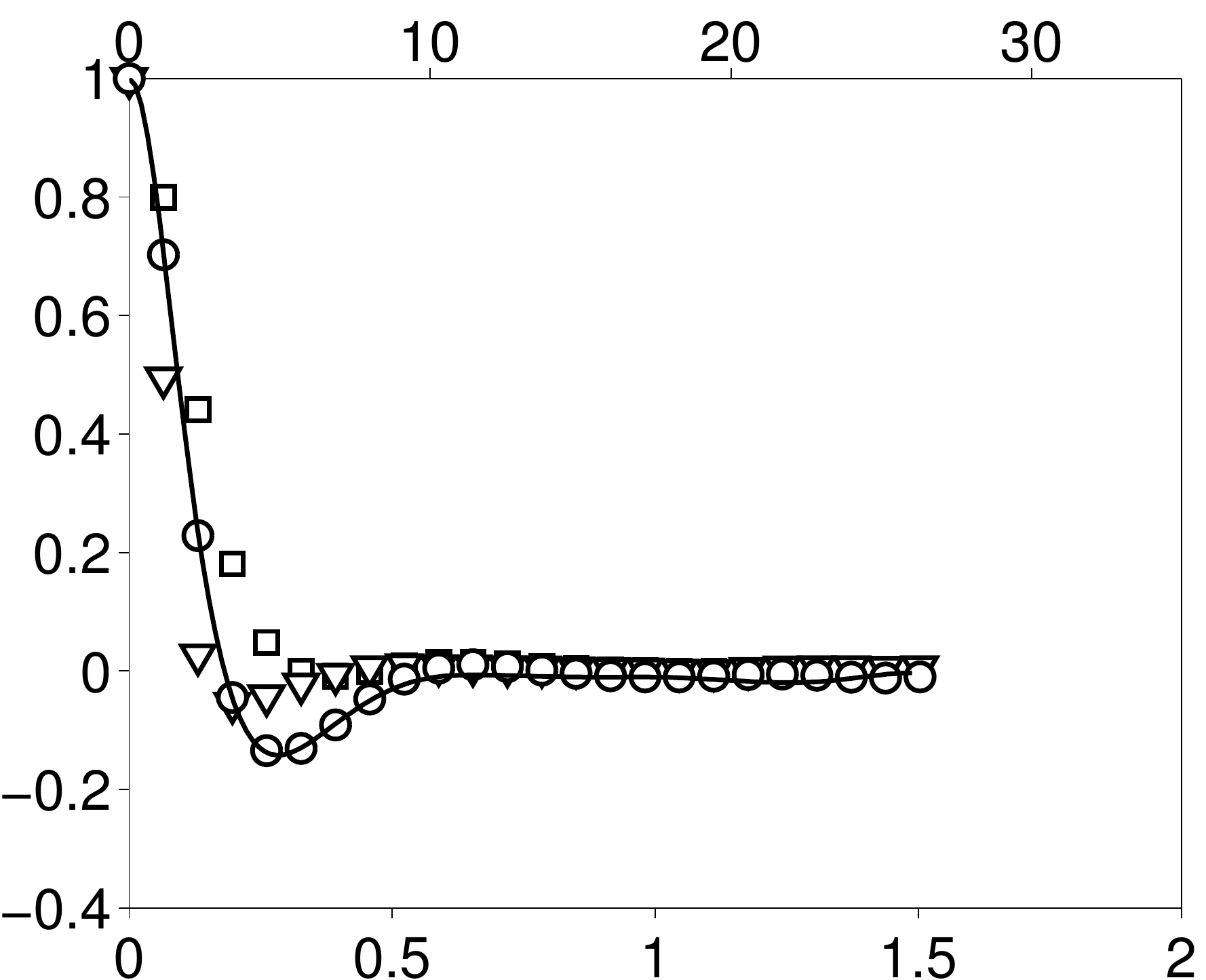}
      \hspace{-0.2\linewidth}\raisebox{0.55\linewidth}{($a$)}
      \\
      \centerline{$\Deltazp / h$} 
    \end{minipage}
    \begin{minipage}{.45\linewidth}
      \centerline{$\Deltazp / D$} 
      \includegraphics[width=0.95\linewidth]
      {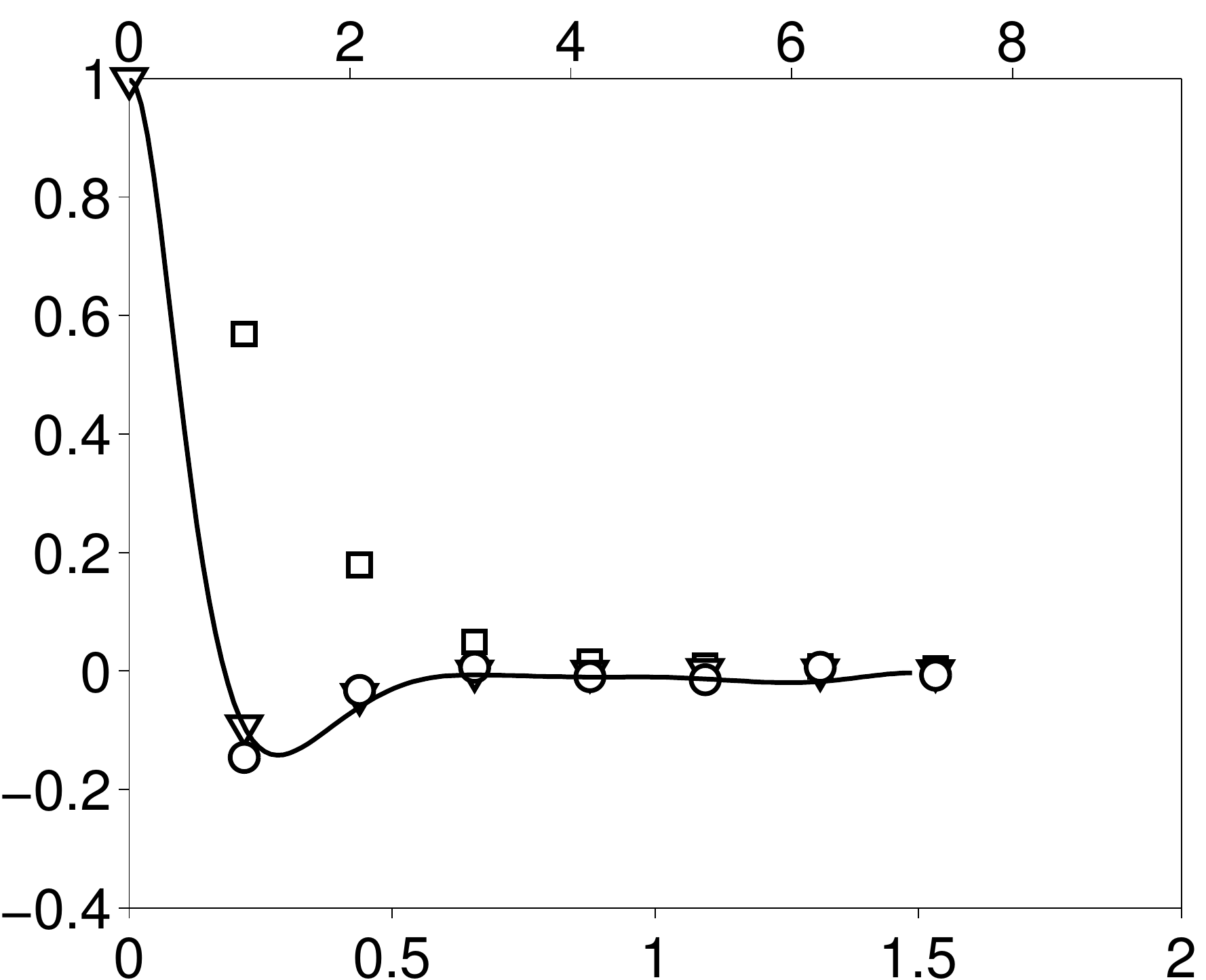}
      \hspace{-0.2\linewidth}\raisebox{0.55\linewidth}{($b$)}
      \\
      \centerline{$\Deltazp / h$} 
    \end{minipage}
    \caption{\label{fig:auto_corr_z}%
      Two-point correlation, $R_{\phi\phi} (\Deltazp)/\sigma_\phi^2$,
      of drag ($F^\prime_x$, $\square$), 
      lift ($F^\prime_y$, $\triangledown$) 
      and spanwise torque ($T^\prime_z$, $\circ$) 
      fluctuations on a particle for
      spanwise separations $\Deltazp$ in case of  
      case F10 ($a$) and case F50 ($b$).
      As a reference, the continuous line shows the
      two-point correlation of shear stress fluctuation
      $\tau^\prime_{12}=\mu \left.\partial
        u^\prime /\partial y\right|_{y=0}$ 
      at a smooth wall as a function of
      $\Deltazp / h$.} 
\end{figure}
\revision{%
  First the two-dimensional two-point correlation of spanwise torque,
  drag and lift fluctuations
  are considered as defined by (\ref{eqn:corr_funct})
  with $\phi=\psi$ and $\tlag=0$.}
{%
  First the two-dimensional two-point correlation of spanwise torque,
  drag and lift fluctuations
  are considered as defined by (\ref{eqn:corr_funct})
  with $\psi^\prime=\phi^\prime$,
  $\phi^\prime=\{T_z^\prime,F_x^\prime,F_y^\prime\}$ and $\tlag=0$.} 
Fig.~\ref{fig:auto_corr_2D} shows the correlation
coefficient 
of these quantities 
as a function of streamwise
and spanwise separation, $\Deltaxp$ and $\Deltazp$ normalised by $D$
and $h$ for both 
flow
cases. 
For spanwise torque Fig.~\ref{fig:auto_corr_2D}($a$,$b$) reveals a
streamwise elongated  
region of significant positive correlation which expands over several
particle diameters 
and is of the order of the effective domain height
$h$ in both 
flow
cases. 
Additionally a smaller region of negative correlation can
be identified centred around $\Deltazp/h=0.3$. 
While in case F10 the region is somewhat elongated in the streamwise
direction, in case F50 the negative region is confined to a small
{circular}
area. 
The two-point correlation of drag fluctuations shown in
Fig.~\ref{fig:auto_corr_2D}($c$,$d$) differ 
from 
those of spanwise torque.
In particular, in the spanwise direction an
area of significant negative
correlation is not visible. 
Also the region of positive correlation in case F10 is less elongated
and exhibits a spanwise contraction around $\Deltaxp/h=0.3$. 
In case F50 the two-point correlation exhibits an area of 
significant negative correlation centred at a 
streamwise separation of $\Deltaxp/h=0.3$ which is comparable in size 
to the area of positive correlation.
The two-point correlation 
of lift is the most confined quantity in both cases
(Fig.~\ref{fig:auto_corr_2D}$e$,$f$);   
in particular, in case F50 it vanishes after spanwise separations of 2
to 3 particle diameters. 
%

%
The spatial coherence of spanwise torque, drag and lift can be
further investigated by considering one-dimensional correlations 
along the coordinate axes. 
In this spirit, Fig.~\ref{fig:auto_corr_x} 
(Fig.~\ref{fig:auto_corr_z})
shows the same two-point correlations as a function of streamwise
(spanwise) separation at zero spanwise (streamwise) separation. 
%
Two 
scalings 
of the spatial separation are shown, 
using $h$ and $D$ as length scales. 
Additionally the two-point correlation of 
the wall shear-stress 
$\tau^\prime_{12} =\mu \left.\partial u^\prime /\partial y\right|_{y=0}$ 
in an equivalent flow with a smooth wall 
($\Reb=2900$, $\Retau=183$) 
is included in the figures 
as a function of $\Deltaxp / h$
for reference.
%
%
Fig.~\ref{fig:auto_corr_x} 
reveals that, when scaled by $h$,
the two-point correlation of spanwise torque fluctuation in case F10
overlaps with the one from the 
shear stress component $\tau_{12}^\prime$ in the smooth wall reference case.
In contrast, the  two-point correlation of spanwise torque in case F50 
is 
significantly lower.
However,the difference between the curves is small  
when scaled in viscous units (plot omitted). 
%
As discussed before with respect to Fig.~\ref{fig:auto_corr_2D}, the
two-point correlation of drag and lift fluctuations 
differ from the one of spanwise torque fluctuation, 
decreasing more rapidly with streamwise separations. 
In case F10 the  two-point correlations of drag
and lift fluctuations remain positive for the range of streamwise
separations shown. 
Here, the  two-point correlation of drag exhibits a local minimum located
at about $\Deltaxp/h=0.3$ 
after 
which the two-point correlation coefficient first increases and then
slowly approaches zero (Fig.~\ref{fig:auto_corr_x}$a$). 
The  two-point correlation coefficient of lift rapidly 
decays
and is essentially
zero for $\Deltaxp/h>0.5$.  
In case F50 (Fig.~\ref{fig:auto_corr_x}$b$) 
the two-point correlation coefficients of drag and lift
exhibit negative values at streamwise separations of
one and two particle diameters and are close to zero for
larger separations. 

Fig.~\ref{fig:auto_corr_z} shows the 
two-point correlation as a function of spanwise separation $\Deltazp$
again
with two axes showing the scaling by $D$ and by $h$.  
%
It is found that in 
both cases (F10 and F50),   
the  two-point correlation of spanwise torque
fluctuations 
overlaps with the two-point correlation of the 
smooth wall shear stress. 
The 
curves
rapidly drop towards a local minimum with negative
correlation coefficients. 
In all cases the
spanwise torque is essentially uncorrelated for spanwise separations
$\Deltazp/h>0.6$. 
In contrast to spanwise torque, the  two-point correlation of drag on
particles decreases monotonically with spanwise separation and
is essentially zero for $\Deltazp/D>5$. 
In case F10 the two-point correlation of lift fluctuation 
decreases faster in spanwise 
direction 
than drag or spanwise torque. 
%
In case F50, on the other hand, 
the two-point correlations of lift 
and spanwise torque nearly coincide, both exhibiting a significantly
more rapid initial decorrelation than particle drag. 
However, this does not imply that drag and lift might not be related
to similar flow structures, which will be discussed in more detail in
\S~\ref{ssec:crosscorr_drag_lift} below.
\subsection{Temporal correlation of force and torque on a particle} 
\label{sec_temporal}
\subsubsection{Auto-correlation of drag, lift and spanwise torque} 
\label{ssec:1D_autocorr}
\begin{figure}
  \begin{center}
    \begin{minipage}{3ex}
      \rotatebox{90}{$R_{\phi\phi}/\sigma_\phi^2$}
    \end{minipage}
    \begin{minipage}{.45\linewidth}
      \includegraphics[height=0.7\linewidth]
      {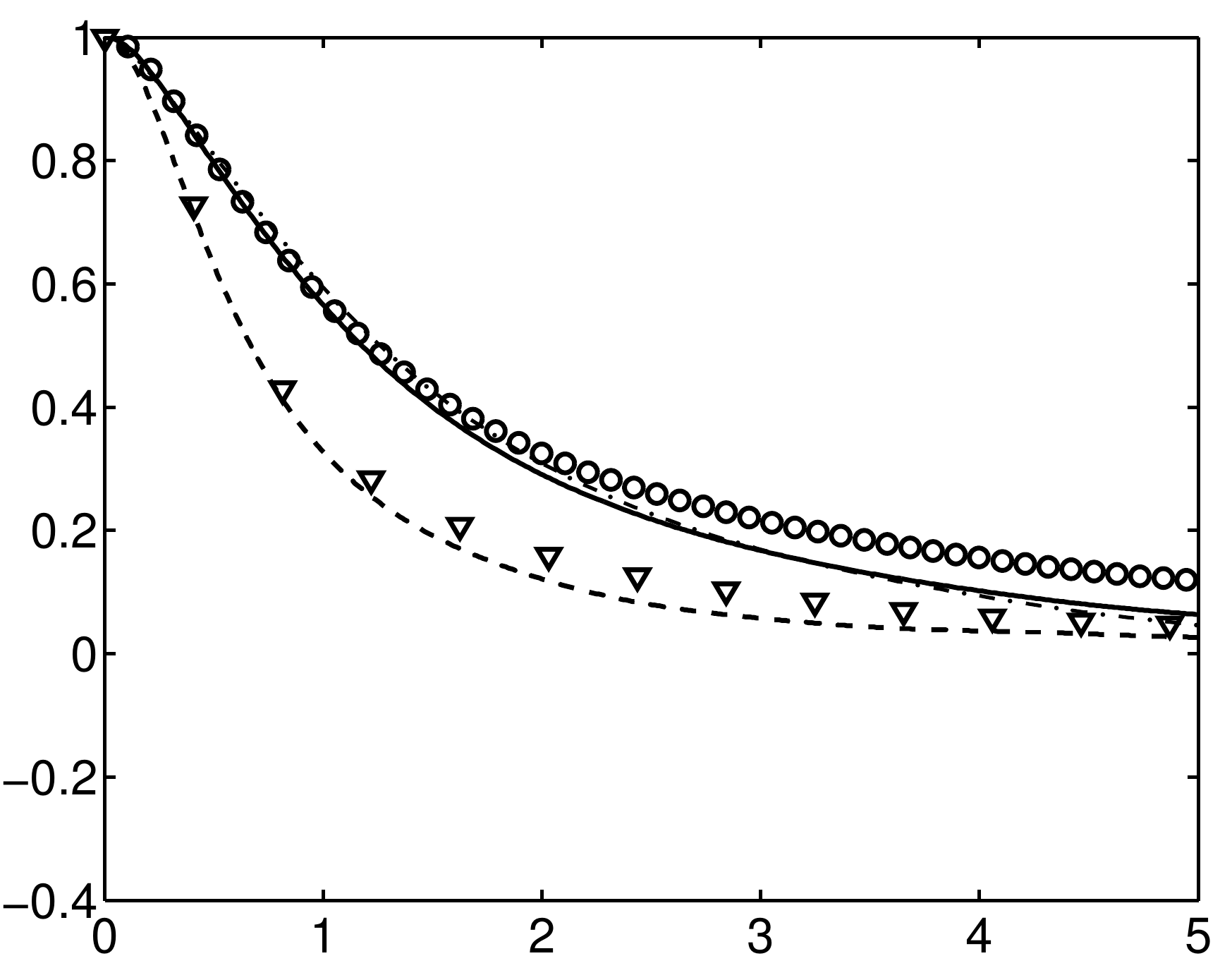}
      \hspace{-0.2\linewidth}\raisebox{0.55\linewidth}{($a$)}
      \\
      \centerline{$\tlag \Ubh/h$} 
    \end{minipage}
    \\[1ex]
    \begin{minipage}{3ex}
      \rotatebox{90}{$R_{\phi\phi}/\sigma_\phi^2$}
    \end{minipage}
    \begin{minipage}{.45\linewidth}
      \includegraphics[height=0.7\linewidth]
      {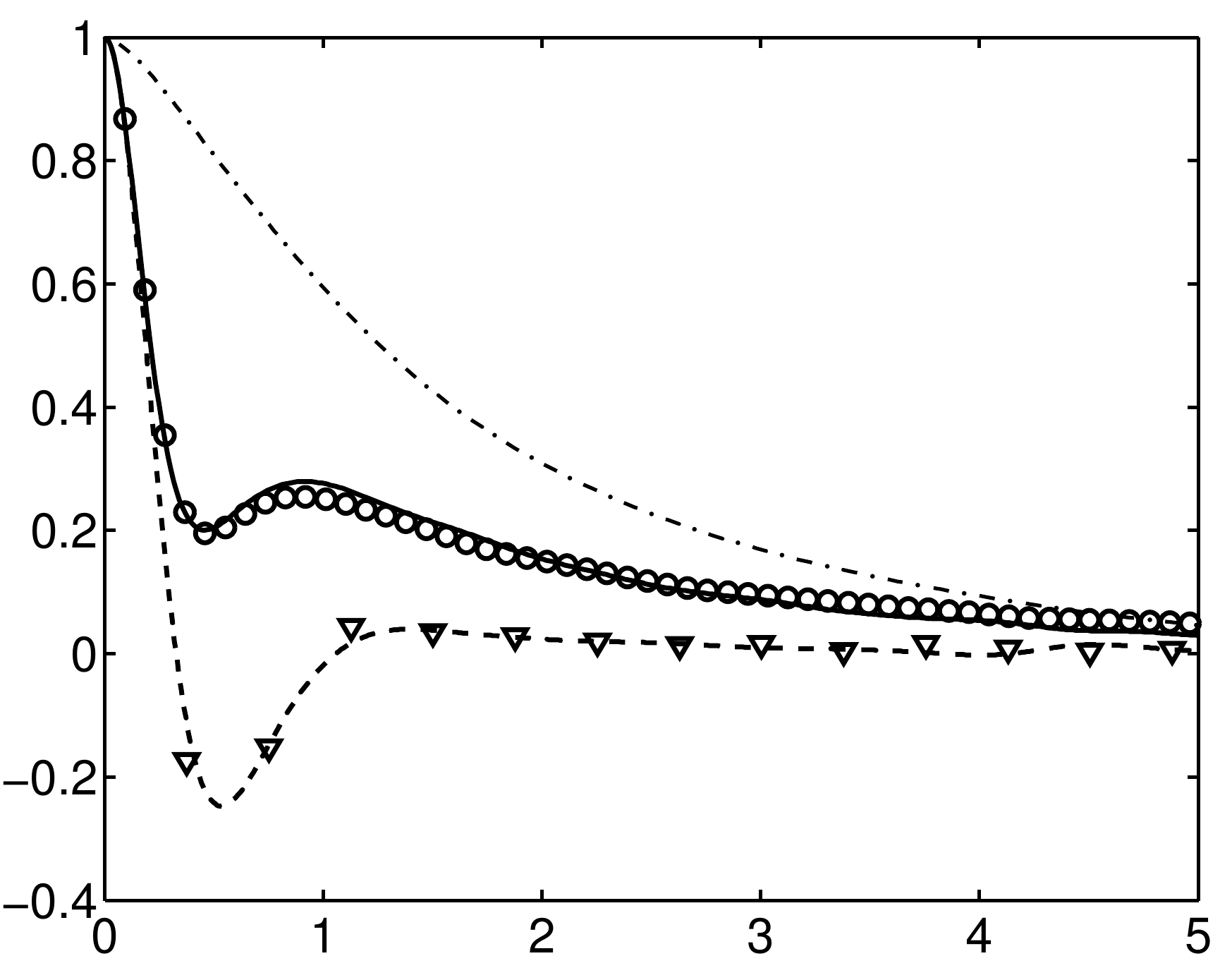}
      \hspace{-0.2\linewidth}\raisebox{0.55\linewidth}{($b$)}
      \\
      \centerline{$\tlag \Ubh/h$} 
    \end{minipage}
    \\[1ex]
    \begin{minipage}{3ex}
      \rotatebox{90}{$R_{\phi\phi}/\sigma_\phi^2$}
    \end{minipage}
    \begin{minipage}{.45\linewidth}
      \includegraphics[height=0.7\linewidth]
      {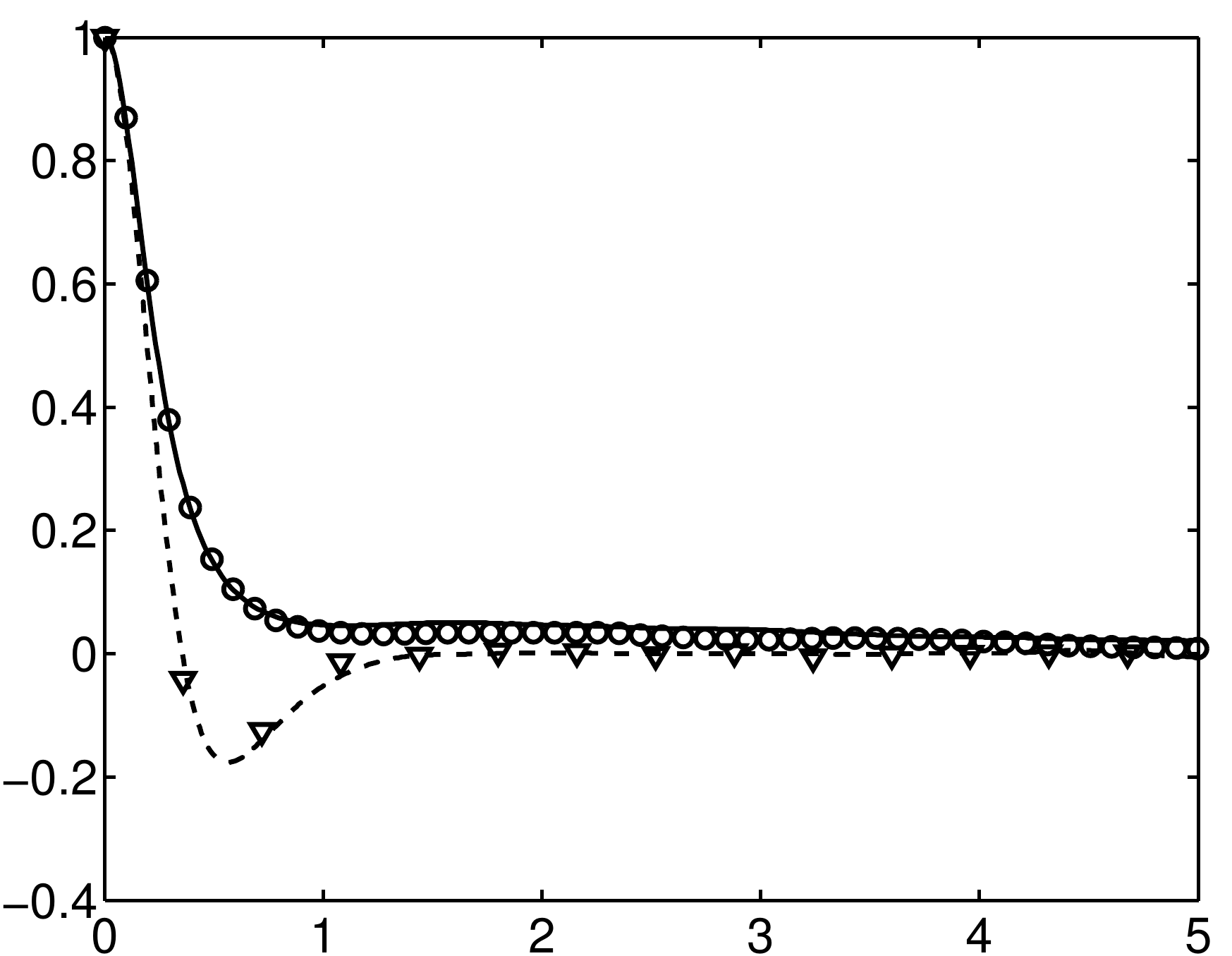}
      \hspace{-0.2\linewidth}\raisebox{0.55\linewidth}{($c$)}
      \\
      \centerline{$\tlag \Ubh/h$} 
    \end{minipage}
    \caption{\label{fig:auto_corr-all}
      Auto-correlation 
      of
      ($a$) spanwise torque fluctuations, $T^\prime_z$,
      ($b$) drag fluctuations, $F^\prime_x$ and
      ($c$) lift fluctuations, $F^\prime_y$. 
      Lines show the temporal auto-correlation 
      of case F10 (\coline) and  case F50 (\daline)
      as a function of the 
      time lag, $\tlag$. 
      Additionally, 
      the auto-correlation of the shear stress fluctuation
      $\tau^\prime_{12}=\mu \left.\partial u^\prime /\partial y\right|_{y=0}$  
      in the smooth wall reference simulation (\dadoline) is provided as a
      reference in ($a$) and ($b$).
      Symbols show the approximate 
      auto-correlation computed 
      from the spatial correlations in case F10 ($\circ$) 
      and F50 ($\triangledown$) by applying a Taylor 
      approximation using the convection velocity 
      $\Uc$, as discussed in \S~\ref{sec:stcorr}. 
      In all figures the bulk time is used to normalize the
      temporal separation.
      }
  \end{center}
\end{figure}
\begin{figure}
  \begin{center}
    \begin{minipage}{3ex}
      \rotatebox{90}{$\omega\widehat{R}_{\phi\phi}(\omega)/ \int
        \widehat{R}_{\phi\phi} \mrd \omega$}
    \end{minipage}
    \begin{minipage}{.45\linewidth}
      \includegraphics[height=0.7\linewidth]
      {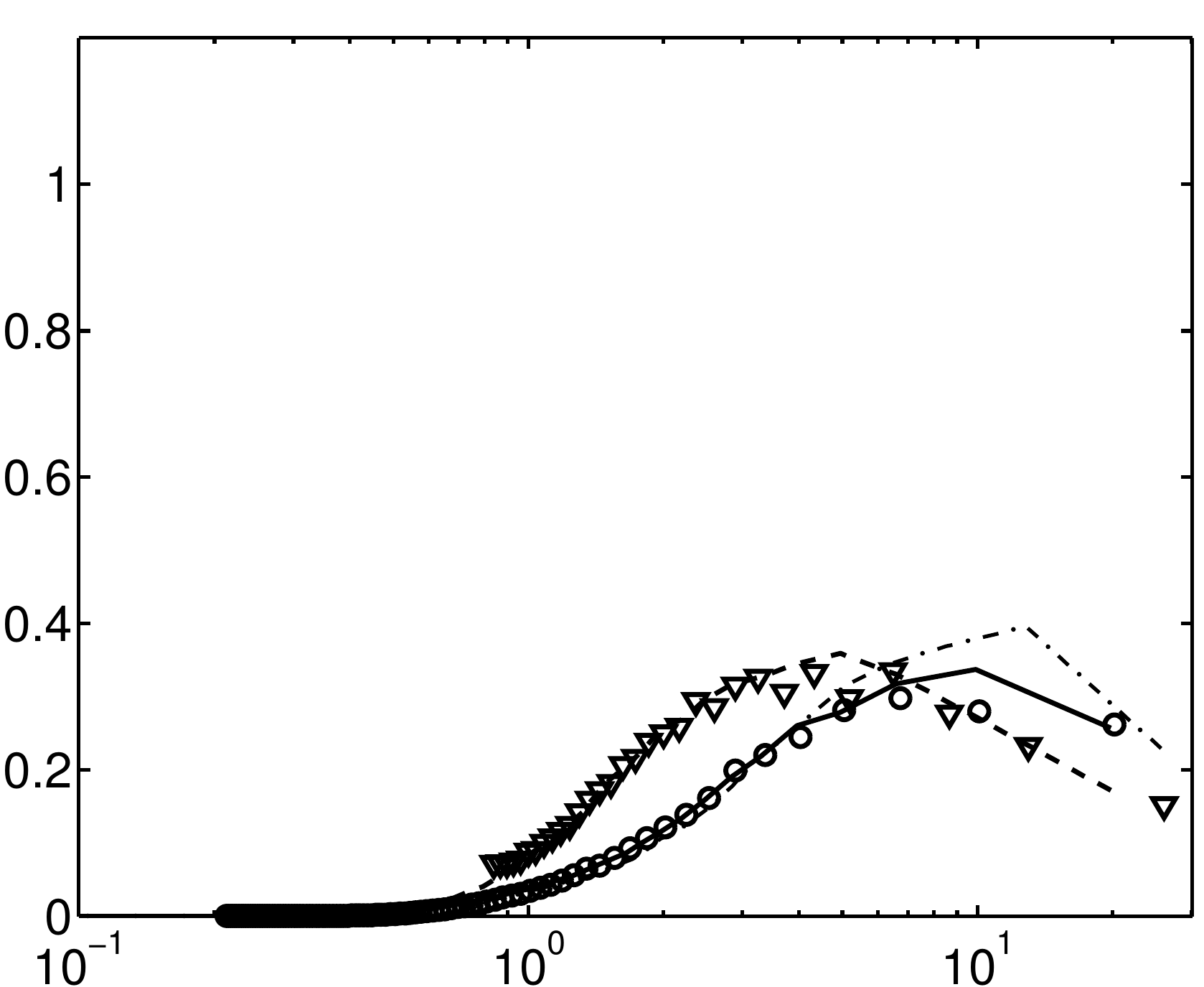}
      \hspace{-0.2\linewidth}\raisebox{0.55\linewidth}{($a$)}
      \\
      \centerline{$\period \Ubh/h$} 
    \end{minipage}
    \\[3ex]
    \begin{minipage}{3ex}
      \rotatebox{90}{$\omega\widehat{R}_{\phi\phi}(\omega)/ \int
        \widehat{R}_{\phi\phi} \mrd \omega$}
    \end{minipage}
    \begin{minipage}{.45\linewidth}
      \includegraphics[height=0.7\linewidth]
      {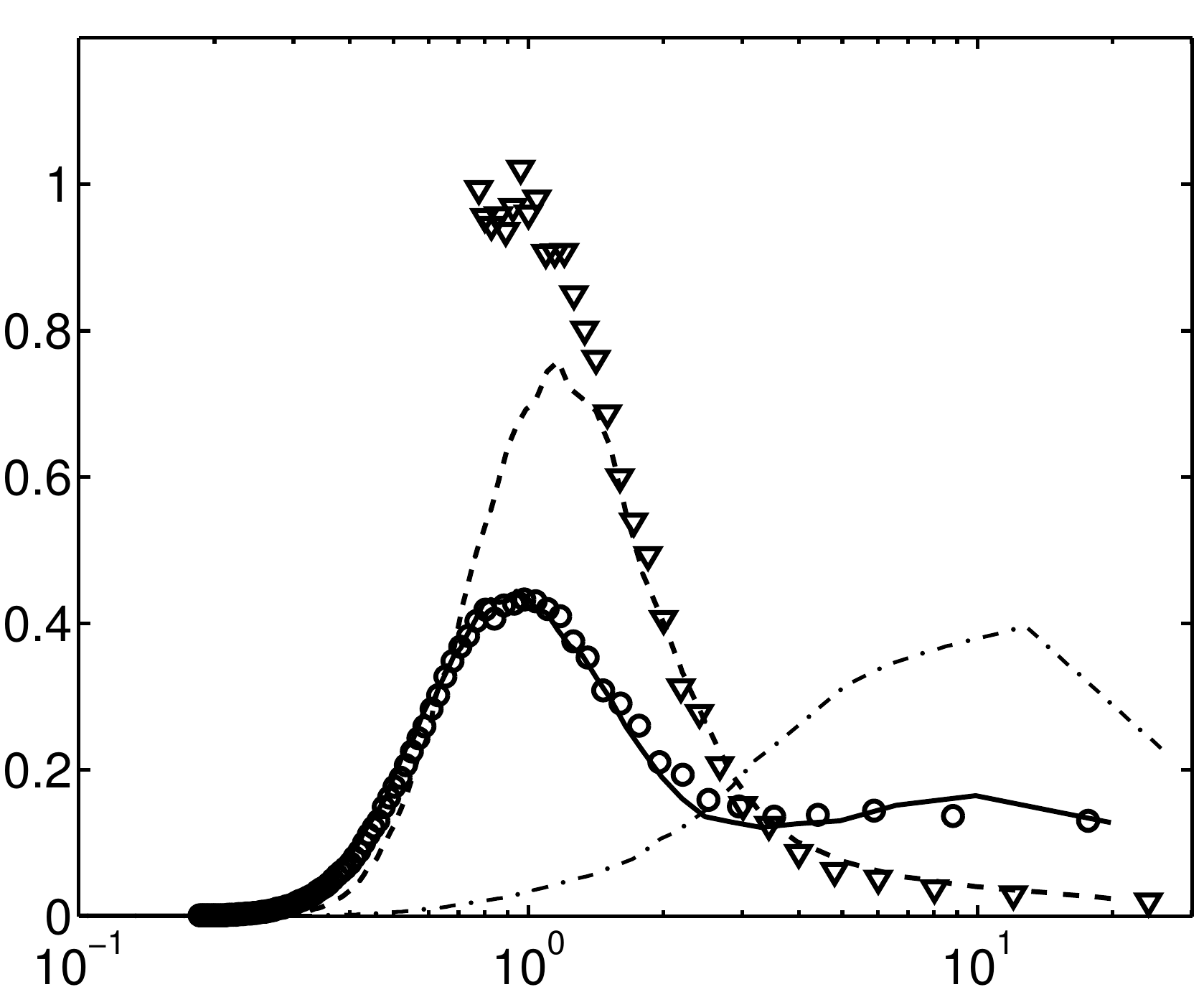}
      \hspace{-0.2\linewidth}\raisebox{0.55\linewidth}{($b$)}
      \\
      \centerline{$\period \Ubh/h$} 
    \end{minipage}
    \\[3ex]
    \begin{minipage}{3ex}
      \rotatebox{90}{$\omega\widehat{R}_{\phi\phi}(\omega)/ \int
        \widehat{R}_{\phi\phi} \mrd \omega$}
    \end{minipage}
    \begin{minipage}{.45\linewidth}
      \includegraphics[height=0.7\linewidth]
      {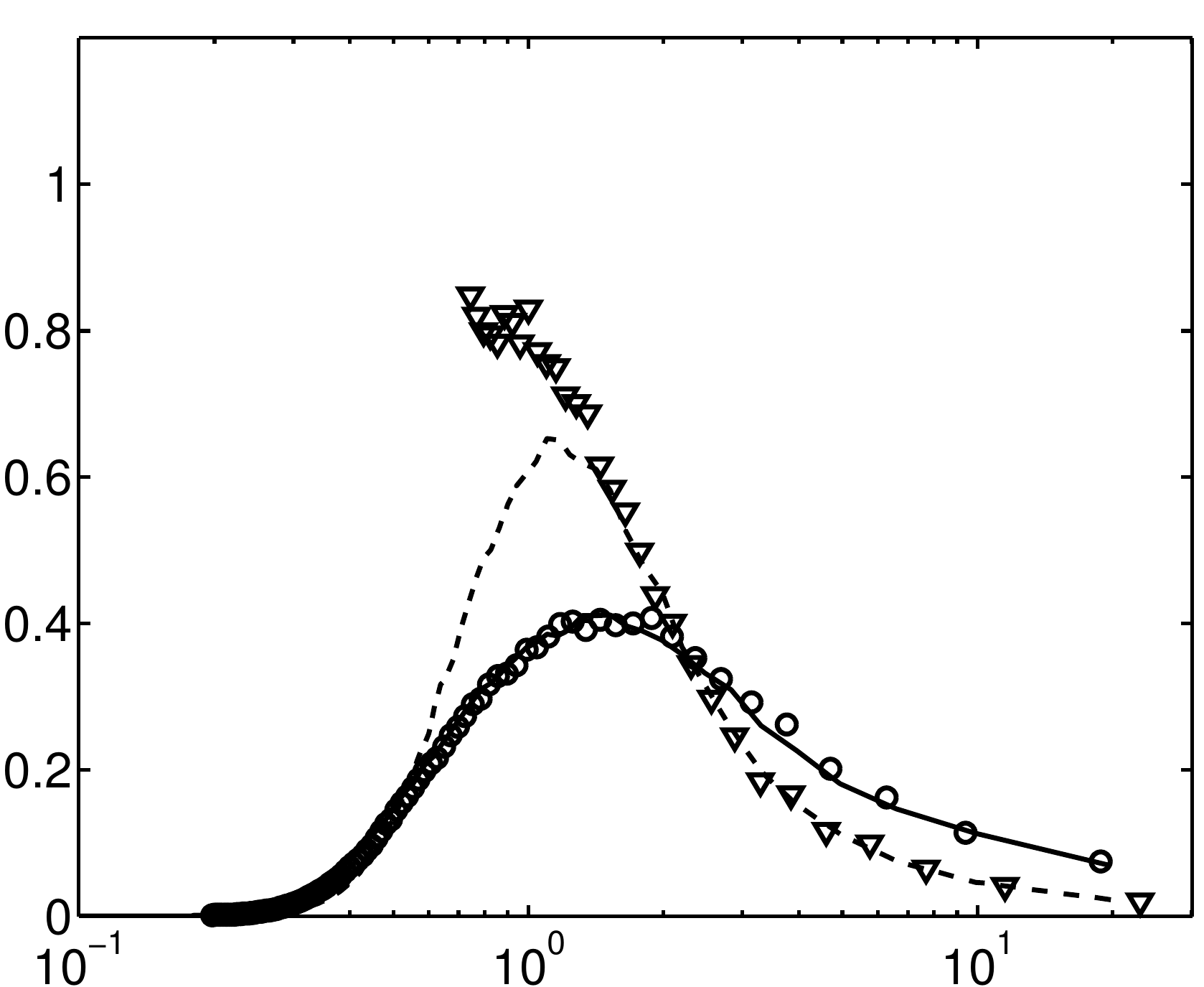}
      \hspace{-0.2\linewidth}\raisebox{0.55\linewidth}{($c$)}
      \\
      \centerline{$\period \Ubh/h$} 
    \end{minipage}
    \caption{\label{fig:premult_spec-all}%
      Pre-multiplied spectrum~$\kappa_x \phihat\phihatast$ of
      ($a$) spanwise torque fluctuations, $T^\prime_z$,
      ($b$) drag fluctuations, $F^\prime_x$ and
      ($c$) lift fluctuations, $F^\prime_y$, 
      shown as a function of the period $\period$ normalised by the
      bulk time scale. 
      Line styles are as in Fig.~\ref{fig:auto_corr-all}.
      %
      %
    } 
  \end{center}
\end{figure}
In the following the auto-correlations in time 
{at zero spatial separation}
are considered (i.e.\ $\phi=\psi$
and $\Deltaxp=\Deltazp=0$ in equation \ref{eqn:corr_funct}). 
Fig.~\ref{fig:auto_corr-all}$(a)$ shows the
auto-correlation in
time of spanwise torque fluctuation, $T^\prime_z$, for case F10
and F50 in comparison with the auto-correlation
of 
$\tau^\prime_{12}$ 
in the smooth-wall reference simulation. 
%
%
%
As in the case of the two-point correlation, 
it is found that the temporal auto-correlation of spanwise torque
almost overlaps with the  auto-correlation of the smooth-wall shear
stress $\tau^\prime_{12}$ in case F10.   
The auto-correlation of spanwise torque in case F50 exhibits 
smaller values for a separation in time scaled in outer flow units
(Fig.~\ref{fig:auto_corr-all}$a$).   
When scaled in inner flow units the 
difference is smaller \citep{Chanbraun_PHD_2012}.
%
This is reflected by the integral time scale which is 
defined as the integral of the  auto-correlation, 
here computed in the range  $\tau_\ell  \utau^2/\nu \in
[0,100]$ (approximately corresponding to $\tau_\ell  \Ubh/h  \in [0,5]$). 
Table~\ref{tab:intscale} shows that the integral time scale of
spanwise torque in case F10 is 15\% (65\%) larger than in case F50
when scaled in inner (outer) scales.
For completeness 
the table also provides 
other
measures that can be  
used 
to characterise 
the time scales of the
torque and drag signals, which will be further discussed below: 
the integral time scale, $\tau_{\ell,abs}$, 
defined by the integral of the absolute value of the
auto-correlation, 
and the temporal Taylor micro-scale, $\tau_\lambda$, defined by the
zero-crossing of the parabola osculating the correlation function at
zero separation.  
%
%

%
In turbulent wall-bounded flows the convection velocity, 
$\Uc$, 
is sometimes used to relate length scales to time scales. 
To this aim, it is often assumed that the flow is 
not significantly modified as it is convected downstream with a
constant convection velocity, which is commonly referred to as
Taylor's frozen turbulence hypothesis \citep{Taylor_RSLA_1938}. 
A detailed discussion on the issue can be found in
\citet{delAlamo_Jimenez_JFM_2009} and the references therein.
%
Here we have applied Taylor's hypothesis to the spatial correlation
data discussed in the previous section in order to compute an
approximate temporal auto-correlation. To this aim we use the
value of the convection velocity as defined in Eq.~(\ref{eqn:def_uconvec})
and further discussed in \S~\ref{sec:stcorr} below. 
The resulting auto-correlations for the spanwise torque are also included in
Fig.~\ref{fig:auto_corr-all}$(a)$. 
It can be observed that the agreement between the actual
auto-correlations and the corresponding data obtained from the spatial
correlations via Taylor's hypothesis is satisfactory, in particular
for small separation times.  
Thus, in the present cases the Taylor approximation appears
to be appropriate to relate scales of spanwise torque in time to the
corresponding spatial scale, 
implying that in both cases the flow structures relevant for spanwise
torque indeed vary more slowly than the convective time scale. 
%


Similar conclusions as from the auto-correlations in time,
$R_{\phi\phi}(\tlag)$, discussed above can be drawn from
the 
temporal spectra,
$\Rhat_{\phi\phi}(\omega)$,
where $\omega$ is the frequency.
%
In case of infinite or periodic time signals, 
$\Rhat_{\phi\phi}(\omega)$ can be defined as the
Fourier transform of the auto-correlation
or alternatively as
$\Rhat_{\phi\phi}(\omega)=
\widehat{\phi}(\omega)\widehat{\phi}^\ast(\omega)$, 
where 
$\widehat{\phi}(\omega)$ is the Fourier transform of $\phi(t)$,
and $\widehat{\phi}^\ast(\omega)$ the conjugate complex of
$\widehat{\phi}$. 
%
In the present case, however, 
the time signals 
are finite and non-periodic, and
$\Rhat_{\phi\phi}(\omega)$ is approximated by
the method of \citet{Welch_IEEE_1967} with 50\%
overlap and applying a Hamming window
\citep{Oppenheim_Schafer_1989} 
to the original
signal $\phi$, prior of transferring it into spectral space. 
Fig.~\ref{fig:premult_spec-all} provides the pre-multiplied spectra
$\omega\widehat{R}_{\phi\phi}(\omega)$ 
normalised with the integral of the spectra over all frequencies, shown
as a function of the period, 
$\period = 2\pi / \omega$.
Concerning spanwise torque fluctuations
(Fig.~\ref{fig:premult_spec-all}$a$), 
the majority of contributions stems from periods 
roughly centered around $\period \Ubh/h =10$ in case F10 and around
$\period\Ubh/h =5$ in case F50, in both cases corresponding to
approximately $\period \utau^2/\nu =100$. An approximate match
between the spectra in both flow cases is reached when inner scaling
is used (plot omitted).   

Turning now to the auto-correlations of drag and lift
forces 
on a particle, $F_x^\prime$ and $F_y^\prime$
(Fig.~\ref{fig:auto_corr-all}), it is found that they differ between
case F10 and case F50. 
In case F50 the curves of drag and lift
fluctuations exhibit 
pronounced local minima of negative value (drag: $-0.25$ at
$\tlag \Ubh/h =0.55$, lift: $-0.18$ at $\tlag \Ubh/h =0.57$).
Conversely, in case F10 the auto-correlation of drag has a clear local
minimum, albeit with a positive correlation value 
($0.20$ at $\tlag \Ubh/h =0.47$). 
For the auto-correlation of lift in case F10 there exists no visible 
local minimum. 
%
%
In order to quantify the short-time behavior of the auto-correlation
functions one can refer to the temporal
Taylor micro-scale $\tau_\lambda$ provided in
Table~\ref{tab:intscale}. 
%
It can be observed that the temporal Taylor micro scales of
spanwise torque are larger than those of drag and
lift.
The  temporal Taylor micro scales of drag and lift
compare best between cases F10 and F50 when
scaled in outer flow units, i.e.\
$\tau_\lambda\Ubh/h = 0.24\pm0.01$.
When scaled in inner units the time scales of drag and lift
are  
approximately 50\% larger in case F50 than in case F10. 
%
In contrast to the temporal Taylor micro-scale, the integral time scales
$\tau_\ell$ and $\tau_{\ell,abs}$
are smaller in case F50 than in case F10 revealing a shortening
of the largest time scales in the time signals. 


The pre-multiplied temporal spectra of particle drag and lift 
(Fig.~\ref{fig:premult_spec-all}$b,c$) 
differ significantly 
from those of spanwise torque on the particles (cf.\ 
Fig.~\ref{fig:premult_spec-all}$a$). 
In particular, the largest spectral contributions
to force fluctuations are
shifted to smaller periods.
%
It is found that in case F10 the spectrum of drag is bi-modal, 
exhibiting a second (weaker) local maximum at 
larger periods ($\period \Ubh /h\approx10$), 
coinciding with the range of maximum contribution to the
spectrum of the smooth wall shear stress component $\tau_{12}$. 
It also corresponds to the range of temporal periods where the torque
spectrum has its peak. 
%
%
This second peak is not present in case F50, neither is it present for
any case in the spectrum of particle lift. 
%
%

Finally, Fig.~\ref{fig:premult_spec-all} also
includes  the spectrum  of drag and lift  
reconstructed from spatial data by way of 
Taylor's hypothesis. 
In case F10, both spectra agree over the entire range of
periods. 
In case F50 the agreement for large periods is similarly
good, whereas a pile-up of energy in the smallest periods
{is exhibited by the spectrum based upon the data 
converted from the spatial correlations}. 
This results from the finite size of particles
yielding a spatially discrete nature 
of the
force and torque data.
As a consequence, the smallest wavelength that
can be resolved in space is limited to the distance between the  
particle centres 
and, upon conversion to temporal data,
the smallest resolvable period is 
$\period_{\rm min} \Ubh/h= (2D /\Uc)( \Ubh/h)$. 
In particular this yields $\period_{\rm min} \Ubh/h\approx0.16$ in
case F10 and $0.8$ in case F50.  
In case F50 this leads to an aliasing of the contribution
to force fluctuations in the small periods for the spectra in space 
as seen in Fig.~\ref{fig:premult_spec-all}.  
\begin{table}
  \begin{center}
    \begin{tabular}{ccrrr ccrrrr}
      case&quantity
      &  $T_z^\prime$ & $F^\prime_x$ &
      $F^\prime_y$
      \quad & \quad & 
      quantity
      &  $T^\prime_z$ & $F^\prime_x$ &
      $F^\prime_y$ \\
      F10&
      $\tau_{\ell} \utau^2/\nu $&
       21.9 &  10.8 &   6.1 &
       \quad &  
      $\tau_{\ell}  \Ubh/h $&
      1.77 &  0.87 &  0.50 
      \\
      F50 &
      $\tau_{\ell} \utau^2/\nu $&
      19.1 &   2.6 &   1.9 &
      \quad & 
      $\tau_{\ell}  \Ubh/h $&
      1.01 &  0.14 &  0.10 
      \\[1ex]
      F50 &
      $\tau_{\ell,abs} \utau^2/\nu $&
      19.1 &   6.3 &   5.2 &
      \quad & 
      $\tau_{\ell,abs}  \Ubh/h $&
      1.01 &  0.34 &  0.28 
      \\[1ex]
      F10 &
      $\tau_\lambda  \utau^2/\nu $&
      10.3 &   2.9 &   3.1 &
      \quad &  
      $\tau_\lambda   \Ubh/h $ &
      0.83 &  0.24 &  0.25 &
      \\
      F50 &
      $\tau_\lambda  \utau^2/\nu $&
      11.4 &   4.5 &   4.5 &
      \quad & 
      $\tau_\lambda   \Ubh/h $ &
      0.61 &  0.24 &  0.24
    \end{tabular}
    \caption{Integral time scale $\tau_{\ell}$ (integrated over $100$
      viscous time units)
      and temporal Taylor micro-scale $\tau_\lambda$  
      of particle torque and force in case F10 and F50. 
      Note that the integral of the absolute value of the respective
      auto-correlation, denoted as $\tau_{\ell,abs}$, is also given in
      case F50.  
    }
    \label{tab:intscale}
  \end{center}
\end{table}
\subsubsection{Cross-correlation between drag and lift}
\label{ssec:crosscorr_drag_lift}
\begin{figure}
  \begin{center}
    \begin{minipage}{3ex}
      \rotatebox{90}{\hspace{2ex}$R_{Fx\,Fy}(\tlag) / (\sigma_F^x \sigma_F^y)$}
    \end{minipage}
    \begin{minipage}{.9\linewidth}
      \includegraphics[width=0.9\linewidth]
      {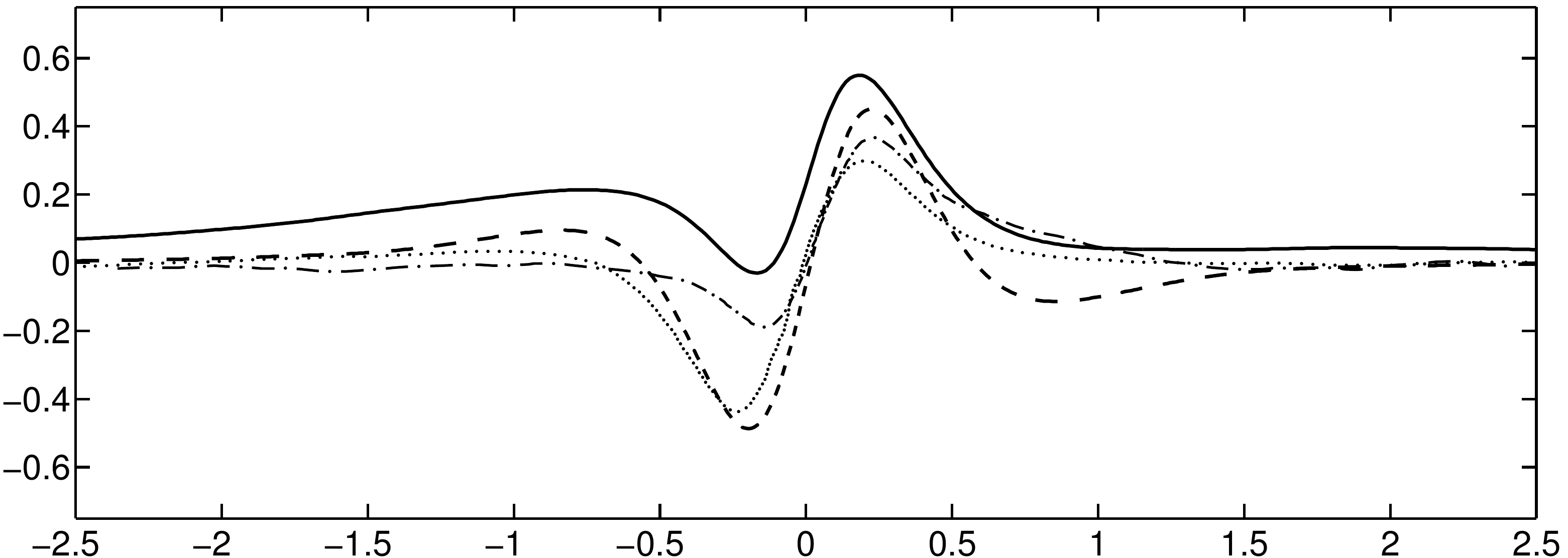}
      \\
      \centerline{$\tlag \Ubh / h $} 
    \end{minipage}

    \caption{Cross-correlation coefficient of drag and lift
      fluctuations, $R_{Fx\,Fy}(\tlag) / (\sigma_F^x \sigma_F^y)$, 
      as a function of separation in time  $\tlag h / \Ubh$
      in  case F10 (\coline) and case F50 (\daline). 
      Additionally, the figure provides the
      results of the indirect 
      measurements of \citet{Hofland_diss_05} (his Fig.~6.5($a$), lowest
      protrusion) for a
      cube within natural gravel with a height of 3300$\deltanu$ at
      $\Reb=1.3\cdot 
      10^5$, 
      (\dodoline) 
      and the direct measurements of
      \citet{Dwivedi_diss_10} (his Fig.~7.23($a$), zero protrusion) of
      spheres with $D^+=3100$ 
      at $\Reb=1.68 \cdot 10^5$ (\dadoline).
    }
    \label{fig:xcorr_drag_lift}
  \end{center}
\end{figure}

Here we are addressing the question whether
drag and lift are correlated in time
and, therefore, related to similar flow structures.
Fig.~\ref{fig:xcorr_drag_lift} presents the 
temporal cross-correlation
of drag and lift 
defined by (\ref{eqn:corr_funct}) with
 $\phi^\prime=F_x^\prime$, $\psi^\prime=F_y^\prime$ and $\Deltaxp=\Deltazp=0$.
For $\tlag=0$, the cross-correlation of drag and lift
fluctuations is small, i.e.\ $R_{Fx\,Fy}(\tlag) /
(\sigma_F^x \sigma_F^y) =0.23$ ($-0.06$) in case F10 (F50).
%
However, a significant correlation occurs at a 
finite temporal separation. 
%
The cross-correlation reaches a maximum value of $R_{Fx\,Fy}(\tlag) /
(\sigma_F^x \sigma_F^y) = 0.55$ (0.45) for a
separation in time of $\tlag \Ubh /h = 0.19$ (0.24) in case F10 (F50).
Thus, on average lift fluctuations follow drag fluctuations of
equal sign with a time lag which approximately matches 
the values of the Taylor micro-scale related to drag and lift (cf.\
Table~\ref{tab:intscale}). 
%
%
The cross-correlation coefficient in Fig.~\ref{fig:xcorr_drag_lift}
reaches a local minimum of value $-0.03$ ($-0.49$) for a negative 
separation in time of $\tlag \Ubh /h = -0.17$ ($-0.20$) in case F10 (F50).
While in case F10 the correlation is essentially zero
at the local minimum, in case F50 it reaches significant negative 
values,
indicating that drag and lift fluctuations are on average of
opposite sign at a separation in time of $\tlag \Ubh/h=-0.20$. 
As the separation in time 
takes increasingly large negative values 
the correlation becomes positive once more, 
then decaying to zero from above. 

In addition to the results of case F10 and F50, the results of
two experimental studies are presented in
Fig.~\ref{fig:xcorr_drag_lift}.  
The first data set 
 corresponds to one of the experiments of
\citet{Hofland_diss_05}
(his figure~6.5$a$),  
who indirectly measured drag and lift by
the difference of two simultaneous pressure measurements.
These measurements were taken on both sides (upstream and downstream)
of a cube with a protrusion height of 3300 viscous units at
$\Reb=1.3\cdot 10^5$.
The ratio of channel width to open channel height was 3.0 and the
ratio of cube height to open channel height  5.6.
The other data set is taken from
\citet{Dwivedi_diss_10} (his figure~7.23$a$), who
directly measured drag and lift on a spherical particle with
$D^+=3100$ in a hexagonal packing of spheres with zero
protrusion at $\Reb=1.68 \cdot 10^5$. 
The ratio of $H/D$ was 5.4 and thus comparable to the one of 
the present case F50
and to the measurements of \citet{Hofland_diss_05}.  
The width to height ratio 
of the open channel flow cross-section in the work of
\citet{Dwivedi_diss_10} 
was $2.1$, which is also comparable
to the one of \citet{Hofland_diss_05}.
From the low value of this ratio 
it seems reasonable to assume that secondary flow might
influence the results in both experiments.
%
In spite of the different Reynolds numbers and particle shapes
considered in the experiments, as well as the differences with 
respect to how the data was obtained,
the drag/lift cross-correlations 
of these authors 
are qualitatively similar to 
the results of both our present flow cases. 
The main differences are in the amplitudes of the local maxima and
minima.  
However, the local extrema appear roughly at the same time separations under
the present scaling. Note that other possible scalings have been
explored \citep[cf.\ Table~3.11 of][]{Chanbraun_PHD_2012}, but were
found less satisfactory. 

\subsection{Space-time correlation and convection velocities of
  spanwise torque, drag and lift} 
\label{sec_spatiotemporal}
\label{sec:stcorr}
%
%
\begin{figure}
  \begin{center}
    \begin{minipage}{3ex}
      \rotatebox{90}{\hspace{4ex}$\omega h / \Ubh$}
    \end{minipage}
    \begin{minipage}{.36\linewidth}
      \includegraphics[width=1.0\linewidth]
      {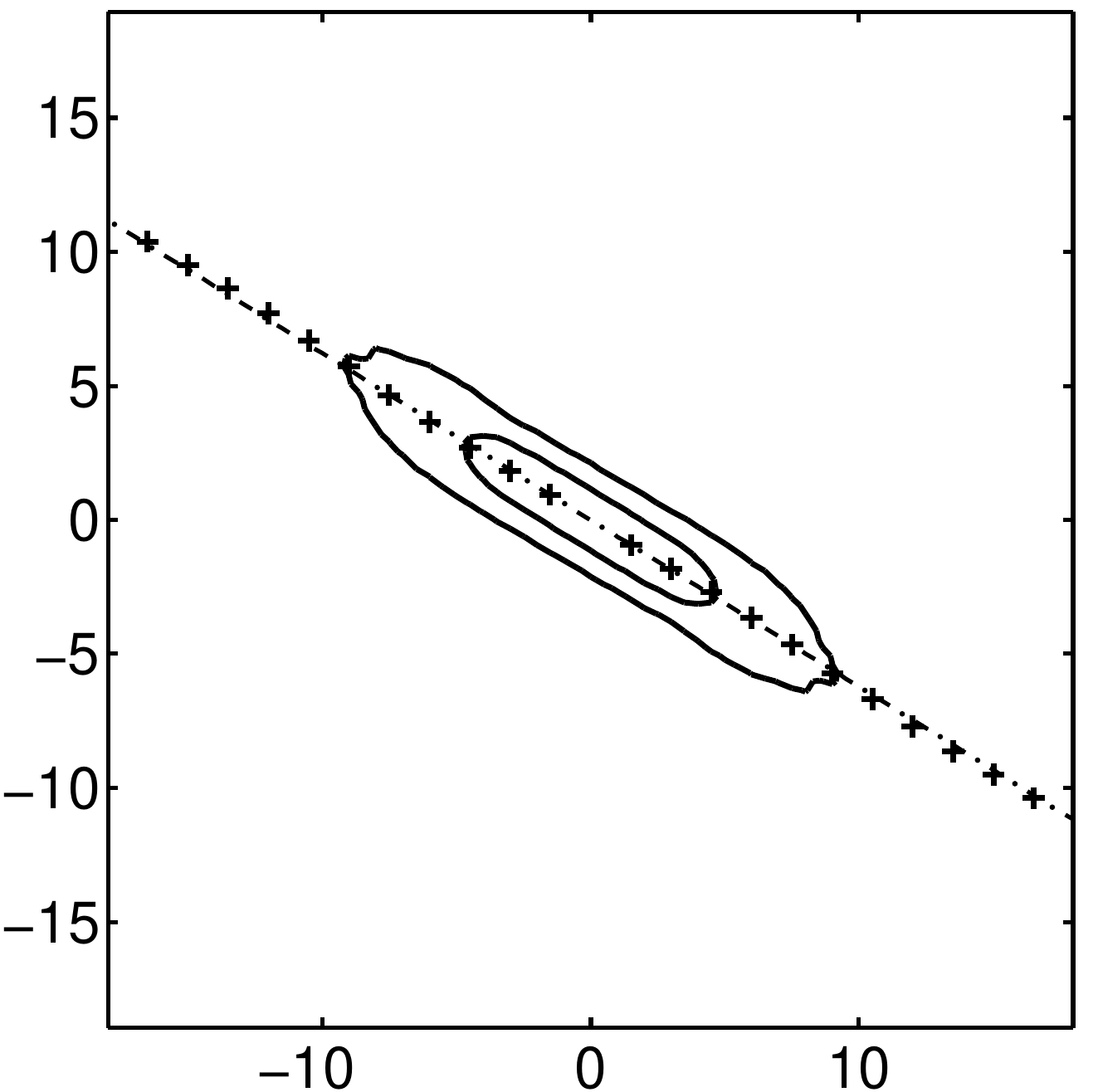}%
      \hspace{-0.25\linewidth}\raisebox{0.85\linewidth}{\colorbox{white}{($a$)}}
    \end{minipage}
    \begin{minipage}{.36\linewidth}
      \includegraphics[width=1.0\linewidth]
      {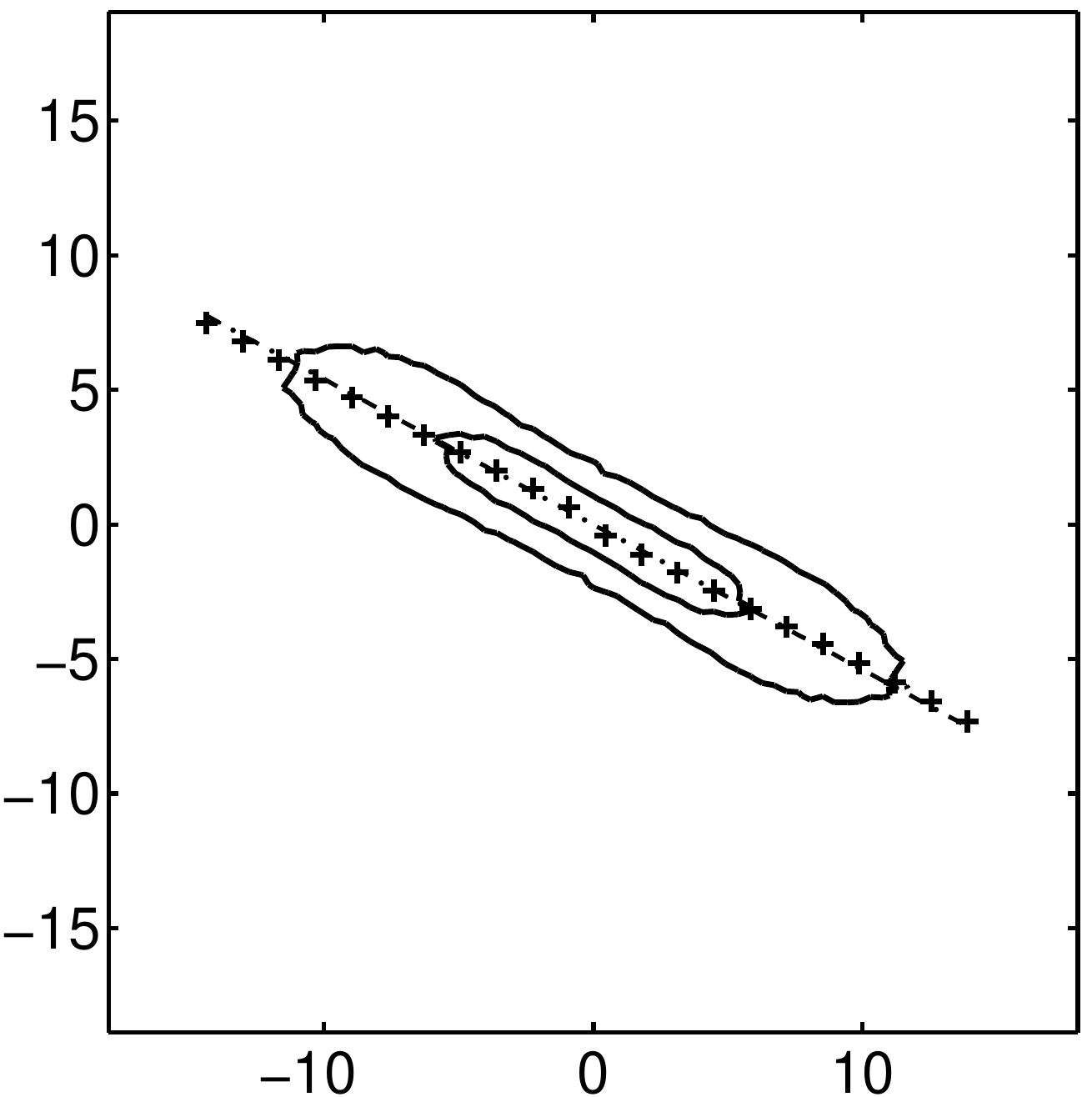}%
      \hspace{-0.25\linewidth}\raisebox{0.85\linewidth}{\colorbox{white}{($b$)}}
    \end{minipage}\\

    \begin{minipage}{3ex}
      \rotatebox{90}{\hspace{4ex}$\omega h / \Ubh$}
    \end{minipage}
    \begin{minipage}{.36\linewidth}
      \includegraphics[width=1.0\linewidth]
      {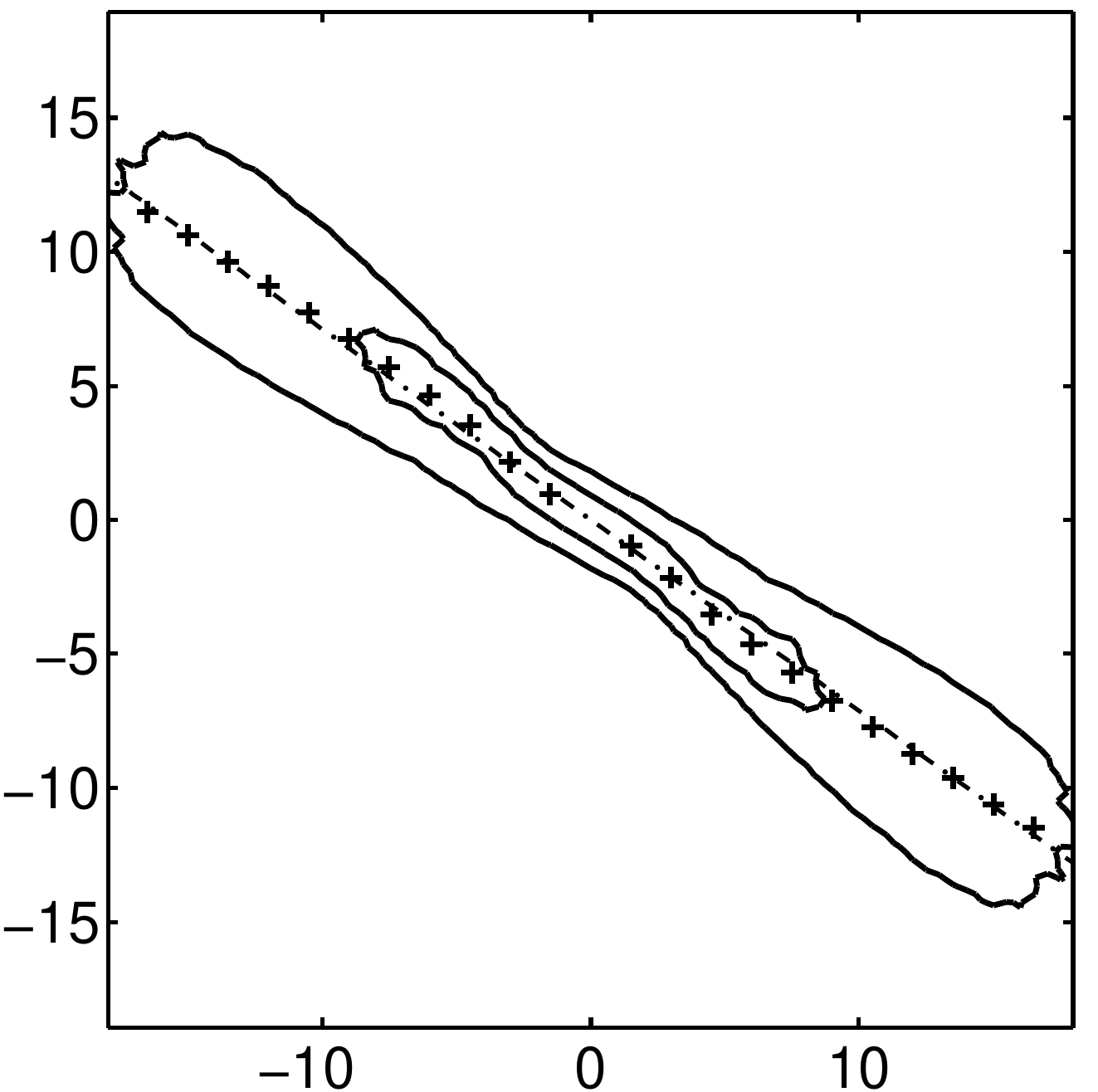}%
      \hspace{-0.25\linewidth}\raisebox{0.85\linewidth}{\colorbox{white}{($c$)}}
    \end{minipage}
    \begin{minipage}{.36\linewidth}
      \includegraphics[width=1.0\linewidth]
      {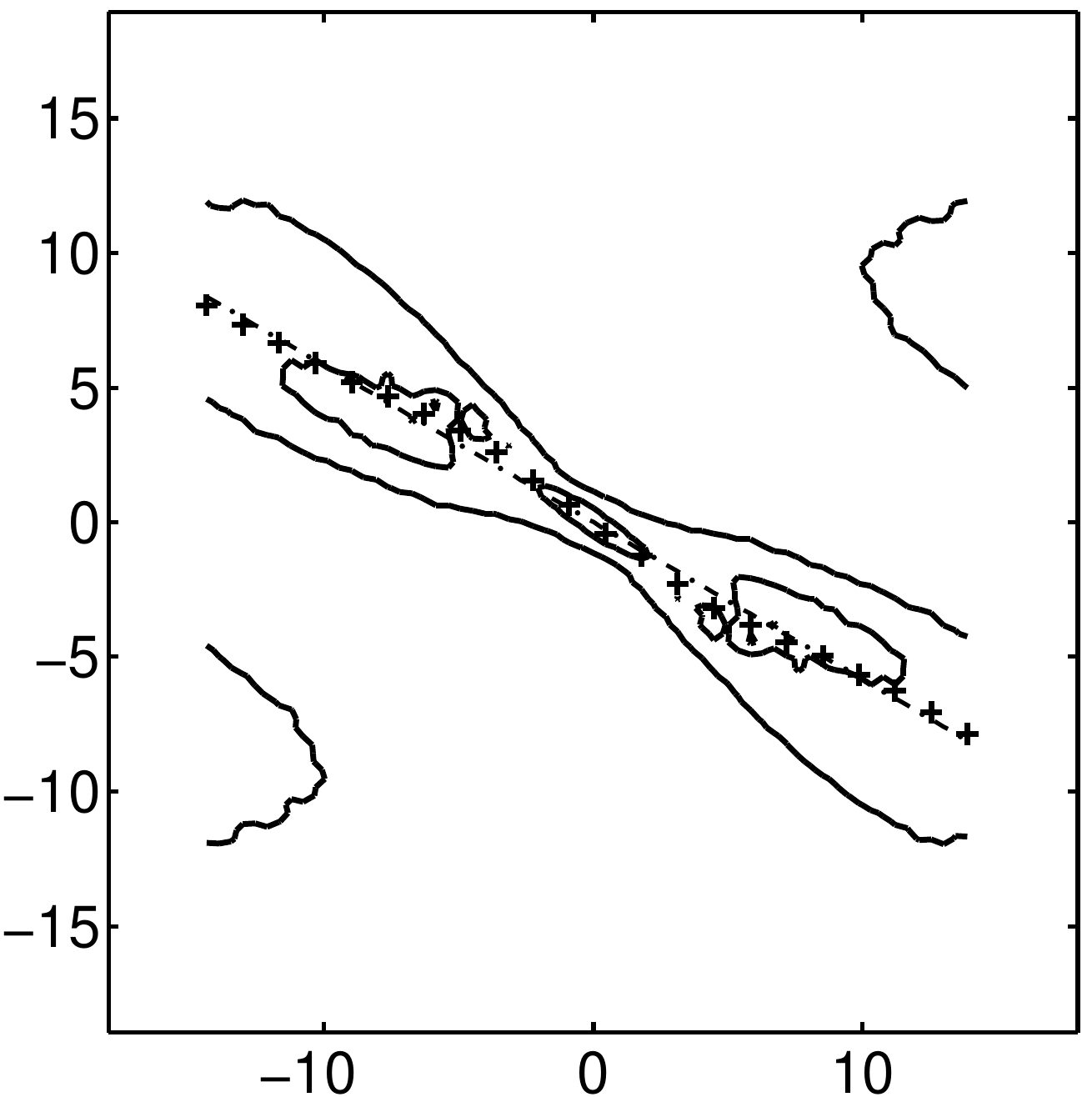}%
      \hspace{-0.25\linewidth}\raisebox{0.85\linewidth}{\colorbox{white}{($d$)}}
    \end{minipage}\\

    \begin{minipage}{3ex}
      \rotatebox{90}{\hspace{4ex}$\omega h / \Ubh$}
    \end{minipage}
    \begin{minipage}{.36\linewidth}
      \includegraphics[width=1.0\linewidth]
      {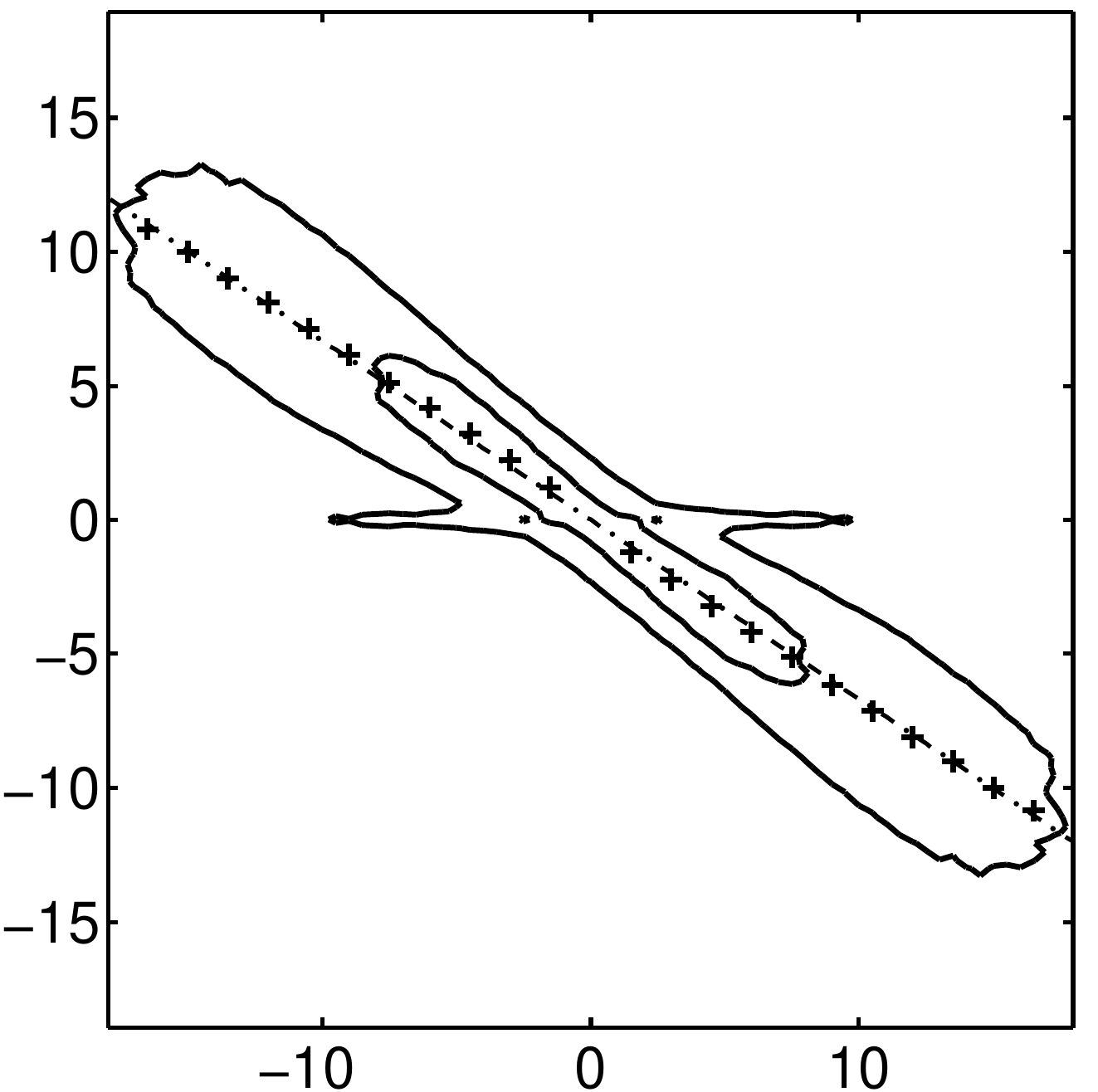}%
      \hspace{-0.25\linewidth}\raisebox{0.85\linewidth}{\colorbox{white}{($e$)}}
      \centerline{$\kappa_p^x h$} 
    \end{minipage}
    \begin{minipage}{.36\linewidth}
      \includegraphics[width=1.0\linewidth]
      {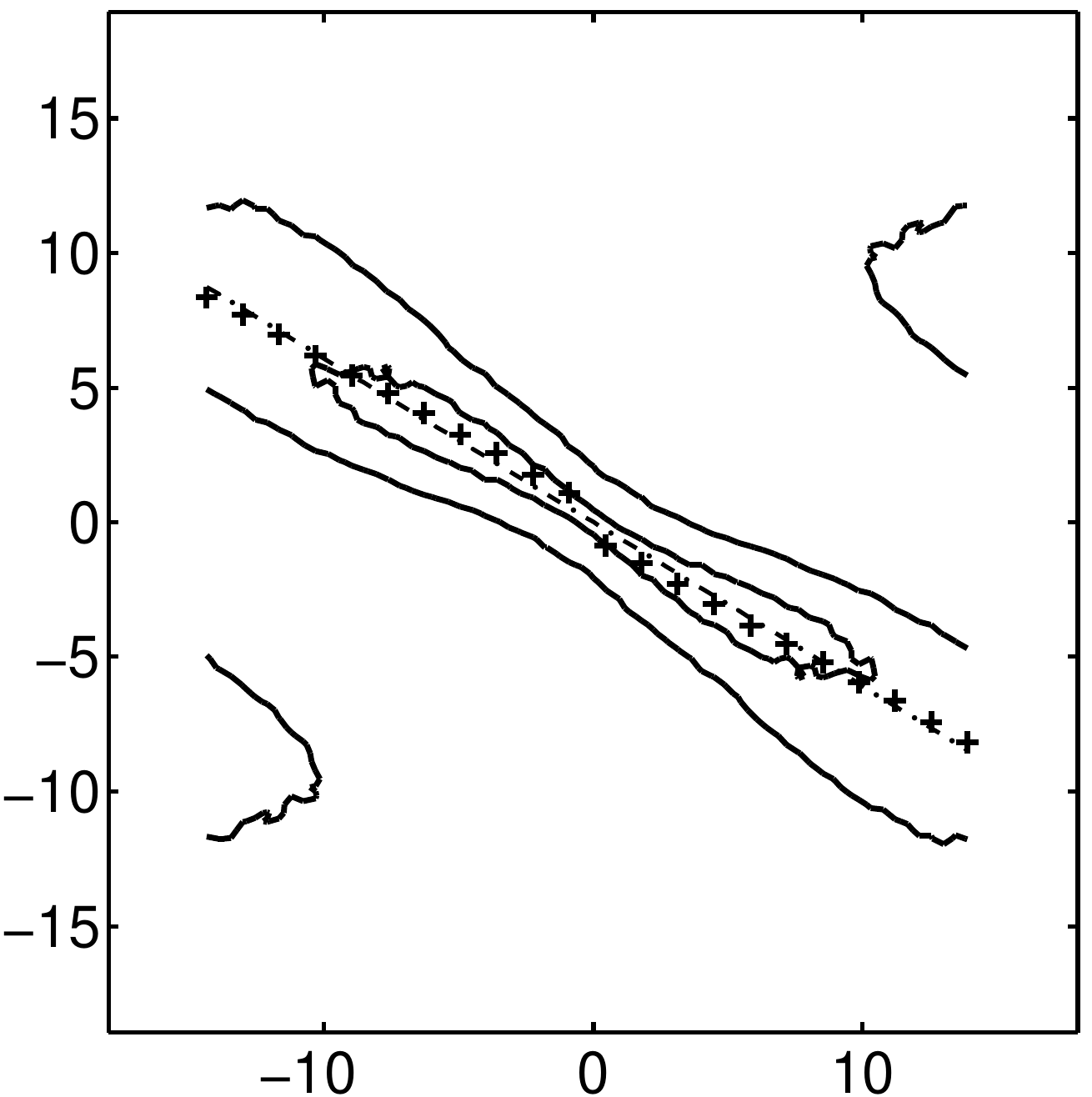}%
      \hspace{-0.25\linewidth}\raisebox{0.85\linewidth}{\colorbox{white}{($f$)}}
      \centerline{$\kappa_p^x h$} 
    \end{minipage}
  \end{center}
    \caption {
      Auto-correlation of different particle-related quantities 
      in case F10~($a$,$c$,$e$) and case F50~($b$,$d$,$f$). 
      Graphs~($a$,$b$) show spanwise torque fluctuations, 
      ($c$,$d$)~show drag fluctuations,
      and ($e$,$f$)~show lift fluctuations. 
      The correlation 
      (normalised by its integral value)
      is plotted as a function of
      frequency, $\omega h / \Ubh $, and streamwise wave number, 
      $\kappa_p^x  h$. 
      Lines show iso-contours at values 
      of [0.001 0.01] (\coline). 
      The symbols ($+$) mark $\omegac (\kappa_p^x)$ as defined in the
      text (only every 4th point is shown). 
      In all panels the slope of the dash-dotted lines corresponds to
      $-\Uc$, where $\Uc$ is the convection velocity defined in
      (\ref{eqn:def_uconvec}). 
      }
  \label{fig:st_spec}
\end{figure}
\begin{table}
\begin{center}
\begin{tabular}{ c ccc cc ccc }
 $\Uc / \Ubh$ & $T_z^\prime$& $F_x^\prime$ & $F_y^\prime$ &\quad&
$\Uc/\utau$ & $T_z^\prime$& $F_x^\prime$ & $F_y^\prime$ \\
F10 &    0.62 & 0.71 &   0.67  &&
F10 &  9.5 & 10.8 &   10.2 
\\ 
F50 &  0.54 & 0.58 &   0.61 &&
F50 &  6.7 &    7.2 &    7.5
%
\end{tabular}
\end{center}
\caption{Convection velocities, $\Uc$, of force and torque fluctuations on
  particles in case F10 and F50
  normalised by $\Ubh$ and $\utau$}\label{tab:convvel} 
\end{table}
%
%
%
This section focuses on the convection velocities of force and
torque fluctuations as a result of turbulent flow structures. 
The definition of the convection velocity used in the present
context requires a streamwise-space/time correlation of force or
torque fluctuations,  
which in physical space can be defined by (\ref{eqn:corr_funct}) with
$\phi=\psi$ and $\Deltazp=0$.  
For reasons of efficiency the space-time correlations were not
computed as above in physical space but in spectral space 
by employing the method of \citet{Welch_IEEE_1967} in time 
and a Fourier transform in the 
streamwise 
direction (cf.\ \S\ref{ssec:1D_autocorr}). 
The space-time correlation in spectral space is denoted by
$\Rhat_{\phi\psi}(\kappa^x,\omega)$, as a 
function of streamwise wave number, $\kappa^x$, and frequency, $\omega$.

Several definitions of the convection
velocity in turbulent shear flows have been proposed 
\citep{quadrio_luchini_pof_03, delAlamo_Jimenez_JFM_2009}. 
It should be noted, that a convection velocity of a flow
field component cannot be defined unambiguously as scales with
different lengths might travel at different speeds.
%
%
Here, the convection velocity, $\Uc$, of a particle-related 
quantity (force, torque) is defined 
as a weighted average of the convection velocity
$\uc$ related to a given wave number, $\kappa_p^x$ 
(Ref.\ \onlinecite{delAlamo_Jimenez_JFM_2009}) as
\begin{equation}
  \Uc = - \frac
  {\int_{\kappa_p}\int_{\omega}
    \uc \,
    |\Rhat_{\phi\psi}(\kappa_p^x,\omega)| \,(\kappa_p^x)^2 \,\mrd
    \omega \mrd \kappa_p} 
  {\int_{\kappa_p}\int_{\omega}
    |\Rhat_{\phi\psi}(\kappa_p^x,\omega)|
    \,(\kappa_p^x)^2 \,\mrd \omega \mrd
    \kappa_p} \,.
  \label{eqn:def_uconvec}
\end{equation}
This average weights the convection velocities of each wave number
with the energy contained in the respective wave number. Thus the higher
energetic wave numbers contribute more to the global convection
velocity defined by (\ref{eqn:def_uconvec}). 
For a given streamwise wavenumber, $\kappa_p^x$,
\citet{delAlamo_Jimenez_JFM_2009} propose to define
$\uc$ by the centre of mass 
of the profile $\Rhat_{\phi\psi}(\kappa_p^x,\omega)$ denoted as
$\omegac$ in the following, i.e.\
$\uc=-\omegac(\kappa_p^x)/\kappa_p^x$. 
However, in the present case such a definition leads to a strong bias 
due to the aliasing of energy in the large wave numbers (small wave
lengths) as discussed in the context of Fig.~\ref{fig:premult_spec-all}. 
The aliasing results in the existence of contour lines at positive and
negative frequencies for wave numbers of large magnitude in 
Fig.~\ref{fig:st_spec}($d$,$f$).  
In order to overcome the bias due to the aliasing,
$\omegac(\kappa_p^x)$ is defined by the centre of
mass of the one-sided frequency spectrum in the present work, i.e.\ 
only the range for which $-\sgn(\kappa_p^x) \omegac > 0$ is
considered.  
This leads to a bias for very small wave numbers (which do not
contribute much to the global convection velocity) 
but reduces the otherwise dominant bias from large wave numbers.

The above definitions 
can be illustrated with the aid of
Fig.~\ref{fig:st_spec}
which
shows the space-time correlation of drag, lift and spanwise torque fluctuation
in spectral 
space. 
The figure contains parts of the information discussed in the previous
chapters, 
i.e.\ the pre-multiplied spectra
in time and in streamwise direction are obtained  
by summation over the horizontal and vertical axes,
respectively, in Fig.~\ref{fig:st_spec}.
 %
Therefore the discussion in the following is limited to
new aspects.
The symbols ``$+$'' in Fig.~\ref{fig:st_spec} show
$\omegac(\kappa_p^x)$ as 
defined above. 
As can be seen, the definition of $\omegac(\kappa_p^x)$ results in 
an approximately linear variation of
$\omegac(\kappa_p^x)$ 
and qualitatively agrees with the expectations. 

The convection velocities $\Uc$ 
which are obtained by (\ref{eqn:def_uconvec}) 
are illustrated in form of straight dash-dotted lines (with slope
$-U_c$) in Fig.~\ref{fig:st_spec}, 
and the corresponding numerical values are given in Table~\ref{tab:convvel}.
Let us first consider case F10, where the roughness elements are small
enough to yield a hydraulically smooth flow. 
In this case the values of the convection velocity are smallest for
$T_z^\prime$ ($\Uc/U_{bh}=0.62$, i.e.\ $\Uc/\utau=9.5$) and largest
for $F_x^\prime$ ($\Uc/U_{bh}=0.71$, $\Uc/\utau=10.8$).  
The values are in the range which is commonly reported for the
convection velocity of flow quantities in smooth
wall flows. 
For example, for smooth wall channel flow
\citet{delAlamo_Jimenez_JFM_2009} reported 
values of $\Uc / \Ubh$ in the range of 0.5 to 0.8
and 0.6 to 0.8 for the streamwise and spanwise velocity fluctuations
close to the wall, respectively. 
Other studies in smooth-walled channels and boundary layers have
reported similar values 
\citep{quadrio_luchini_pof_03, krogstad_haspesen_rimestad_PoF_98}. 
The convection velocity of shear stress fluctuations and pressure
fluctuations in smooth-wall channel flow have been studied by
\citet{jeon_etal_pof_99}.  
The authors report values of $9.6$ for the streamwise shear stress and
$13.1$ for the pressure fluctuations.
The present results for the convection velocity of force and torque
fluctuations follow a similar trend. Both drag and lift are by
definition influenced by pressure fluctuations, and their convection
velocity is found to be higher than the one for spanwise torque to
which pressure does not contribute.

Next, let us compare the convection velocity of the transitionally
rough flow case F50 to the above results for F10. 
Table \ref{tab:convvel} 
shows that the convection velocities of spanwise torque, drag and lift 
decrease from case F10 to case F50. 
In particular, $\Uc/\Ubh$ ($\Uc/\utau$) in case F50 is on average 
13\% (30\%) smaller than in case F10 for the particle properties above.
As in case F10 the values of $\Uc/\Ubh$ and $\Uc/\utau$
in case F50 are smallest for $T_z^\prime$. 
Interestingly, in case F50 the convection velocity of lift exceeds the
one of drag. 
%
Data on the convection velocities of flow in the vicinity of a rough
wall is scarce. 
From the simulations and the analysis of \citet{Flores_Jimenez_JFM_06}
it can be deduced that the convection velocity tends to decrease due
to the influence of wall roughness. 
In the experiments of \citet{Krogsstad_antonia_JFM_94} a reduction by
18\% of the smooth-wall value for the convection velocity was assumed
in the rough wall case.  
%
%

A feature of the space-time correlation deserves further explanation,
namely, the contributions of small streamwise wavenumbers to lift
fluctuations at zero frequency in case F10 which appear in
Fig.~\ref{fig:st_spec}($e$) in the form of two needle-like, horizontal
excursions of the lowest-valued contour at zero frequency.  
The lift force acting on the particles, when averaged in time (not over the
particles), $\langle F_y^{(i)}(x_p^{(i)},z_p^{(i)})\rangle_t$, 
is not fully converged in the present data-set; 
instead it varies with a a standard deviation of $0.03\,F_R$.

\section{Correlation between particle-related
  quantities and the flow  field}
\label{sec_corr_flow_for}
%
\label{ssec:paflo}
The purpose of the present section is to infer the flow structures
which significantly contribute to the particle force and torque
statistics. To this end 
we have computed correlation functions between the flow field (either
$u$, $v$, $w$ or $p$) on one
hand, and the particle-related quantities (the six components of
force and torque) on the other.
%
%
%
\revision{%
  It is beyond the scope of this paper to present all of the
  correlations. Therefore, the analysis is limited to the correlations 
  $(T^\prime_z\,,u^\prime)$, $(F^\prime_x\,,u^\prime)$,
  $(F^\prime_x\,,p^\prime)$.}
{%
  It is beyond the scope of the present paper to discuss the
  correlations  between all possible combinations of these
  quantities. Our selection is based upon the following
  considerations.  
  The net particle drag force $F_x$ consists of two
  contributions, one from viscous 
  stresses 
  (which mainly stem from the
  streamwise velocity field) and one from 
  the pressure. 
  Therefore, we choose to discuss first the correlations between
  $F_x^\prime$ and the streamwise velocity fluctuations $u^\prime$,
  and secondly those between $F_x^\prime$ and the fluctuating pressure
  field $p^\prime$. 
  %
  %
  Concerning the torque, it is of interest to analyze its correlation
  with the velocity field, since -- by definition -- this quantity is
  not affected by the pressure field. 
  Finally, we are omitting the correlations involving the lift force
  $F_y^\prime$, since they are found to be similar to the
  corresponding correlations involving drag, and since we have 
  found that drag and lift are significantly correlated with a lag in
  time (cf.\ figure~\ref{fig:xcorr_drag_lift}). 
  In summary, the following discussion is limited to the correlations  
  $(F^\prime_x\,,u^\prime)$, $(T^\prime_z\,,u^\prime)$, 
  $(F^\prime_x\,,p^\prime)$. Additional correlation pairs can be found
  in Ref.~\onlinecite{Chanbraun_PHD_2012}.  
}
%
%

The correlation function between the fluctuation of a scalar flow quantity
$\phi^\prime(\xvec,t)$ and a quantity $\psi^\prime(\xvecp^{(l)},t)$ associated
with the $l$th particle is defined as
\revision{%
  \begin{equation}
    \label{eqn:fix_scales_paflo_corr}
    R_{\phi\psi} (\Deltax,y,\Deltaz) = 
    \frac{1}{N_t N_p}
    \sum_{i=1}^{N_t} \sum_{l=1}^{N_p}  \phi^\prime(x^{(l)}_p +
    \Deltax, y,z^{(l)}_p+\Deltaz,t_i) \
    \psi^\prime(x^{(l)}_p,y^{(l)}_p,z^{(l)}_p, t_i)\,. 
  \end{equation}
}{%
  \begin{equation}
    \label{eqn:fix_scales_paflo_corr}
    R_{\phi\psi} (\Deltax,y,\Deltaz) = 
    \frac{1}{N_{t2} N_p}
    \sum_{i=1}^{N_{t2}} \sum_{l=1}^{N_p}  \phi^\prime(x^{(l)}_p +
    \Deltax, y,z^{(l)}_p+\Deltaz,t_i) \
    \psi^\prime(\mathbf{x}^{(l)}_p, t_i)\,. 
  \end{equation}
}
Note two main differences with respect to definition
(\ref{eqn:corr_funct}): first, the resulting correlation function
$R_{\phi\psi}$ is a three-dimensional field; second, the correlation
function is not restricted to a discrete grid related to the
inter-particle spacing, but to the finite-difference grid (which is 
considerably finer).
%
The sums in (\ref{eqn:fix_scales_paflo_corr})
run over the number of particles
$N_p$ and 
\revision{%
  temporal records $N_t$ 
  distributed over the observation interval 
  ($N_t=67$ in case F10 and $N_t=114$ in case F50).
}{%
  over a number of snapshots $N_{t2}$ 
  distributed over the observation interval 
  (cf.\ Table~\ref{tab:setup_param}).
}
%
%
%
The statistical convergence of a correlation
can be judged by deviations from the symmetry (or 
anti-symmetry)
with respect to the $\Deltax$ or the $\Deltaz$-axis and is found
to be satisfactory
for 
the correlations shown. 
%

%
\begin{table}
\begin{center}
\begin{tabular}{l cc ddd ddd}
   & \multicolumn{1}{c}{F10} &\multicolumn{1}{c}{F50}   &
  \multicolumn{1}{c}{$\Delta_{x, \rm max}^{\rm F10}/D$} & 
  \multicolumn{1}{c}{$y_{\rm max}^{\rm F10}/D$} & 
  \multicolumn{1}{c}{$\Delta_{z, \rm max}^{\rm F10}/D$} & 
  \multicolumn{1}{c}{$\Delta_{x, \rm max}^{\rm F50}/D$} & 
  \multicolumn{1}{c}{$y_{\rm max}^{\rm F50}/D$} & 
  \multicolumn{1}{c}{$\Delta_{z, \rm max}^{\rm F50}/D$} \\ 
  %
%
%
 $(T_z^\prime \,, u^\prime)$ & 2.19 & 1.53 & 4.86 & 1.86 & -0.07 & 0.54 & 1.11 & -0.02 \\
 $(F_x^\prime \,, u^\prime)$ & 1.50 & 0.48 & 6.29 & 1.86 & -0.07 & -1.13 & 1.11 & -0.02 \\
%
%
%
 $(F_x^\prime \,, p^\prime)$ & 0.83 & 1.26 & -2.57 & 0.86 & -0.07 & -0.74 & 0.98 & -0.02 
\end{tabular}
\end{center}
\caption{\label{tab:corr_paflo_maxval}%
  Maximum amplitude of the correlation between particle
  force or torque and the flow field,
  $\Rmax$, in case F10 and case F50 
  (second and third column). 
  $\Rmax$ is normalised by $\utau$ ($\rhof \utau^2$) as a
  characteristic measure of the velocity (pressure) fluctuation
  and by the standard deviation of the force/torque 
  fluctuation. 
  Additionally the position of $\Rmax$ with respect to
  $\Deltax/D$, $y/D$ and $\Deltaz/D$ are provided in columns 
  four to nine.} 
\end{table}
%
The maximum 
absolute values of the considered correlations, $\Rmax$,
are given in Table~\ref{tab:corr_paflo_maxval}. 
Here, $\Rmax$ is normalised by $\utau$ and $\rhof\utau^2$
as a characteristic scale for the
velocity and pressure fluctuations, respectively, and by the standard
deviation of the respective force or torque fluctuation.
%
%
Note that under the normalization used, the values of $R_{\phi\psi}$
reported do not correspond to correlation coefficients, i.e.\ $\Rmax$ is not
bounded from above by unity. Instead, the upper 
bound
is given by
the maximum of the standard deviation of the flow quantity
$\phi^\prime$: e.g.\ under the current normalisation the correlation
between $F_x^\prime$ and $u^\prime$ is bounded by the maximum of
$u_{rms}/u_\tau$.   
%
\begin{figure}
  \begin{center}
    \begin{minipage}{2.6ex}
      \rotatebox{90}{$ (y-\ynull) / h$}
    \end{minipage}
    \begin{minipage}{0.45\linewidth}
      \hspace{-0.9\linewidth}
      \includegraphics[width=1.\linewidth,clip=true]
      {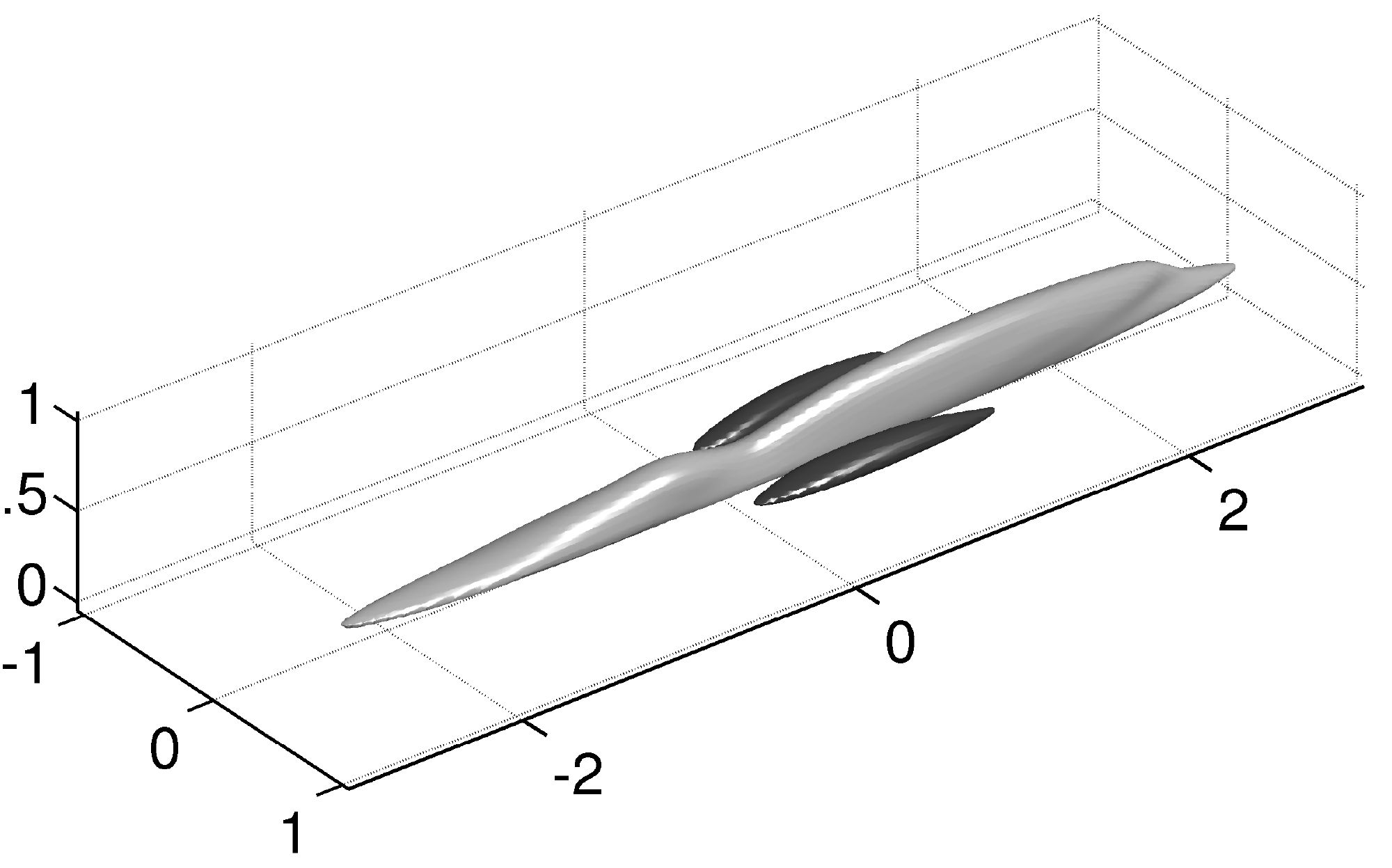}
      \hspace{-0.4\linewidth}\raisebox{0.1\linewidth}{$\Deltax/h$}
      \hspace{-0.6\linewidth}\raisebox{0.4\linewidth}{($a$)}
      \hspace{-0.25\linewidth}\raisebox{0.0\linewidth}{$\Deltaz/h$}
    \end{minipage}
    \begin{minipage}{2.6ex}
      \rotatebox{90}{\hspace{0ex}$ (y-\ynull) / h$}
    \end{minipage}
    \begin{minipage}{0.45\linewidth}
      \hspace{-0.9\linewidth}
      \includegraphics[width=1.\linewidth,clip=true]
      {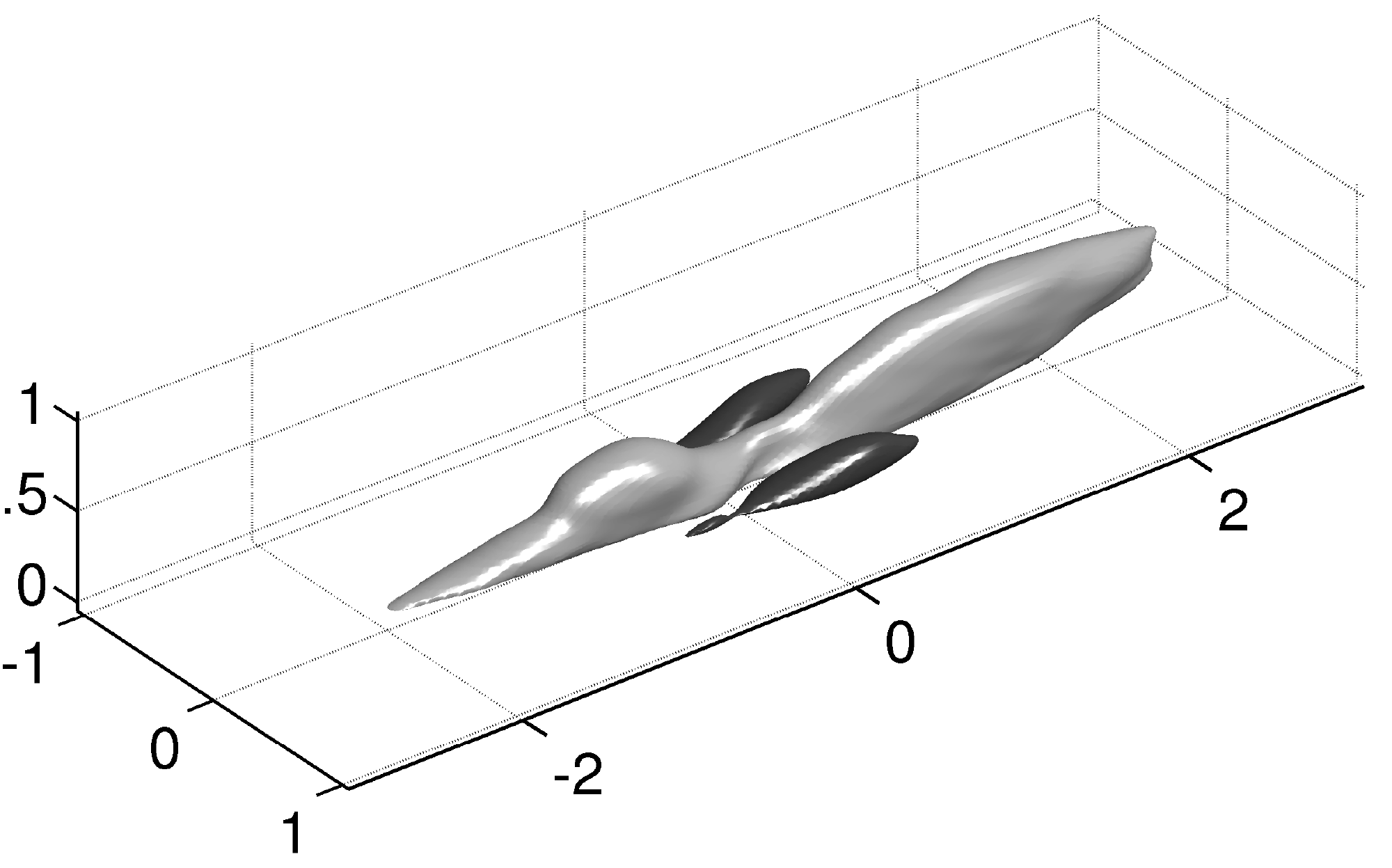}
      \hspace{-0.4\linewidth}\raisebox{0.1\linewidth}{$\Deltax/h$}
      \hspace{-0.6\linewidth}\raisebox{0.4\linewidth}{($b$)}
      \hspace{-0.25\linewidth}\raisebox{0.\linewidth}{$\Deltaz/h$}
      \end{minipage}
  \end{center}
    \caption{\label{fig:corr_paflo_forx_u_3D}%
      Iso-surfaces of 
      correlation, $R_{\phi\psi}(\Deltax , y , \Deltaz) /
        \Rmax$, of the streamwise particle force fluctuation,  
        $F^\prime_x $, and streamwise velocity
        fluctuation, $u^\prime$ at values  $0.15$ (light) and $-0.15$
        (dark). Panels show case F10 ($a$) and case F50 ($b$).}
\end{figure}
\begin{figure}
  \begin{center}
    \begin{minipage}{2.6ex}
      \rotatebox{90}{\hspace{4ex}$ (y-\ynull) / h$}
    \end{minipage}
    \begin{minipage}{0.66\linewidth}
      {($a$)}\\
      \includegraphics[width=1.\linewidth,clip=true]
      {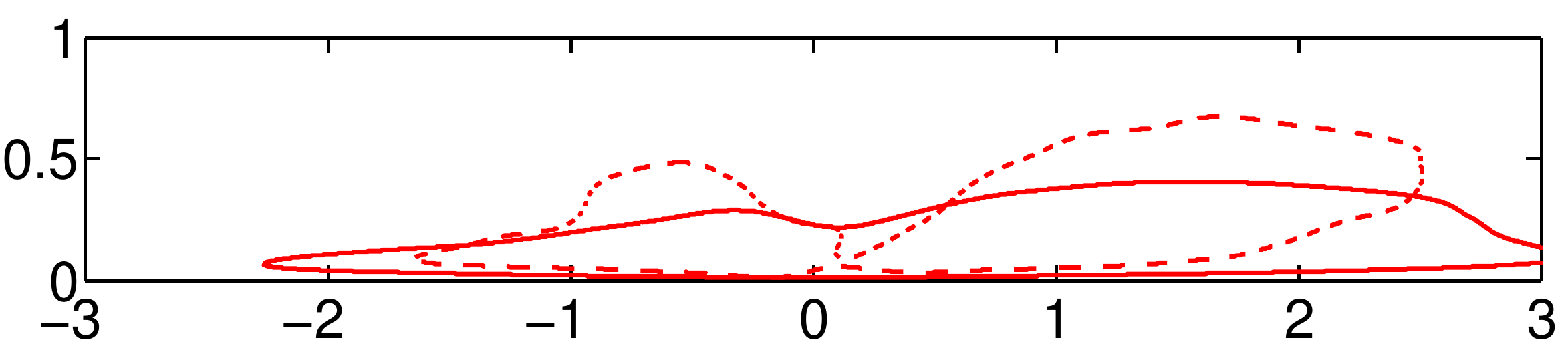}
      \centerline{$ \Deltax / h$} 
    \end{minipage}
    \begin{minipage}{.25\linewidth}
      {($b$)}\\
      \includegraphics[width=1.\linewidth,clip=true]
      {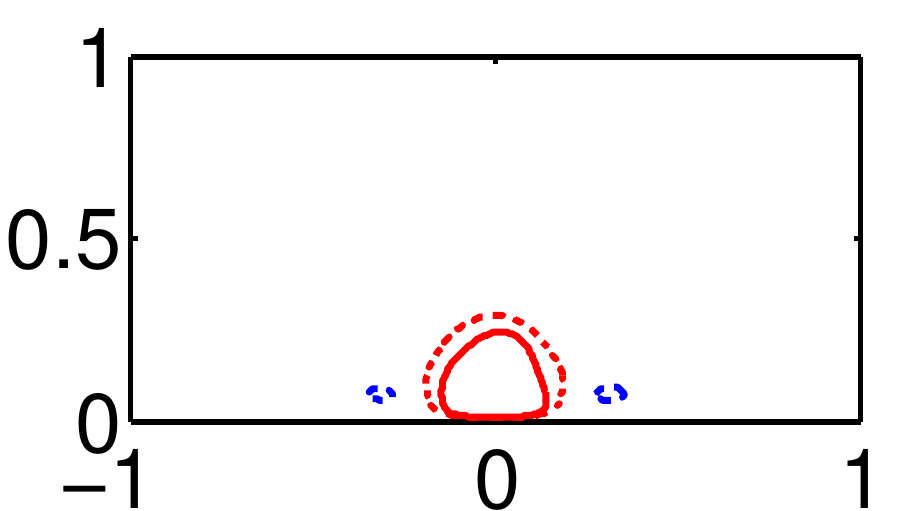}
      \centerline{$\Deltaz / h$} 
    \end{minipage}
  \end{center}
  \caption{\label{fig:corr_paflo_forx_u}%
      Correlation, $R_{\phi\psi}(\Deltax , y , \Deltaz) / \Rmax$, of the  
      streamwise particle force fluctuation, 
      $F^\prime_x $, and streamwise velocity
      fluctuation, $u^\prime$.   
      ($a$) Plane at zero spanwise shift, $\Deltaz=0$, as function of
      $(y-\ynull) / h $ and streamwise shift $\Deltax /h $;
      ($b$) plane at zero streamwise shift, $\Deltax=0$, as function
      $(y-\ynull) / h$ and 
      spanwise shift $\Deltaz / h$.
      Red and blue coloured lines show $0.15$ and $-0.15$,
      respectively, of the maximum amplitude in each case. 
      Continuous (dashed) lines indicate results for case F10 (F50).
      %
    }   
\end{figure}
\begin{figure}
  \begin{center}
    \begin{minipage}{2.6ex}
      \rotatebox{90}{\hspace{2ex}$(y-\ynull) / D$}
    \end{minipage}
    \begin{minipage}{0.9\linewidth}
      {($a$)}\\
      \includegraphics[width=1.\linewidth,clip=true]
      {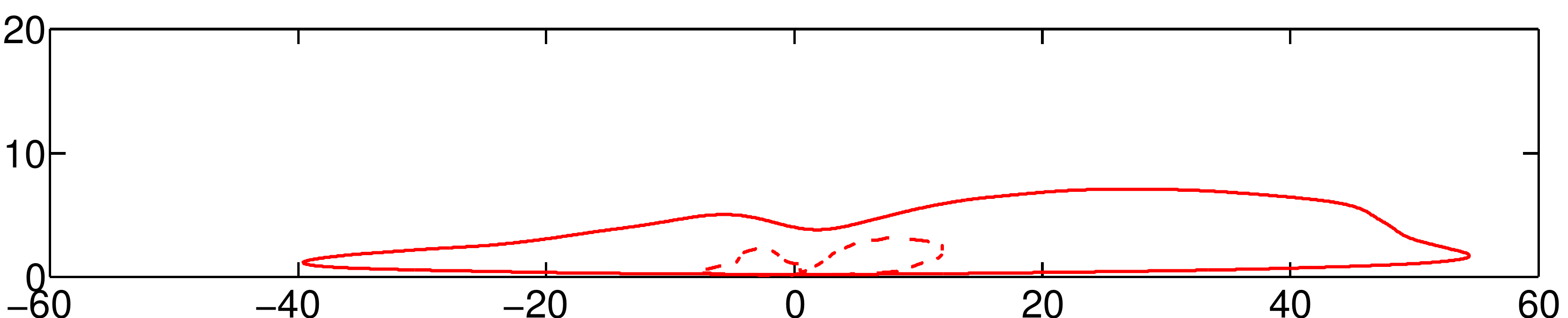}
      \centerline{$ \Deltax / D$} 
    \end{minipage}\\
    \begin{minipage}{2.6ex}
      \rotatebox{90}{\hspace{4ex}$ (y-\ynull) / D$}
    \end{minipage}
    \begin{minipage}{.175\linewidth}
      {($b$)}\\
      \includegraphics[width=1.\linewidth,clip=true]
      {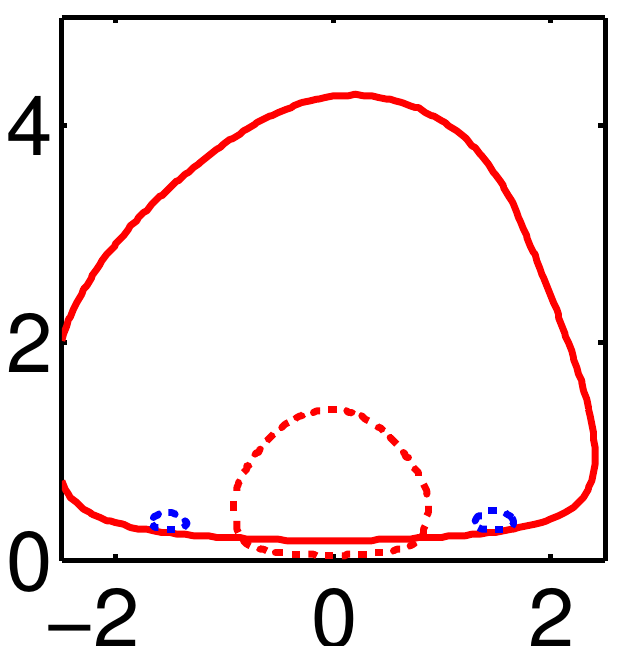}
      \centerline{$\Deltaz / D$} 
    \end{minipage}
  \end{center}
    \caption{\label{fig:corr_paflo_forx_uD}%
      As Fig.~\ref{fig:corr_paflo_forx_u} but axes
      scaled with particle diameter $D$. }   
\end{figure}

Figures~\ref{fig:corr_paflo_forx_u_3D} to \ref{fig:corr_paflo_forx_uD}  
present the
correlation of
particle 
drag fluctuation, 
$F_x^\prime$, with the streamwise
fluid velocity fluctuations, $\uprime$, in case F10 and case F50.
Fig.~\ref{fig:corr_paflo_forx_u_3D} illustrates the 
correlations in form of iso-surfaces at $\pm0.15\Rmax$,
with lengths
scaled by $h$.
Additionally, Fig.~\ref{fig:corr_paflo_forx_u}
and Fig.~\ref{fig:corr_paflo_forx_uD} display the correlation in
form of  
iso-contours of value $\pm0.15\Rmax$ 
in cross-sections of zero streamwise and spanwise separation, 
respectively.  
In Fig.~\ref{fig:corr_paflo_forx_u}
the axes are normalised by $h$, in figure
\ref{fig:corr_paflo_forx_uD} the axes are normalised by $D$.  
%
All figures exhibit positive values of the correlation in the vicinity
of the particle centre.
This reveals that on average structures of positive (negative) 
$\uprime$ in the vicinity of a particle relate to positive (negative)
drag fluctuations. 
The 
surfaces visualized by the chosen isovalue
are of similar size
when scaled with 
$h$ (cf.\ Fig.~\ref{fig:corr_paflo_forx_u_3D} and
Fig.~\ref{fig:corr_paflo_forx_u}) and differ 
between cases F10 and F50 when scaled in $D$
(cf.\ Fig.~\ref{fig:corr_paflo_forx_uD}).  
In the streamwise direction the positively-valued isosurface 
extends over approximately 
$5.5\,h$
(4.5$h$) in 
case F10 (F50), cf.\ Fig.~\ref{fig:corr_paflo_forx_u}$a$,
which corresponds to approximately 100$D$ (22$D$), cf.\
Fig.~\ref{fig:corr_paflo_forx_uD}$a$. 
%
In both cases the positively-valued isosurface of the correlation
between drag and the streamwise velocity extends over a substantial
fraction of the channel height, while larger wall distances are
reached in case F50 (up to $(y-\ynull)/h=0.4$ in case F10 and
$(y-\ynull)/h=0.6$ in case F50).  
In the latter case, the isosurface appears to be somewhat lifted away
from the wall. 
The spanwise size of the region of positive correlation in figure
\ref{fig:corr_paflo_forx_u}($b$) is similar in both cases and 
measures approximately $0.25h$. 
This corresponds to 
$4.5D$ ($2D$) in case F10
(F50) in Fig.~\ref{fig:corr_paflo_forx_uD}($b$).

The uplifted shape of the correlation function in case
F50 is intriguing. 
\citet{Krogsstad_antonia_JFM_94} reported that roughness alters flow
structures close to the roughness surface in that they are more
inclined with respect to the horizontal axis, i.e.\ the angle of the two-point
correlation function was 10$^\circ$ in 
their considered boundary layer above a smooth wall,  
and 38$^\circ$ in the boundary layer above a rough wall.
%
This is thought to be a local effect, limited to the roughness layer
at a wall-distance up to five times the roughness scale and thus does
not contradict Townsend’s wall similarity hypothesis
\citep{Nakagawa_Hanratty_POF_2001,
  Flores_Jimenez_JFM_06, 
  Volino_Schultz_Flack_JFM_2009,
  krogstad_efros_POF_2012}. 
%
On the other hand, it has been observed 
that the large-scale structures of the streamwise 
velocity 
are damped in case F50 compared to the case of a smooth wall 
\citep{Chanbraun_PHD_2012}.
These two observations might be related to
%
the lower correlation 
for $\Deltax/h > 1.5h$ and $(y - \ynull)/h<0.4$ in case F50 (cf.\
figure \ref{fig:corr_paflo_forx_u}$a$) as compared to case F10. 
%

%
Fig.~\ref{fig:corr_paflo_forx_u_3D} shows that
in both cases regions of negative correlation values exist. The shapes
of the negative-valued iso-surfaces are similar between those two
cases, the one of case F10 being slightly more elongated.
In both cases the 
negative-valued iso-surface 
intersects little 
with the plane at 
$\Deltax=0$, 
which leads to the small 
contours at $-0.15\Rmax$  observed
in Fig.~\ref{fig:corr_paflo_forx_u}($b$). 
The centre of the negative-valued region 
is located downstream of the particle
position and at a spanwise distance of
approximately $\Deltaz/h=\pm0.3$.
This corresponds to approximately 55 (70) wall
units in case F10 (F50) and compares well to the 
average distance 
between high-speed and low-speed streaks
commonly found in smooth wall flows
\citep[cf.][]{Smith_Metzler_JFM_1983, Kim_Moin_Moser_JFM_1987}. 

Please observe that the shape of the visualized isosurfaces of the correlation
$(F_x^\prime,u^\prime)$ agrees relatively well for cases F10 and F50, while
the respective maxima of the correlations
  differ by roughly a factor of three from case F50 to case F10 
 under the presently chosen normalization using 
   the friction velocity and the standard deviation
  of drag (cf.\ Table~\ref{tab:corr_paflo_maxval}).  

A clear difference between the characteristics of the correlations in
case F10 and case F50 is the pronounced drop
around the origin
visible in
Fig.~\ref{fig:corr_paflo_forx_u_3D}($b$) and in 
the contour lines in Fig.~\ref{fig:corr_paflo_forx_u}($a$).
A drop is also present in case F10, however to a much smaller
extent. 
%
\begin{figure}
  \begin{center}
    \begin{minipage}{2.6ex}
      \rotatebox{90}{$ (y-\ynull) / h$}
    \end{minipage}
    \begin{minipage}{0.45\linewidth}
      \hspace{-0.9\linewidth}
      \includegraphics[width=1.\linewidth,clip=true]
      {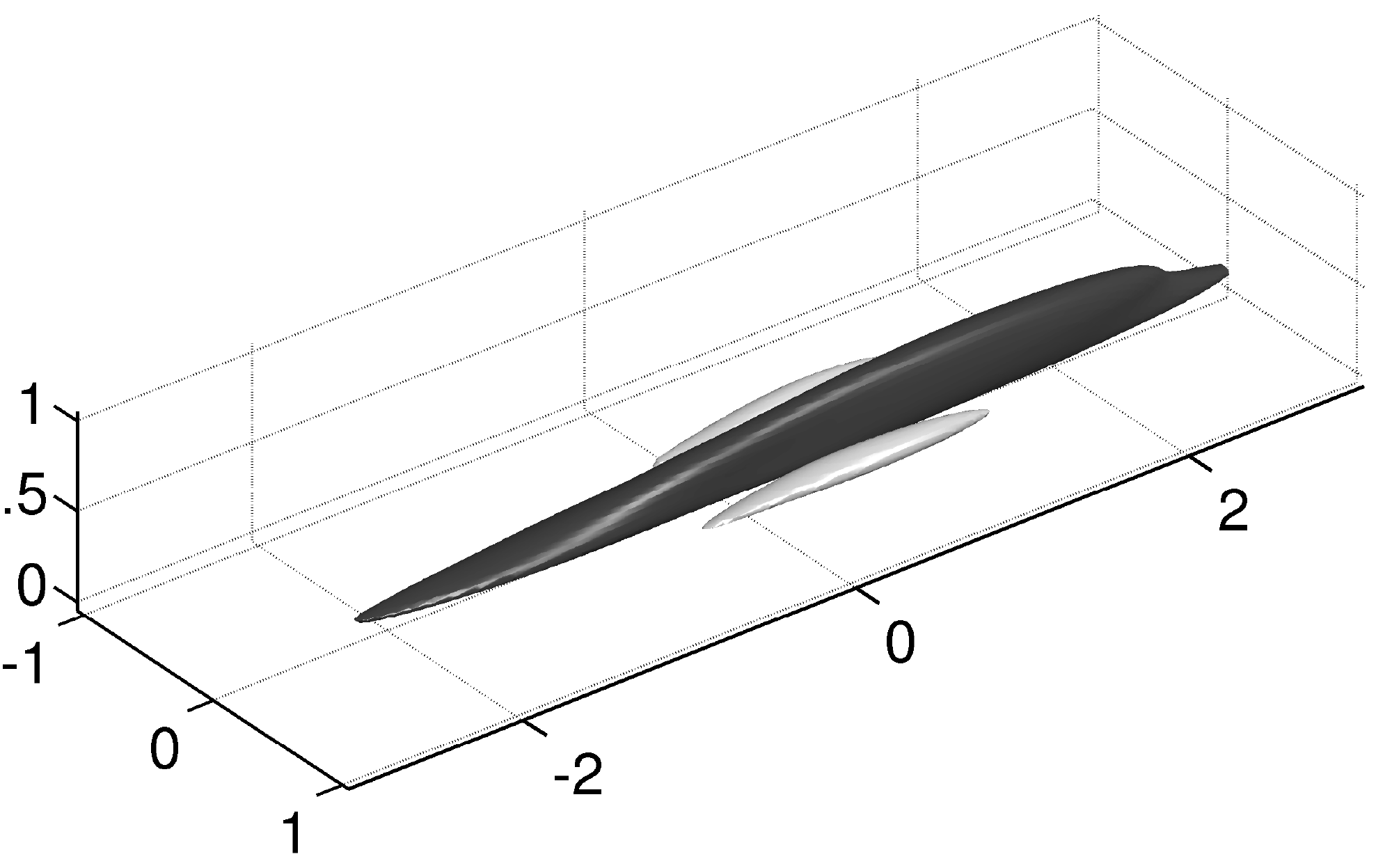}
      \hspace{-0.4\linewidth}\raisebox{0.1\linewidth}{$\Deltax/h$}
      \hspace{-0.6\linewidth}\raisebox{0.5\linewidth}{($a$)}
      \hspace{-0.25\linewidth}\raisebox{0.\linewidth}{$\Deltaz/h$}
    \end{minipage}
    \begin{minipage}{2.6ex}
      \rotatebox{90}{\hspace{0ex}$ (y-\ynull) / h$}
    \end{minipage}
    \begin{minipage}{0.45\linewidth}
      \hspace{-0.9\linewidth}
      \includegraphics[width=1.\linewidth,clip=true]
      {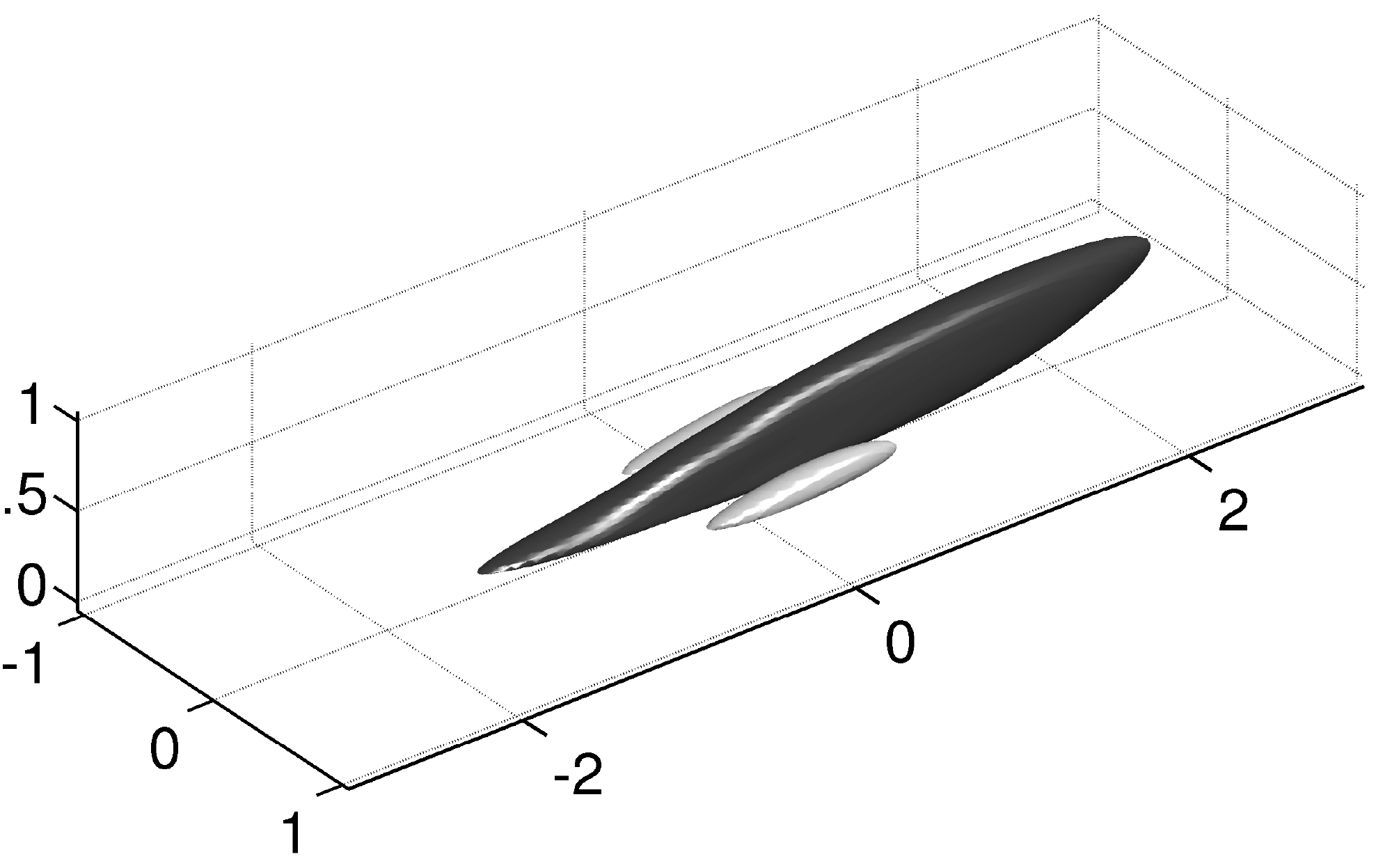}
      \hspace{-0.4\linewidth}\raisebox{0.1\linewidth}{$\Deltax/h$}
      \hspace{-0.6\linewidth}\raisebox{0.5\linewidth}{($b$)}
      \hspace{-0.25\linewidth}\raisebox{0.\linewidth}{$\Deltaz/h$}
      \end{minipage}
    \end{center}
      \caption{\label{fig:corr_paflo_torz_u_3D}%
        Iso-surfaces of
        correlation, $R_{\phi\psi}(\Deltax , y , \Deltaz) / \Rmax$, of the  
      spanwise particle torque fluctuation,  
      $T^\prime_z $, and streamwise velocity
      fluctuation, $u^\prime$ at values  $0.15$ (light) and $-0.15$
      (dark). Panels show case F10 ($a$) and case F50 ($b$).}
\end{figure}
%
\begin{figure}
  \begin{center}
    \begin{minipage}{2.6ex}
      \rotatebox{90}{\hspace{2ex}$ (y-\ynull) / h$}
    \end{minipage}
    \begin{minipage}{0.66\linewidth}
      {($a$)}\\
      \includegraphics[width=1.\linewidth,clip=true]
      {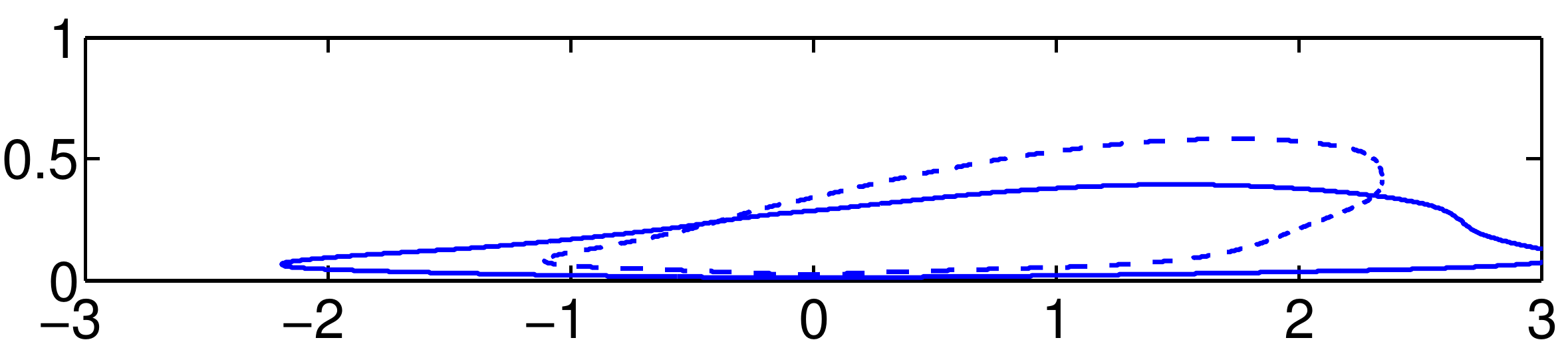}
      \centerline{$ \Deltax / h$} 
    \end{minipage}
    \begin{minipage}{.25\linewidth}
      {($b$)}\\
      \includegraphics[width=1.\linewidth,clip=true]
      {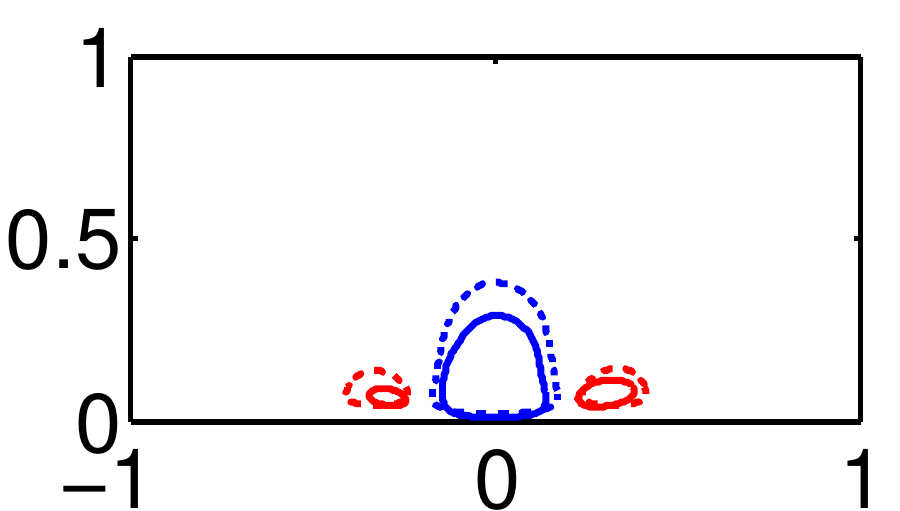}
      \centerline{$\Deltaz / h$} 
    \end{minipage}
  \end{center}
  \caption{\label{fig:corr_paflo_torz_u}%
    Correlation, $R_{\phi\psi}(\Deltax , y , \Deltaz) / \Rmax$, of the  
    spanwise particle torque fluctuation,
    $T^\prime_z $ and streamwise velocity
    fluctuation $u^\prime$.  
    Panels and line styles as in Fig.~\ref{fig:corr_paflo_forx_u}.}   
\end{figure}

Figs.\ \ref{fig:corr_paflo_torz_u_3D} and \ref{fig:corr_paflo_torz_u}
show the 
correlation between spanwise torque fluctuations acting on the
particle, $T_z^\prime$, and streamwise fluid velocity fluctuations,
$\uprime$.  
From Fig.~\ref{fig:corr_paflo_torz_u_3D} it can be observed 
that the visualized iso-surface of the correlation has a similar shape
as the 
correlation of
$F_x^\prime$ and  $\uprime$ shown in
Fig.~\ref{fig:corr_paflo_forx_u_3D}, albeit at 
opposite sign. 
The shown iso-surfaces are again elongated in the 
streamwise direction and of
sizes much larger than the particle diameter. 
%
Therefore, 
structures of positive (negative) 
$\uprime$ relate on average to negative (positive) spanwise torque
fluctuations on a particle.
%
Likewise, the regions of 
negative correlation are flanked by smaller
streamwise elongated regions of correlation with
a positive 
sign.
%
Again,
the identified region in figure
\ref{fig:corr_paflo_torz_u}($a$) 
shortens in its streamwise 
extent
from case F10 to case F50 and
appears to be more inclined to
the $x$-axis in case F50.  
The main difference between the correlation
$(T_z^\prime,u^\prime)$
(Fig.~\ref{fig:corr_paflo_torz_u})
 and the correlation  
$(F_x^\prime,u^\prime)$
(Fig.~\ref{fig:corr_paflo_forx_u}) 
is the absence of the drop around the origin in the former.
\begin{figure}
  \begin{center}
    \begin{minipage}{2.6ex}
      \rotatebox{90}{$ (y-\ynull) / h$}
    \end{minipage}
    \begin{minipage}{0.43\linewidth}
      \hspace{-0.9\linewidth}
      \includegraphics[width=1.\linewidth,clip=true]
      {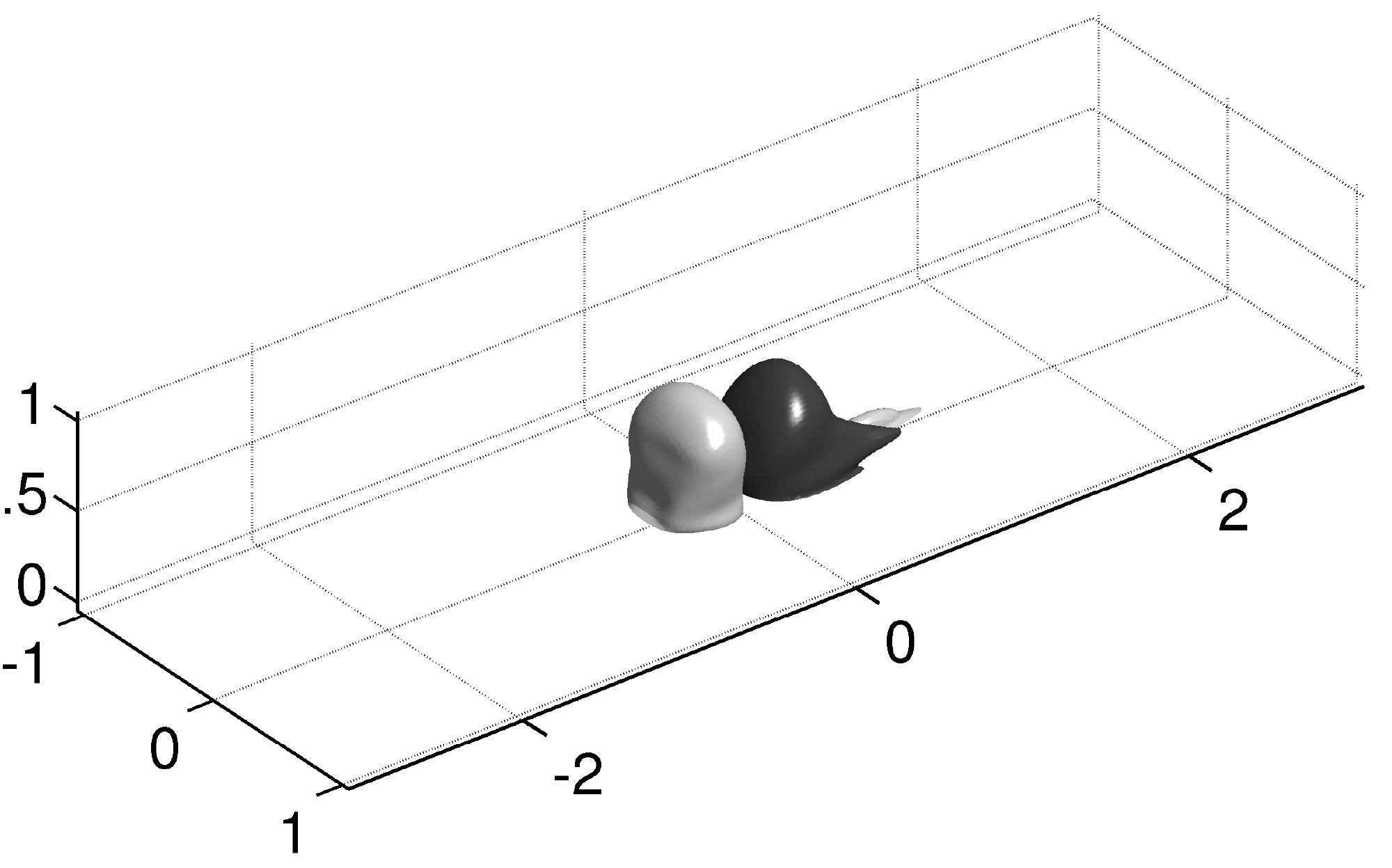}
      \hspace{-0.4\linewidth}\raisebox{0.1\linewidth}{$\Deltax/h$}
      \hspace{-0.6\linewidth}\raisebox{0.5\linewidth}{($a$)}
      \hspace{-0.25\linewidth}\raisebox{0.\linewidth}{$\Deltaz/h$}
    \end{minipage}
    \begin{minipage}{2.6ex}
      \rotatebox{90}{\hspace{0ex}$ (y-\ynull) / h$}
    \end{minipage}
    \begin{minipage}{0.43\linewidth}
      \hspace{-0.9\linewidth}
      \includegraphics[width=1.\linewidth,clip=true]
      {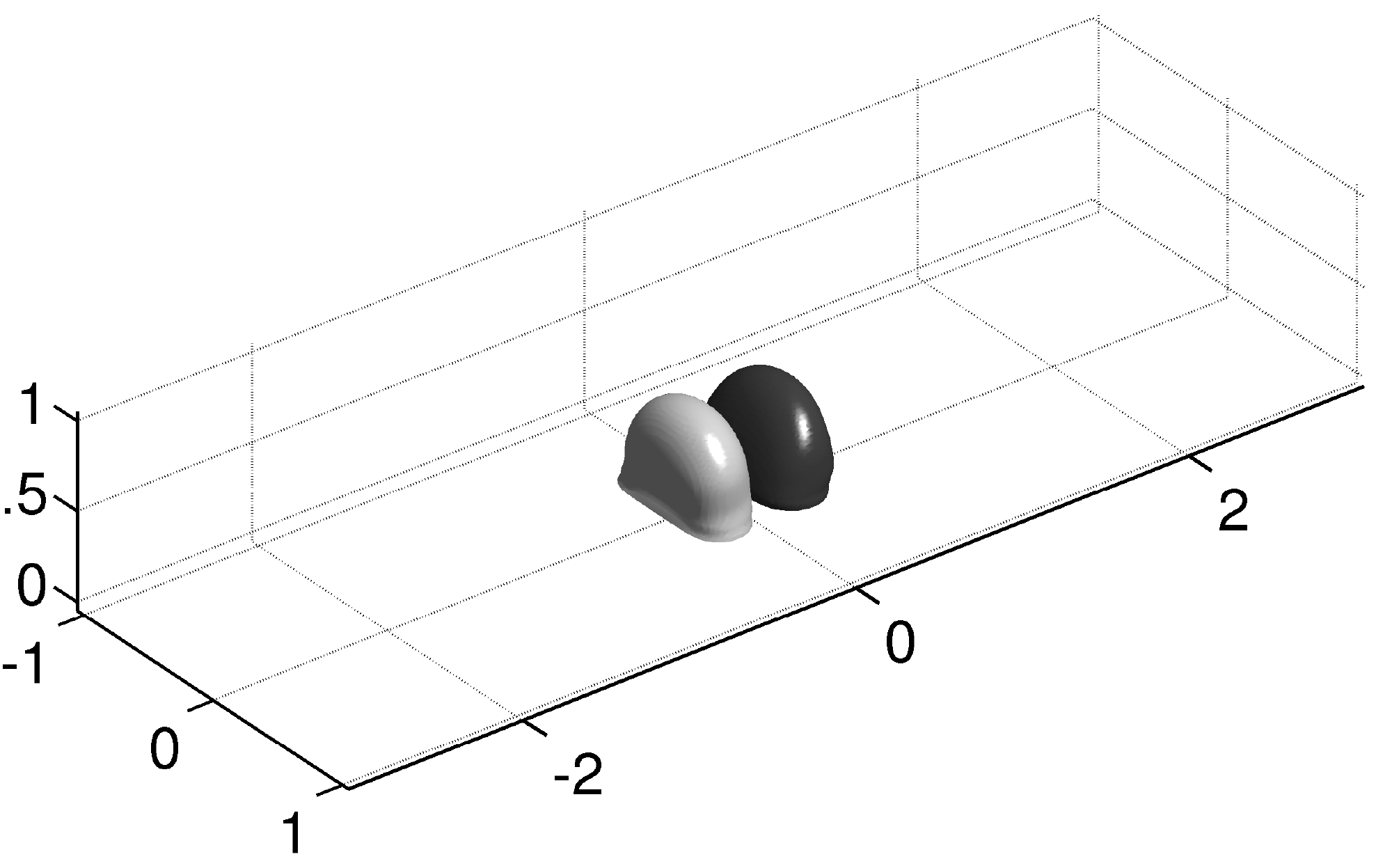}
      \hspace{-0.4\linewidth}\raisebox{0.1\linewidth}{$\Deltax/h$}
      \hspace{-0.6\linewidth}\raisebox{0.5\linewidth}{($b$)}
      \hspace{-0.25\linewidth}\raisebox{0.\linewidth}{$\Deltaz/h$}
      \end{minipage}
  \end{center}
    \caption{Iso-surfaces of
      correlation, $R_{\phi\psi}(\Deltax , y , \Deltaz) / \Rmax$, of the  
      streamwise particle force fluctuation,  
      $F^\prime_x $, and pressure fluctuation, $\pprime$ at values
      $0.15$ (light) and $-0.15$ 
      (dark). Panels show case F10 ($a$) and case F50 ($b$).}   
    \label{fig:corr_paflo_forx_p_3D}
\end{figure}
%
\begin{figure}
  \begin{center}
    \begin{minipage}{2.6ex}
      \rotatebox{90}{\hspace{2ex}$ (y-\ynull) / h$}
    \end{minipage}
    \begin{minipage}{0.66\linewidth}
      \includegraphics[width=1.\linewidth,clip=true]
      {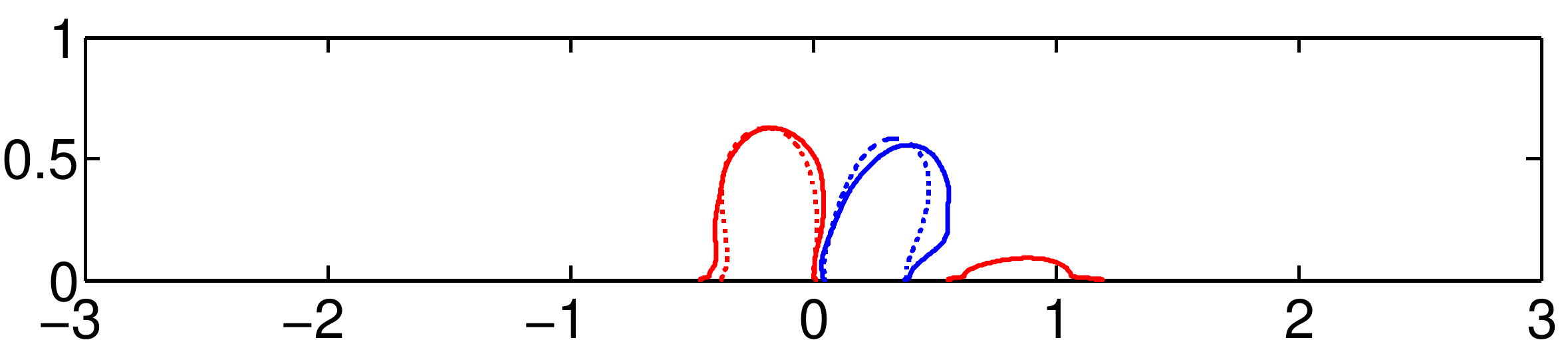}
      \centerline{$ \Deltax / h$} 
    \end{minipage}
  \end{center}
    \caption{\label{fig:corr_paflo_forxz_p}
      Correlation, $R_{\phi\psi}(\Deltax , y , \Deltaz) /
      \Rmax$, of the   
      streamwise particle force fluctuation, 
      $F^\prime_x $, and pressure
      fluctuation, $p^\prime$,
      at zero spanwise shift, $\Deltaz=0$.
      Line styles as in Fig.~\ref{fig:corr_paflo_forx_u}.
    }   
\end{figure}

Fig.~\ref{fig:corr_paflo_forx_p_3D} shows the correlation of 
particle drag fluctuations, $F_x^\prime$, 
with 
the fluctuating pressure field, $\pprime$. 
%
Fig.~\ref{fig:corr_paflo_forxz_p}($a$) shows the corresponding
iso-contours at zero spanwise separation, i.e.\ $\Deltaz=0$,
in outer length scales.
%
%
In both flow cases, we observe two bulges corresponding to iso-surfaces
of opposite sign located upstream (positive correlation value)
and downstream of the particle (negative value). In contrast to the 
previously presented correlations, the observed shapes are not
streamwise elongated and do not present significant differences
between flow cases F10 and F50.
%
%
Therefore, it is found that, on average, a negative (positive) pressure
gradient across the particle relates to positive (negative) drag
fluctuations.  
%
%
%
%
%
Note, that in the three-dimensional time-averaged three-dimensional 
pressure field around a particle a similar formation of 
high and low values upstream and
downstream of the particle exists
{as a consequence of the mean streamline curvature very close to
  the roughness elements}
\citep[][]{chanbraun_garciavillalba_uhlmann_JFM_2011}. 
The dimensions of the present iso-surfaces, however,
 are significantly larger than the
particle diameter, e.g.\
10$D$ (3$D$) in wall-normal direction for case F10 (F50) and therefore
do not stem from the time-averaged pressure field around the
particles.  
Note that the bulge-like iso-surfaces found here for the correlation
$(F_x^\prime,u^\prime)$ are approximately located in the region where a notable
drop (case F50) in the correlation amplitudes was observed when
correlating the same force component $(F_x^\prime)$ with the streamwise
velocity $(u^\prime)$ (cf.\ Fig.~\ref{fig:corr_paflo_forx_u}).  

\section{Conclusion\label{sec_concl}}
We have further analyzed the results of the direct numerical
simulation of open channel flow over a geometrically rough wall 
of Ref.~\onlinecite{chanbraun_garciavillalba_uhlmann_JFM_2011},  
focusing on the spatial and temporal structure of
the hydrodynamic force and torque acting on the wall-mounted spheres 
as well as on their correlation with the surrounding flow field. 

In the simulations of
Ref.~\onlinecite{chanbraun_garciavillalba_uhlmann_JFM_2011}
two flow cases were considered. 
In the first one, F10, the spheres are small and the flow regime is 
hydraulically smooth. 
Here it has been shown that the spatial and the temporal statistics of
the spanwise torque acting on the spheres agree well with 
those of the shear stress on a smooth wall.  
In the second case, F50, the spheres are approximately three times
larger, leading to a transitionally rough flow regime. 
It is found that in this case the similarity between the spatial and
temporal characteristics of the spanwise torque acting on the
wall-mounted spheres and those of the shear stress on a smooth wall is
less pronounced. 

The spatial and temporal structure of drag presents discrepancies with
the structure of the shear stress on a smooth wall already for case F10. 
The differences are more evident
in case F50, 
as can be expected. 
The auto-correlations of drag present local minima for both
 cases, F10 and
F50, minima which  
are not present in the correlation of shear stress on a smooth wall. 
Also for the lift, for case F50, a local minimum is present.  

Lift and drag are correlated with a shift in time. 
This result appears to be very robust, since 
we have compared the temporal cross-correlation from the two flow
cases considered in the  
present study to experimental data
obtained at different flow conditions (higher Reynolds number) and
different roughness geometries 
(cube positioned in natural gravel, spheres in hexagonal packing). The
temporal location of the 
local extrema of the cross-correlation 
between lift and drag  agrees well in all cases, when scaled with the
bulk velocity and the flow depth.  
Some differences are observed in the amplitude 
of the extrema, 
which might be linked to the
differences between the setups and the way measurements were
performed.   

By using the convection velocity of the force (or torque)
fluctuations, it has been possible to  
convert, using the Taylor hypothesis, the spatial two-point
correlation 
into a 
temporal auto-correlation. 
It has been shown that the Taylor hypothesis works relatively well in
both flow cases. The convection 
velocities computed in the present study follow trends reported in
previous studies.  
First, it has been shown in flow over smooth walls that pressure
fluctuations travel somewhat  
faster than shear stress fluctuations. In the present case, drag and
lift fluctuations, which are by  
definition influenced by pressure fluctuations, have been found to
travel somewhat faster than  
torque fluctuations, which are unaffected by pressure. 
Second, it can be expected that convection velocities close to the
rough wall become smaller when increasing the
roughness effect.
Consistently, we have observed a reduction of the convection velocity
from case F10 (hydraulically smooth) to F50 (transitionally rough).

Finally, the analysis of the correlation between the 
streamwise velocity
field and the 
drag/spanwise torque
acting on the particles has revealed that the 
spatial structure of
such correlation functions is reminiscent of buffer-layer streaks. 
For both flow cases, the correlation between the streamwise velocity
fluctuations 
and the drag/spanwise-torque presents an elongated shape, which scales
with the flow depth and 
not with the size of the particle. In case F50,
where the effect of roughness
is more significant than in case F10,
 the selected isosurfaces are somewhat lifted away 
from the wall compared to case F10. Additionally, in case F50  
 the correlation between the streamwise velocity
fluctuation and drag presents a drop above the particle which is not
present in the correlation 
between the streamwise velocity fluctuation and torque. 
The correlation between the pressure 
fluctuations and the drag has also been discussed. It has been shown
that this correlation 
is not elongated and consists of two short bulges located  upstream
and downstream of the particle. 
%
\begin{acknowledgments}
This work was supported by the German Research Foundation (DFG) under
project JI~18/19-1.
The computations have been carried out at the Steinbuch Centre for
Computing (SCC) of Karlsruhe Institute of Technology and at the
Leibniz Supercomputing Centre (LRZ) of the Bavarian Academy of
Sciences and Humanities. The support from these institutions is
gratefully acknowledged. 
\end{acknowledgments}
%
\clearpage
%
%

%
\end{document}